\documentclass[%
 aip,
 jcp,
 amsmath,amssymb,
 reprint,%
]{revtex4-1}

\usepackage{graphicx}% Include figure files
\usepackage{dcolumn}% Align table columns on decimal point
\usepackage{bm}% bold math

\usepackage[utf8]{inputenc}
\usepackage[T1]{fontenc}
\usepackage{mathptmx}

\usepackage{gensymb}
\usepackage{placeins} %mine

\usepackage{hyperref}
\hypersetup{
    colorlinks,
    citecolor=black,
    filecolor=black,
    linkcolor=black,
    urlcolor=black
}

\usepackage{xcolor}
\newcommand*{\citen}[1]{%
  \begingroup
    \romannumeral-`\x % remove space at the beginning of \setcitestyle
    \setcitestyle{numbers}%
    \cite{#1}%
  \endgroup   
}

\begin{document}

%\preprint{AIP/123-QED}

\title[Characterising DNA T-motifs by Simulation and Experiment]{Characterising DNA T-motifs by Simulation and Experiment\vspace{0.5em}}% Force line breaks with \\

%\thanks{Footnote to title of article.}

\author{Behnam Najafi}
%\email{behnam.najafi@physics.ox.ac.uk}
\affiliation{ 
Clarendon Laboratory, Department of Physics, University of Oxford, Parks Rd, Oxford OX1 3PU, United Kingdom
}
\author{Katherine G. Young}
%\email{katherine.young@physics.ox.ac.uk}
\affiliation{ 
Clarendon Laboratory, Department of Physics, University of Oxford, Parks Rd, Oxford OX1 3PU, United Kingdom
}
\author{Jonathan Bath}
\affiliation{ 
Clarendon Laboratory, Department of Physics, University of Oxford, Parks Rd, Oxford OX1 3PU, United Kingdom
}
\author{Ard A. Louis}
\affiliation{ 
Rudolph Peierls Centre for Theoretical Physics, University of Oxford, Parks Rd, Oxford OX1 3PU, United Kingdom
}
\author{Jonathan P. K. Doye}
\affiliation{ 
Physical and Theoretical Chemistry Laboratory, Department of Chemistry, University of Oxford, South Parks Road, Oxford OX1 3QZ, United Kingdom
}
\author{Andrew J. Turberfield}
\email{andrew.turberfield@physics.ox.ac.uk}
\affiliation{ 
Clarendon Laboratory, Department of Physics, University of Oxford, Parks Rd, Oxford OX1 3PU, United Kingdom
}
 \homepage{}

\date{\today}% It is always \today, today,
             %  but any date may be explicitly specified

\begin{abstract}

The success of DNA nanotechnology has been driven by the discovery of novel structural motifs with a wide range of shapes and uses. We present a comprehensive study of the T-motif, a 3-armed, planar, right-angled junction that has been used in the self-assembly of DNA polyhedra and periodic structures. The motif is formed through the interaction of a bulge loop in one duplex and a sticky end of another. The polarity of the sticky end has significant consequences for the thermodynamic and geometrical properties of the T-motif: different polarities create junctions spanning different grooves of the duplex. We compare experimental binding strengths with predictions of oxDNA, a coarse-grained model of DNA, for various loop sizes. We find that, although both sticky-end polarities can create stable junctions, junctions resulting from 5$'$ sticky ends are stable over a wider range of bulge loop sizes. We highlight the importance of possible coaxial stacking interactions within the motif and investigate how each coaxial stacking interaction stabilises the structure and favours a particular geometry.

%Valid PACS numbers may be entered using the \verb+\pacs{#1}+ command.
\end{abstract}

%\pacs{Valid PACS appear here}% PACS, the Physics and Astronomy
                             % Classification Scheme.
%\keywords{DNA Nanotechnology, DNA Origami, Self-assembly, DNA T-junctions}%Use showkeys class option if keyword
                              %display desired
\maketitle

%%%%%%%%%%%%%%%%%%%%%%%%%	 INTRODUCTION	%%%%%%%%%%%%%%%%%%%%%%%%%%

\section{Introduction}\label{sec::intro}

\begin{figure*}[]
	\centering
		\includegraphics[width=0.95\textwidth]{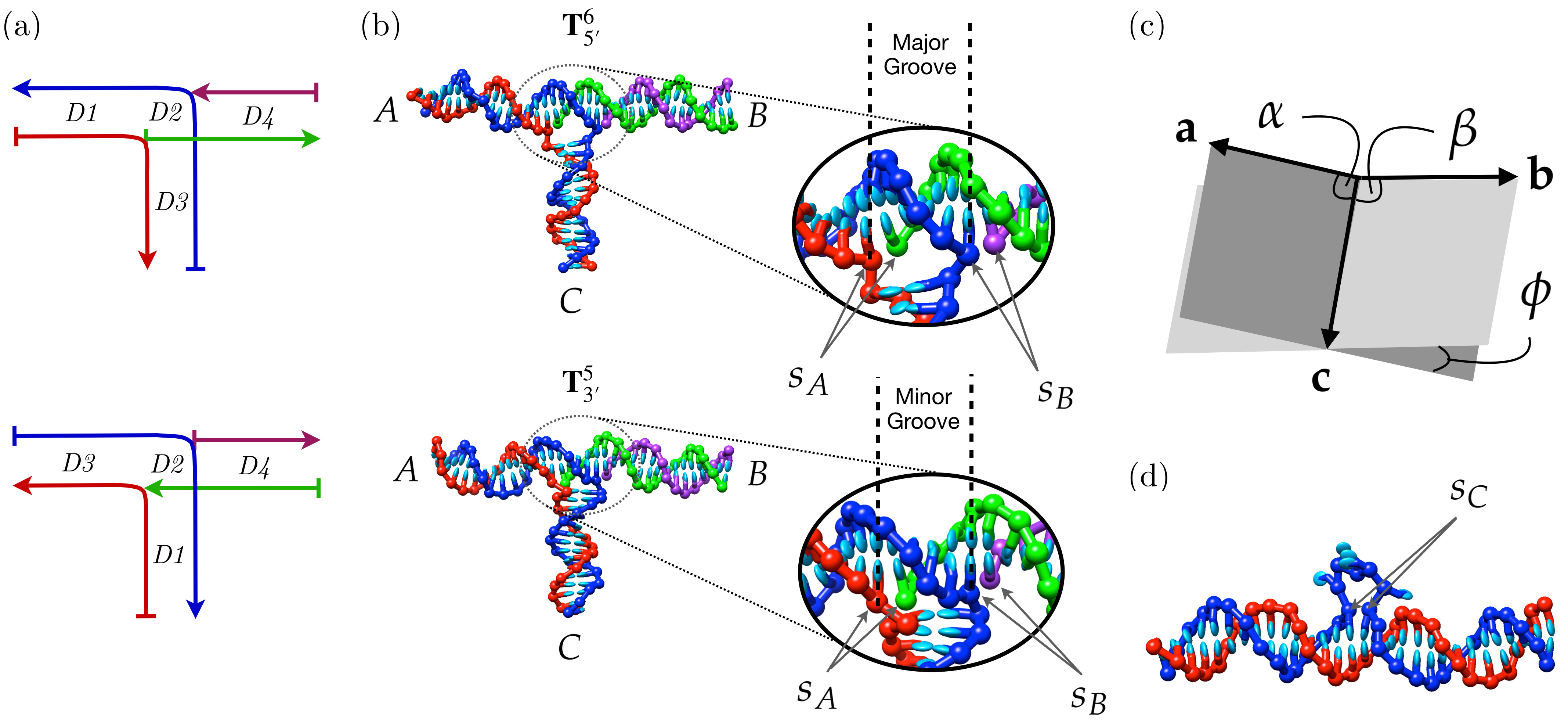}
		\caption{
			(a) Schematic representations of T-motifs, highlighting the four domains of interaction ($D1$-$D4$) between the four component strands. The coloured arrows indicate the polarities of the strands from 5$^\prime$ to 3$^\prime$. The junction is formed between two duplexes, one ($D4$) having a sticky end and one ($D1D3$) incorporating a bulge loop. The T-motif is formed through the hybridization of the sticky end and bulge loop in the junction domain $D2$. The geometry of the junction depends on the polarity of the sticky end. In the case of a 5$^\prime$ sticky end, the side branch of the junction is duplex $D3$, which straddles the major groove of the ``nicked'' duplex formed by coaxially stacked domains $D1 D2 D4$; in the case of a 3$^\prime$ sticky end, side branch $D1$ straddles the minor groove of the ``nicked'' duplex $D3 D2 D4$. We use the notation $\mathbf{T}_{\mathcal{P}}^{L}$ to describe a T-motif with a $D2$ domain of size $L$ base pairs and sticky end polarity $\mathcal{P}\in\{3^\prime,5^\prime\}$.
                    	(b) A T-motif comprises a side-branch duplex $C$ straddling a doubly nicked duplex $AB$; the positions of the two nicks delimit the junction domain $D2$. Snapshots from oxDNA simulations of fully formed T-motifs are shown, highlighting coaxial stacking sites $s_A$ and $s_B$ on either side of $D2$.  
                    	(c) Scheme used to characterise the geometry of T-motifs. Vectors  $\bold{a}$, $\bold{b}$ and $\bold{c}$, which correspond to the orientations of the three arms of the junction, are calculated by summing over base-normal vectors of the nucleotides in the 11 middle base pairs of the corresponding domain. We define angles $\alpha$ and $\beta$ as positive for $(\bold{a} \times \bold{c}) \cdot (\bold{b} \times \bold{c}) < 0$ and $\phi$ as positive for $[(\bold{a} \times \bold{c}) \times (\bold{b} \times \bold{c})] \cdot \bold{c} < 0$. All three angles have positive values in the configuration shown here.
                    	(d) A third coaxial stacking interaction, at site $s_C$, can occur between the base pairs of $D1$ and $D3$ that flank the bulge; this configuration competes with the formation of the T-motif, especially for shorter bulge loops.
		}
	\label{fig::Single}
\end{figure*}

Over the past few decades, DNA has been used extensively as a material for bottom-up nano-fabrication. \cite{seeman2018} The robust Watson-Crick base pairing rules enable a target structure to be encoded as the global free-energy minimum. Inspired by naturally occurring Holliday junctions, \cite{holliday1964} DNA nanotechnology began with the sequence design of short, single-stranded oligonucleotides that hybridize to form immobile four-way junctions: \cite{seeman1982,seeman1983} sticky ends at the end of the branching duplexes allow inter-motif binding. Further advances have led to the creation of more rigid structural motifs such as the double-crossover molecule (DX tile),\cite{seeman1993} the basis of many two-dimensional structures, \cite{winfree1998,liu2005,zheng2009} and DNA polyhedra. \cite{shih2004, goodman2005, zhang2008, he2008, zhang2009, he2010, Iinuma2014} The DNA origami \cite{rothemund2006} and DNA brick \cite{wei2012, ke2012} techniques have enabled the creation of larger structures with intricate twists and curves.\cite{ke2009,douglas2009,dietz2009,han2011,liedl2010} These techniques may be combined, for example in the growth of tile-based crystals using DNA origami seeds. \cite{schulman2009}

The success of DNA nanotechnology has stimulated the search for a greater variety of structural motifs. Current motifs can be broadly classed into tiles,\cite{seeman1993,labean2000} branched junctions,\cite{seeman1983,zheng2009,wang2019,zhang2015,benson2015,hong2016} and two- and three-dimensional blocks of parallel, densely-packed helices.\cite{rothemund2006,douglas2009,dietz2009,ke2012} Most structures make extensive use of reciprocal strand exchange to connect parallel helices and nearly all inter-motif connections are based on sticky-end cohesion, with the notable exceptions of paranemic crossover,\cite{zhang2002,qian2012} edge-sharing,\cite{yan2001} and stacking interactions.\cite{gerling2015} Recently, an interesting motif was presented by Hamada et al \cite{hamada2009} which does not rely on reciprocal strand exchange between parallel duplexes and produces a right-angled geometry resembling the letter T. DNA T-motifs consist of one duplex with a bulge loop and a second duplex with a single-stranded (3$^\prime$ or 5$^\prime$) sticky end. Hybridization of the bulge loop and the sticky end leads to a well-defined, planar geometry (Figure \ref{fig::Single}). T-motifs have been used to create a variety of periodic \cite{hamada2009,li2017} and finite structures \cite{mao2012,mao2015} and to perform computations in algorithmic assembly schemes.\cite{tandon2019} To the best of our knowledge, all of the structures presented in the literature make use of 5- or 6-nucleotide sticky ends with 5$^\prime$ polarity; however, the reason for this choice is not immediately clear. Our aim is to present a comprehensive study of T-motifs to understand the underlying interactions at the junction and provide a useful guide to incorporating these motifs in larger structures.

In this paper, we investigate the thermodynamic stabilities and conformations of T-motifs experimentally and by means of a coarse-grained model, oxDNA.\cite{ouldridge2011,sulc2012,doye2013,snodin2015} While all-atom simulations have recently been used to study a variety of DNA systems,\cite{yoo2013,gopfrich2016,wu2013,lee2017} their extreme computational costs limit their scope to specific cases. On the other hand, coarse-grained models whose fundamental units are larger than a nucleotide\cite{penna2010,castro2011} are unable to capture important processes involved in T-motif formation such as hybridization, duplex fraying and coaxial stacking. OxDNA provides a description of DNA interactions at the nucleotide level and has been highly successful at reproducing structural, mechanical and thermodynamic properties of single-stranded and duplex DNA. The model has been applied successfully to probe the kinetics of hybridization and toehold-mediated strand displacement \cite{ouldridge2013,srinivas2013,mosayebi2014} as well as a variety of dynamic systems including DNA tweezers,\cite{ouldridge2010} walkers,\cite{ouldridge2013_walker,sulc2014} and origami\cite{snodin2016} and brick assembly.\cite{fonseca2018} A recent study \cite{schreck2015} using oxDNA quantified the flexibility induced by bulge-loops of different sizes in DNA duplexes, predicting higher bend angles for larger bulge-loops. Crucially, oxDNA has been recently adapted to include the difference in groove widths of duplex DNA,\cite{snodin2015} making it well suited to study the geometries of T-motifs.

%%%%%%%%%%%%%%%%%%%%%%%%%	     METHODS	   %%%%%%%%%%%%%%%%%%%%%%%%%%%

%\FloatBarrier
%\newpage
\section{Methods}\label{sec::methods}

\subsection{Simulation Methods} \label{sec::oxDNA}

In oxDNA,\cite{snodin2015} each nucleotide is represented as a rigid body that interacts pairwise with other nucleotides through three interaction sites: (a) a backbone connectivity site, (b) a stacking and coaxial stacking site, and (c) a hydrogen-bonding and cross-stacking site. The interaction potential also includes an excluded volume interaction and a Debye-H\"uckel term describing screened electrostatic interactions that takes into account different salt conditions. A more detailed description of the model can be found in Supplementary Material SI A but we discuss some relevant aspects of the model here.

There are several reasons why oxDNA is well suited to this study. First, the model includes an explicit description of stacking and can describe different helical and non-helical conformations of single-stranded DNA\cite{ouldridge2011}. It is crucial that the description of such conformations is accurate when investigating the interactions between the sticky end and the bulge loop at the T-motif junction. Second, the model has been parametrized to reproduce the melting temperatures of thousands of DNA sequences \cite{sulc2012} and can provide a good description of hybridization thermodynamics including the tendency of duplexes to fray close to their melting temperature and the breaking of base pairs caused by extreme stresses. Importantly, parameters describing interactions involving base pairing, base stacking along a strand, and cross stacking along duplexes are all sequence-dependent, allowing direct comparison to our experiments. Third, the most recent version of the model has been adapted to include a distinct assymetry between the major and minor grooves, \cite{snodin2015} providing a more realistic geometric description of duplex DNA. This feature is crucial to studying T-motifs, since their properties are highly sensitive to the geometric fit of the branch duplex to the helical groove that it spans.

The model does incorporate some simplifications that are important for this study. First, coaxial stacking interactions across backbone ``nicks'' are not sequence-dependent; instead, an average-base parametrization is used for these interactions. These interactions are important in T-motif formation but oxDNA cannot provide any insights into the effect of changing the identity of the bases that flank the ends of the junction. Second, the simple treatment of electrostatics in oxDNA cannot capture ion-specific effects. The parametrisation of the Debye-H\"uckel term reproduces the dependence of duplex melting temperatures on monovalent [Na$^+$] ions. Therefore, local effects of charges at the T-motif junctions may not be fully described by the model. Third, only interactions between canonical Watson-Crick (A-T/G-C) pairs are taken into account in the oxDNA potential, ignoring other interactions such as sugar-edge bonds and Hoogsteen pairing. These interactions have been shown to affect the flexibility of bulge loop nucleotides in RNA\cite{macchion2008} and may play a similar role in DNA.

Standard Monte Carlo algorithms are inefficient for sampling systems of strongly interacting particles with short-range interactions. We use the virtual move Monte Carlo method\cite{whitelam2008} (VMMC) to allow trial collective movements of clusters by iteratively adding nearby nucleotides to a growing cluster with probabilities that depend on the resultant changes in energy. The collective movement is then accepted with a probability that ensures correct sampling from the canonical ensemble. The algorithm is particularly suited to DNA nanostructures because collective motion significantly enhances the sampling speed of conformations of a strand or aggregates of strands. In addition, we use the umbrella sampling technique \cite{torrie1977} to apply biases to our simulations in order to flatten energy barriers and sample transitions between meta-stable states. In this work the bias is applied over two order parameters: the end-to-end distance between the bulge loop and sticky-end, to quantify the entropy loss on binding; and the number of base pairs formed between the bulge-loop and the sticky-end, to quantify the free energy change during hybridization (see Supplementary Material SI B).

All simulations were run at 25$\degree$C and salt conditions of [Na$^+$] = 0.5M, with all stands at a concentration of 42 $\mu$M. This relatively high concentration is used for computational efficiency. The results can be extrapolated to other concentrations by a thermodynamically appropriate scaling of the (unbiased) partition function of bound states relative to that of the unbound states. \cite{Ouldridge2010bulk} Strand sequences and precise simulation details can be found in Supplementary Material SI C.%\jpkd{Mention use of Tom's finite system corrections?}

%%%%%%%%%%%%%%%%%%%%%%%%%%%%%
\subsection{Experimental Methods} \label{sec::experiment}

To form a T-motif, the four component strands were cooled from 95 to 25$\degree$C at a rate of 1$\degree$C/min, in a TAE/Mg$^{2+}$ buffer (40mM Tris base, 1mM EDTA, 20mM acetic acid, and 12.5mM MgCl$_2$, pH 8.5$\pm$0.2). Polyacrylamide gel electrophoresis (PAGE) was used to assess the conformations of the annealed complexes. A fluorescent label was conjugated to the 5$'$ end of the strand that contains the bulge loop (see Supplementary Material SII A for full sequences).  

In order to obtain more accurate measurements of changes in electrophoretic mobility, standard reference markers were used. Our protocol (Supplementary Material SII B) corrects for variability between gel lanes and provides a reliable way of measuring the electrophoretic mobility of each sample relative to reference markers as $M_{\text{sample}} = (\mu_{\text{HDM}} - \mu_{\text{sample}}) / (\mu_{\text{HDM}} - \mu_{\text{LDM}})$, where $\mu_{\text{sample}}$, $\mu_{\text{HDM}}$ and $\mu_{\text{LDM}}$ are the mobilities of the sample, and the high-mobility and low-mobility markers, respectively. 

In order to calculate T-motif dissociation constants (Section \ref{sec::ExpRes}) the concentrations of strands within the electrophoresis gel were estimated on the basis of the approximate volume occupied by a gel band (see Supplementary Material SII B).

%%%%%%%%%%%%%%%%%%%%%%%%%	     RESULTS	   %%%%%%%%%%%%%%%%%%%%%%%%%%%

\section{Results and Discussion}\label{sec::results}

\subsection{Simulation Results}\label{sec::oxDNAres}

\begin{figure}[t]
		\includegraphics[trim=0.9cm 1cm 0cm 0cm, width=1.0\linewidth]{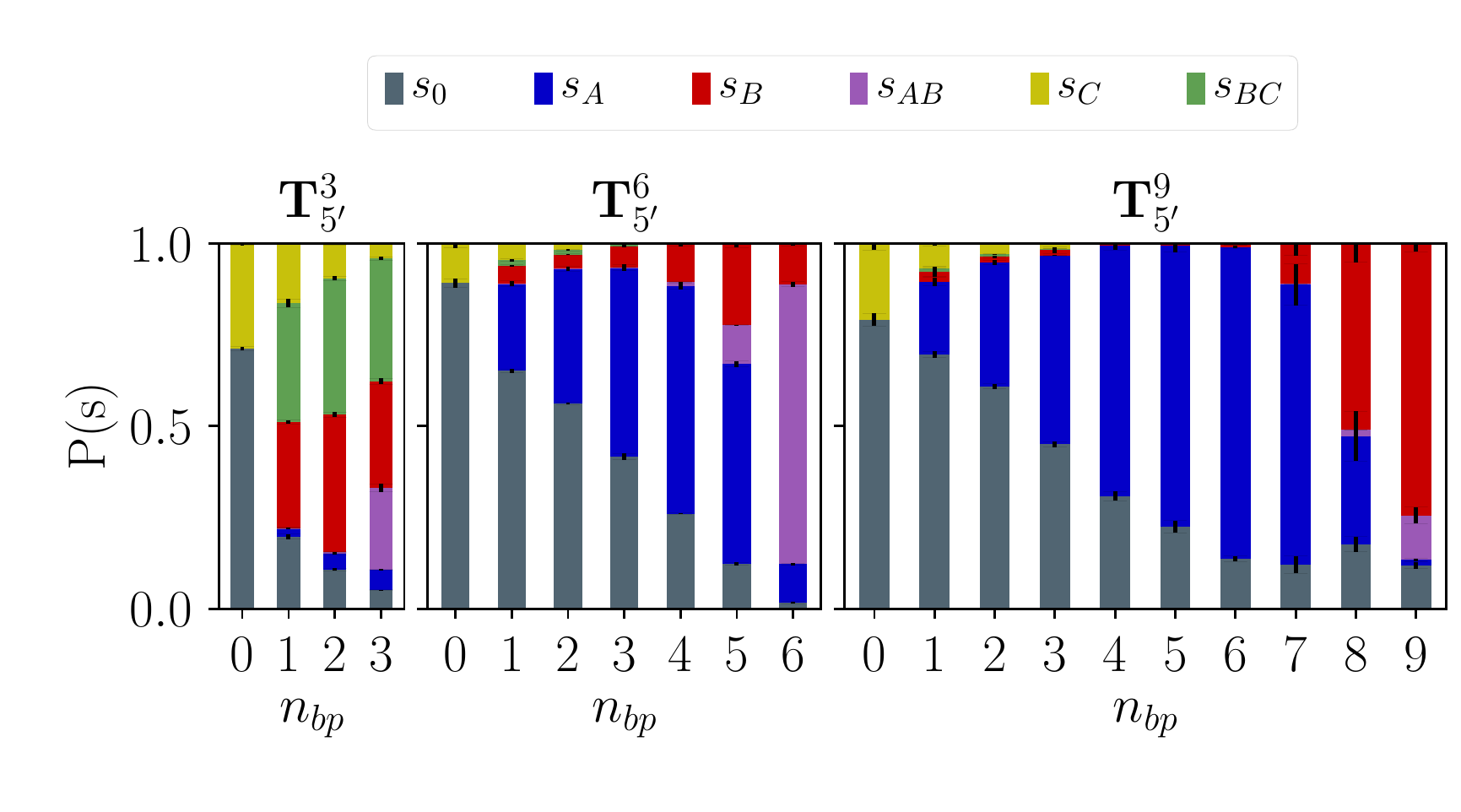}
	\caption{
		Coaxial stacking probabilities, as functions of the number of base pairs formed between sticky end and bulge loop, during the formation of 5$^\prime$ T-motifs with matched bulge / sticky-end lengths. The three cases shown have interaction domains that are relatively short ($L$=3), ideal ($L$=6), and large ($L$=9). Error bars represent the standard error of the mean computed from configurations in five independent VMMC trajectories.
		}
	\label{fig::Stackn}
\end{figure}

As shown in Figure \ref{fig::Single}, a T-motif formed by the hybridization of a bulge loop in one duplex with a terminal sticky end on another creates a three-arm junction. Two of the arms form an approximately continuous duplex. The third arm comprises one of the double-helical domains flanking the bulge: its two component strands branch off at nicks in the backbone on opposite sides of one of the helical grooves of the quasi-continuous duplex. The geometry of the B-DNA double-helix implies that the junction will span a different groove depending on the polarity, $\mathcal{P}\in\{3',5'\}$, of the sticky end. We use the notation $\mathbf{T}_{\mathcal{P}}^{L}[n_{bp},s]$ to specify a T-motif. The stability and geometry of a T-motif depends on its polarity $\mathcal{P}$, on the length of the bulge loop, $L$, and on the number of base pairs within the loop formed with the sticky end, $n_{bp}$. Additional stability is provided by coaxial stacking at the nicks. We describe the coaxial stacking state of the junction by $s\in\{s_0,s_A,s_B,s_C,s_{AB},s_{BC}\}$, where: $s_{AB}$ corresponds to a well-formed T-motif; $s_0$ denotes no coaxial stacking; $s_A$,  $s_B$ are partially folded states that have only one of the sites depicted in Figure \ref{fig::Single}(b) stacked; $s_C$ is the misfolded state shown in Figure \ref{fig::Single}(d) in which domains $D1$ and $D3$ of the bulge duplex are approximately collinear; and $s_{BC}$ has, in addition, stacking between the terminal base pair of a partly-formed loop duplex D2 and the end of sticky-end duplex $D4$. 

\begin{figure}[t]
		\includegraphics[width=1.0\linewidth]{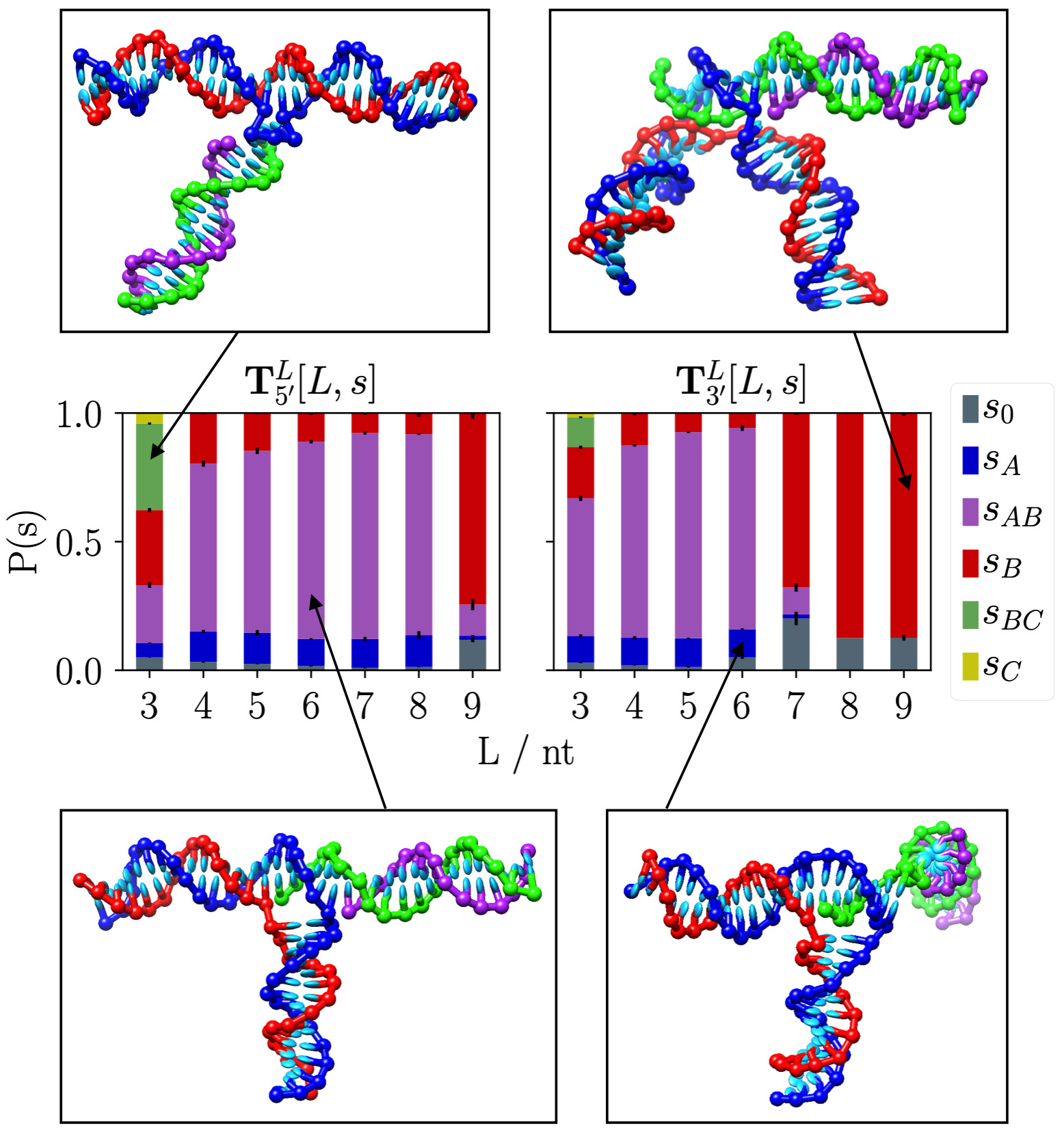}
	\caption{
		Coaxial stacking probabilities for fully-formed 5$^\prime$ and 3$^\prime$  T-motifs. The desired configuration
($s_{AB}$) is most probable for bulge sizes between 4 and 8 nt for 5$^\prime$ T-motifs and for bulge sizes between 4 and 6 nt for 3$^\prime$ T-motifs. When the number of base pairs in the bulge is too large to fit across the major/minor groove of duplex DNA it becomes thermodynamically favourable to break the coaxial stacking at site $s_A$. Error bars represent the standard error of the mean computed from configurations in five VMMC trajectories.
		}
	\label{fig::StackL}
\end{figure}

\begin{figure}[t!]
	\centering
		\includegraphics[trim=0.3cm 0.5cm 0.3cm 0cm, width=1\linewidth]{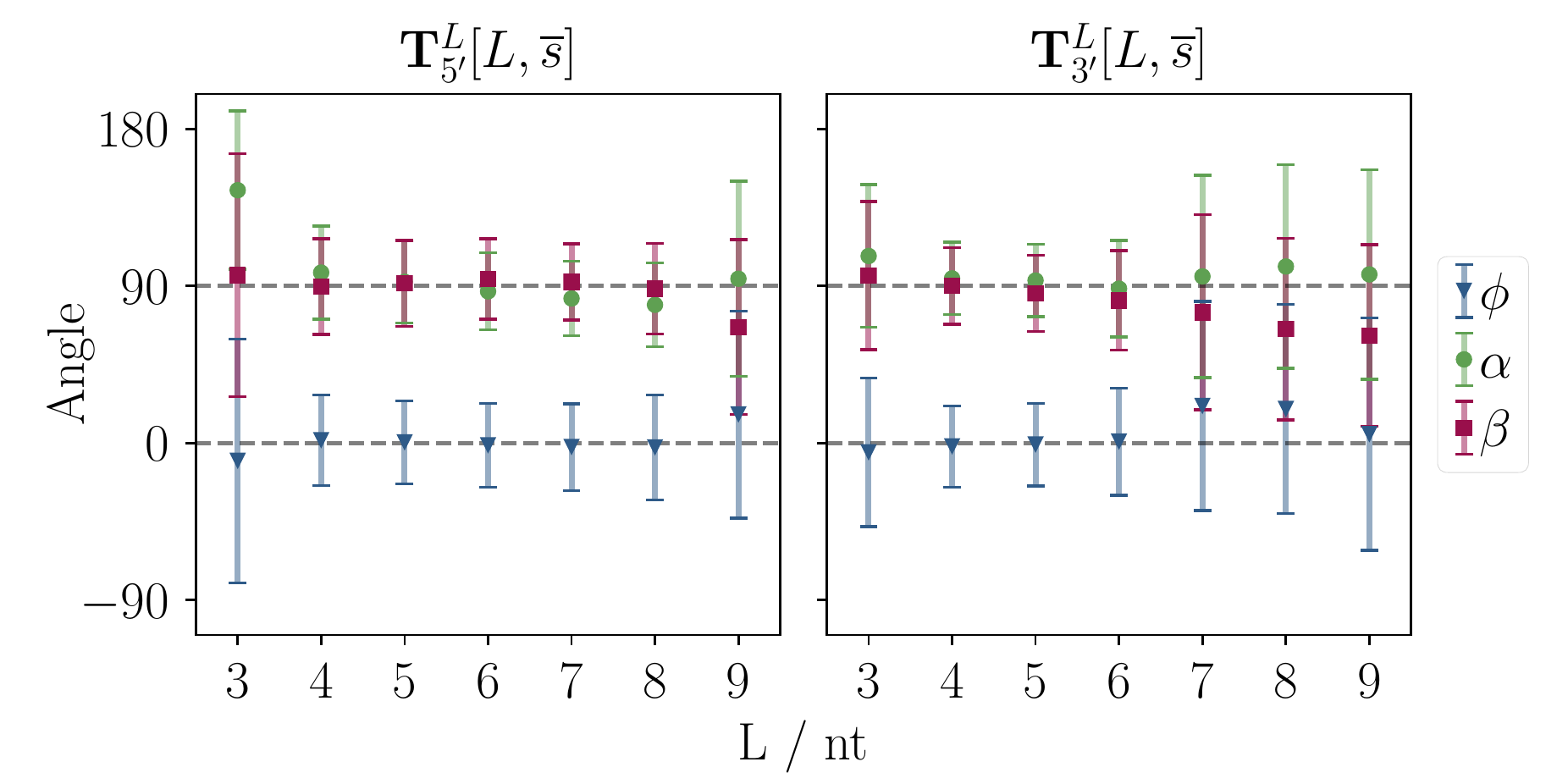}	
	\caption{
		Mean values and standard deviations for angles between arms (see Figure \ref{fig::Single}c for definitions). The data is averaged over all observed stacking states of fully-formed target structures (for which the maximum number of base pairs in the loop is formed, $n_{bp}=L$). Fully-formed 5$^\prime$ T-motifs have a stable, planar, and right-angled geometry ($\alpha \simeq \beta \simeq 90\degree, \phi \simeq 0)$ for bulge sizes $4 \leq L \leq 8$; for 3$^\prime$ T-motifs the range is more limited, $4 \leq L \leq 6$.
		}
	\label{fig::Angle}
\end{figure}

\begin{figure*}[t]
	%\centering
		\includegraphics[trim=2.5cm 3.5cm 3cm 3cm, width=1\linewidth]{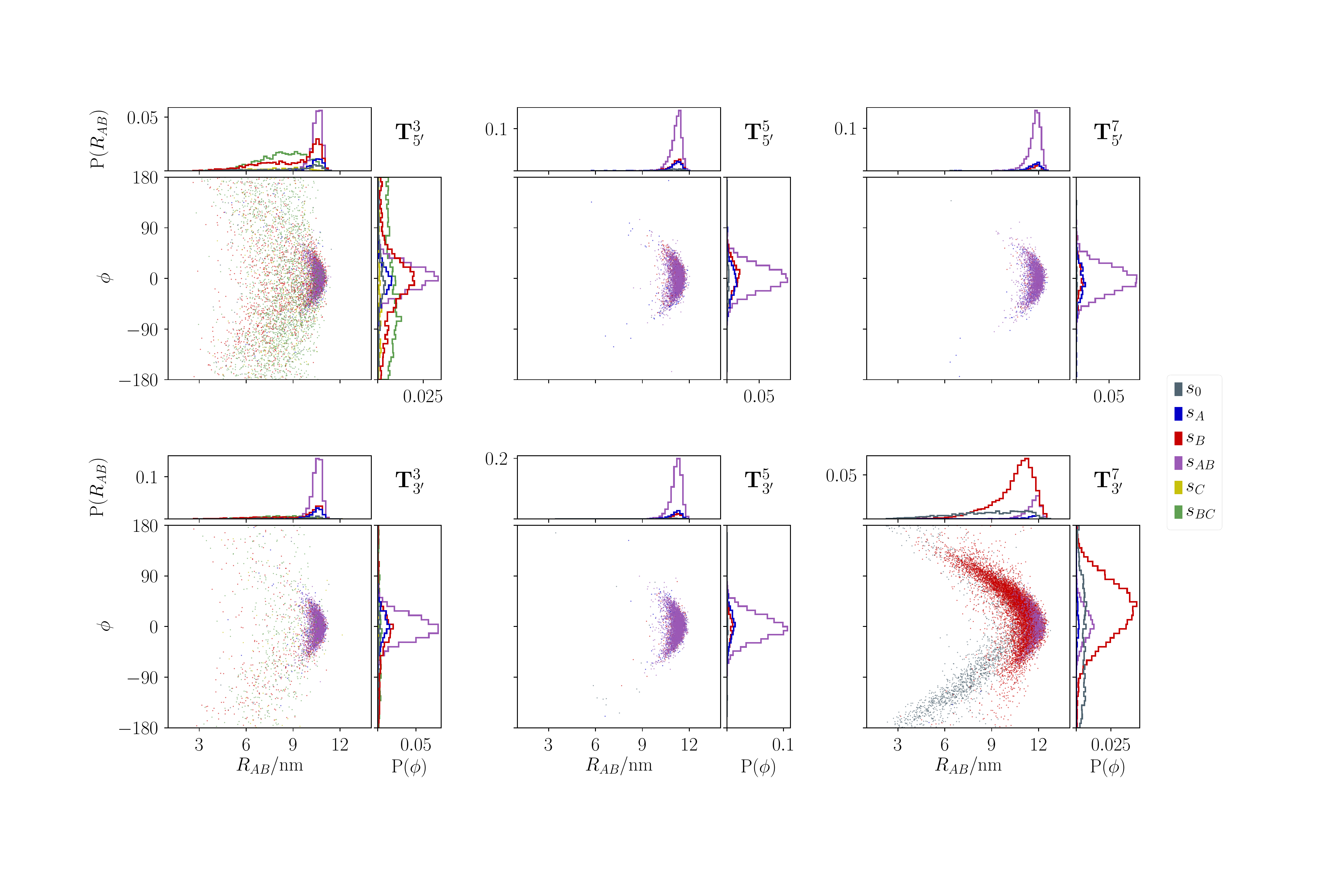}% Which??
	\caption{
		Scatter plots showing the relationship between the dihedral angle, $\phi$, and the end-to-end distance along the ``nicked" duplex, $R_{AB}$, for a selection of T-motifs. The projections in the top and right panels show the probability of states having particular values of $R_{AB}$ and $\phi$, respectively. Sampled configurations include only states in which the bulge loop is fully bound ($n_{bp}=L$); colours identify different stacking states. In the ideal configuration, the motif is in stacking state $s_{AB}$ (purple), leading to a planar geometry in which $\phi=0\degree$ and $R_{AB}$ is maximal.
		}
	\label{fig::AngleDist}
\end{figure*}

A site is characterised as stacked if the magnitude of the coaxial stacking interaction energy between the relevant nucleotides is above a threshold value, taken to be the same as the threshold used to assess hydrogen-bonding of $\sim$0.6 kcal mol$^{-1}$ at 25$\degree$C which corresponds to 0.093 times the well depth of the interaction potential. This cutoff was chosen to avoid counting fleeting and weak interactions.

Figure \ref{fig::Stackn} shows the probabilities of coaxial stacking states for states with a given $n_{bp}$ during the formation of 5$'$ T-motifs with various loop sizes. Equivalent results for 3$^\prime$ T-motifs are shown in Supplementary Material SIII D. For an optimal loop size (eg. $\mathbf{T}_{5'}^{6}$), coaxial stacking at site $s_A$ is observed during initial stages of binding (small $n_{bp}$) and is followed by coaxial stacking at site $s_B$. In the fully formed $\mathbf{T}_{5'}^{6}$, both sites are simultaneously stacked (stacking state $s_{AB}$) with high probability. When the bulge loop and inserted sticky end are too large (eg. $\mathbf{T}_{5'}^{9}$), a similar behaviour is observed in the initial stages of binding but stacking at site $s_A$ is broken in the final stage in order to relieve stress when all base pairs are incorporated (see the top right configuration in Figure \ref{fig::StackL}). In such cases, there is a preference to stack at site $s_B$ rather than $s_A$ as this is geometrically more feasible without breaking base pairs in the side-branch duplex. For short loop sizes (eg. $\mathbf{T}_{5'}^{3}$), there is a high probability of coaxial stacking between arms $A$ and $B$ at site $s_C$ (Figure \ref{fig::Single}d). When all base pairs have formed between loop and sticky end, states in which the duplex incorporating the bulge loop forms a continuous helix ($s_C$, $s_{BC}$) compete with states that have the T-motif geometry ($s_A$, $s_B$, $s_{AB}$). The relative weakness of states $s_C$, $s_{BC}$ for longer bulge loops is consistent with the tendency seen in Ref. \citen{schreck2015} for duplexes with longer bulged loops (and without sticky-end insertions) to adopt configurations with no stacking across the bulge. As will be discussed later, these two classes of configuration are evident in the observed bimodal probability distributions for the angles between arms. 

Figure \ref{fig::StackL} shows the probabilities of coaxial stacking states for T-motifs with fully-bound bulge loops. In an ideal structure, sites $s_A$ and $s_B$ (Figure \ref{fig::Single}(b)) would both be stacked (configuration $s_{AB}$). This ideal case occurs with high probability in fully-base-paired 5$^\prime$ T-motifs with loop sizes of $4 \leq L \leq 8$ and 3$^\prime$ T-motifs with $4 \leq L \leq 6$. The difference between these optimal loop (and sticky end) lengths arises because a 3$^\prime$ T-motif forms a junction spanning the narrower minor groove of the quasi-continuous, coaxially-stacked duplex, whereas a 5$^\prime$ T-motif junction spans the wider major groove (Figure \ref{fig::Single}). 

To characterise the geometries of T-motifs, vectors pointing in the direction of each arm were calculated according to the scheme shown in Figure \ref{fig::Single}c: direction vectors are averages over base-normal vectors of the 11 central base pairs of each arm. The means and standard deviations of the angles between arms, calculated assuming a wrapped normal distribution, are plotted in Figure \ref{fig::Angle}. For an ideal T-motif, we expect a planar geometry in which $\alpha=90\degree$, $\beta=90\degree$ and $\phi=0\degree$. It is clear that 5$^\prime$ T-motifs with loop sizes $4 \leq L \leq 8$ nucleotides and 3$'$ T-motifs with $4 \leq L \leq 6$ nucleotides best satisfy these requirements with low variance in the distribution of angles. These ranges coincide with loop sizes for which coaxial stacking at sites $s_A$ and $s_B$ stabilizes the T-motif.

Figure \ref{fig::AngleDist} examines more closely the relationship between coaxial stacking and T-motif geometries. The distributions of end-to-end distances across the nicked duplex, $R_{AB}$, and the dihedral angle, $\phi$, are shown for each coaxial stacking state. For every loop size, coaxial stacking state $s_{AB}$ corresponds to the target geometry in which $R_{AB}$ is maximised and $\phi \simeq 0$. Partially stacked and unstacked states have much wider distributions of $\phi$ and $R_{AB}$, leading to the large variances observed in Figure \ref{fig::Angle} for extreme loop sizes ($\mathbf{T}_{3'}^{7}$ and $\mathbf{T}_{5'}^{3}$). Stacking states $s_{C}$ and $s_{BC}$, in which stacking within the bulge duplex inhibits the formation of a well-defined T-junction, have a well-defined  geometry of their own. In these cases, the duplex containing the bulge loop is extended, $\alpha \sim 180\degree$, maximising its end-to-end distance $R_{AC}$; the sticky-end duplex hangs off the loop with $\beta \sim 90\degree$. See Supplementary Material SIII D and Supplementary Figures S13 - S20 for all distributions of characterised angles and lengths.

\begin{figure}[]
	\centering
        		\includegraphics[trim=1cm 1cm 0cm 0cm, width=1\linewidth]{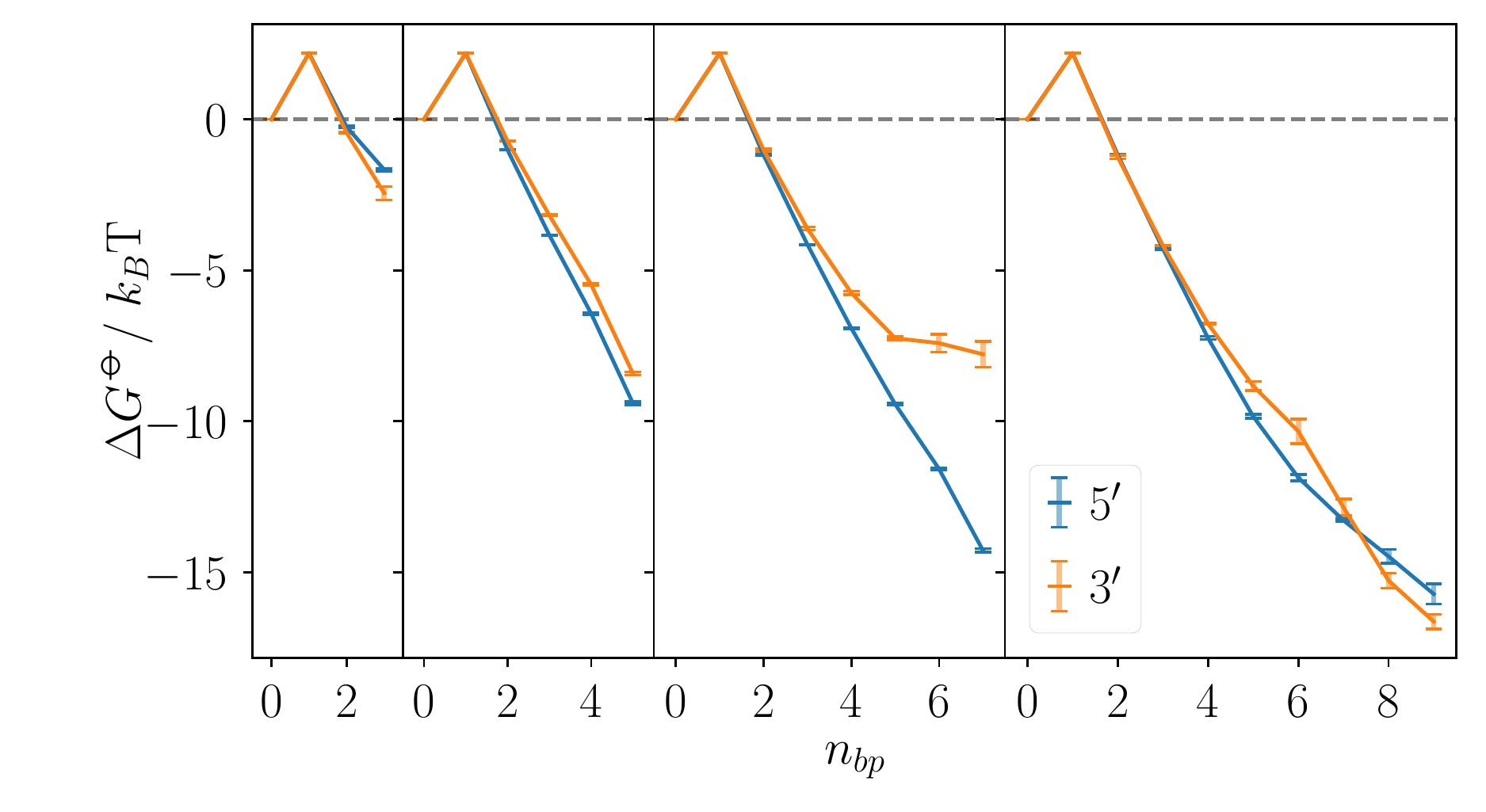}
	\caption{
		Free-energy profiles for the formation of T-motifs with bulge sizes $L\in\{3,5,7,9\}$, derived from oxDNA simulations at 25$\degree$C. Simulations were carried out with all strands at 42 $\mu$M and scaled to molar concentration. In 3$^\prime$ T-motifs the junction spans the minor groove, so creation of more than 5 base pairs between bulge and sticky end prevents assembly of a well-formed junction. 5$^\prime$ T-motifs span the wider major groove and can tolerate a larger number of base pairs in the junction. 
	}
	\label{fig::EnergyLand}
\end{figure}

The thermodynamics of T-motif formation was investigated by using umbrella sampling to speed up transitions between metastable states. The number of base pairs formed between the bulge loop and the sticky end, $n_{bp}$, was used as an order parameter to obtain the free energy $\Delta G(n_{bp}) = -k_BT \ln [Z(n_{bp})/Z(n_{bp}=0)]$, where $Z(n_{bp})$ is the (unbiased) partial partition function of the states with $n_{bp}$ base pairs. Figure \ref{fig::EnergyLand} shows free-energy profiles for T-motifs with different loop sizes $L$. To understand the features of the free energy profiles, let us highlight several events in a hypothetical assembly pathway for a T-motif with ideal loop size. First, there is a concentration-dependent entropic cost associated with bringing the interacting bulge and sticky end into close proximity. Once the first base pair forms, hybridization between sticky end and bulge loop progressively decreases the free energy through enthalpic gains which outweigh further entropic penalties as the duplex ``zips up''. As discussed above, coaxial stacking of bases across the nicks increases the stability of the structure. As the final base pairs form, these stacking-energy gains stabilise the T-junction as it ``clicks'' into shape. For an ideal bulge size (e.g. $\mathbf{T}_{5'}^{7}$), the formation of the final base pair thus results in the highest gain in stability. However, if the duplex formed by hybridization of the bulge with the sticky end is just larger than the groove width (e.g. $\mathbf{T}_{3'}^{7}$), the formation of the final base pairs results in little energy gain as the junction is sterically constrained, inhibiting base stacking. When the bulge size is much larger than the groove width (e.g. $\mathbf{T}_{3'}^{9}$), the junction loses its well-defined T-shape; it is stabilized by the formation of additional base pairs but lacks the contribution of base stacking across the nicks.

\subsection{Experimental Results}\label{sec::ExpRes}

The electrophoretic mobility of a T-motif is lower than those of the component sticky-end and loop duplexes. For a fixed concentration of loop duplex, the mobility of the gel band corresponding to the T-motif was found to be dependent on the concentration of the sticky-end duplex. This behaviour is consistent with fast on- and off-rates for association of the sticky-end with the bulge loop, causing them to bind and unbind while co-migrating through the gel. As shown in Supplementary Material SII D, when the difference in mobilities of the component duplexes is increased, they migrate separately and no band shift is observed.

\begin{figure}[h]
	\centering
		\includegraphics[width=1\linewidth]{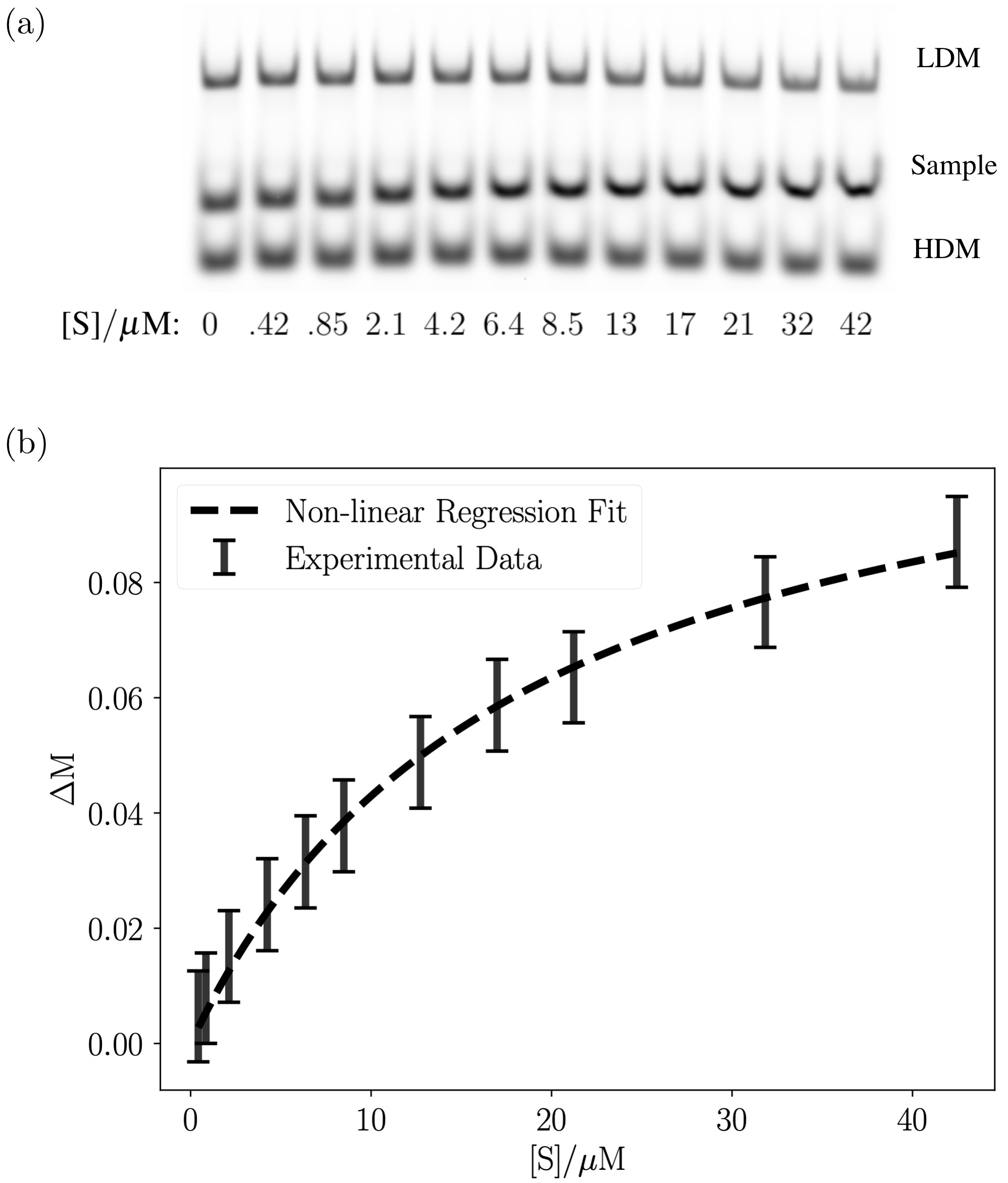}
	\caption{
		Estimation of the dissociation constant, $K_d$, for T-motif $\mathbf{T}_{3'}^{5}$. (a) Gel showing the change in electrophoretic mobility of the loop duplex when annealed with increasing concentration, $\text{[S]}$, of the sticky-end duplex. High- and low-mobility reference markers (HDM and LDM) are also shown. In all lanes, the concentration of loop duplex is estimated to be 0.42 $\mu$M (b) Fitted binding curve corresponding to an estimated dissociation constant $K_d$=$18\pm1\mu$M.% The calculation does not take into account systematic errors arising from estimating the concentration of the duplexes within the gel environment (Supplementary Material \ref{app::marker}).
		}
	\label{fig::ExpCurve}
\end{figure}

To estimate the dissociation constant of a T-motif, $K_d$, the relative electrophoretic mobility of the gel band containing the loop duplex, $M(\text{[S]})$, was measured as a function of sticky-end duplex concentration, [S]. Reference markers were used to enable accurate measurement of $M$, as outlined in Section \ref{sec::experiment}. Dissociation constants were estimated by fitting measurements of $\Delta M(\text{[S]}) = M(\text{[S]}) - M(0)$ to the function $\Delta M(\text{[S]})=\Delta M_{max}\text{[S]}/(K_d+\text{[S]})$, corresponding to the assumptions that the change in relative electrophoretic mobility is proportional to the fraction of loop duplex bound in a T-motif and that this fraction is close to the equilibrium value. Experimental results for motif $\mathbf{T}_{3'}^{5}$ are shown in Figure \ref{fig::ExpCurve}, along with the resulting binding curve.

\begin{figure}[]
	\centering
        		\includegraphics[trim=1cm 1.5cm -0.5cm 0.5cm, width=1\linewidth]{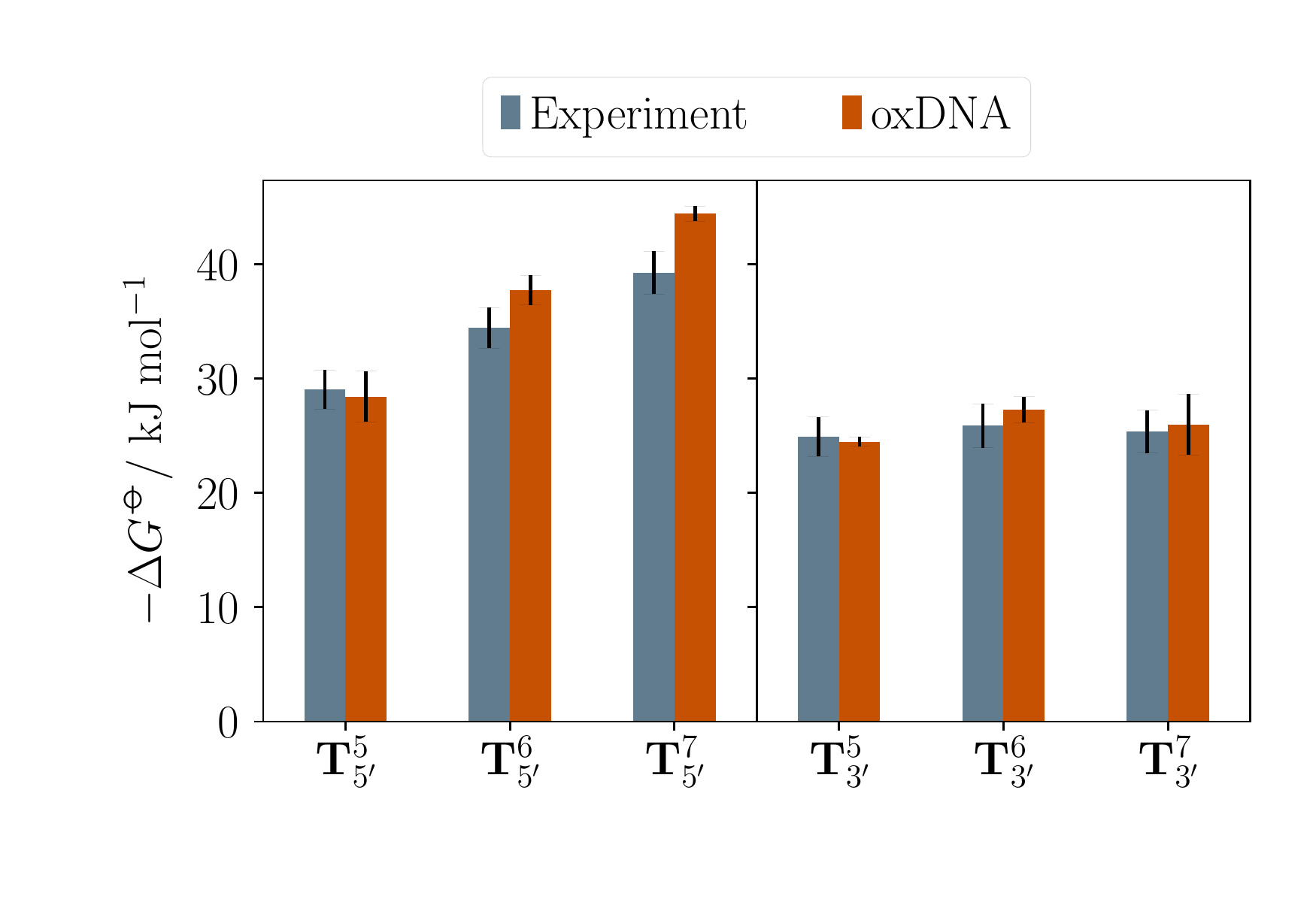}
	\caption{
		Gibbs free energies for the formation of T-motifs with a range of bulge sizes at $4\degree$C and molar concentration. Results are inferred from experimental measurements of dissociation constants at $4\degree$C and from oxDNA simulations. oxDNA errors represent the standard deviations of 5 independent trajectories and uncertainties at extrapolation from the simulation temperature of $25\degree$C. Experimental error bars reflect uncertainties in estimating band shifts and the systematic error associated with estimating local concentrations in gel electrophoresis (Supplementary Material SII B).
		}
	\label{fig::Energy}
\end{figure}

Binding affinities, corresponding to the fitted dissociation constants, for T-motifs with bulge sizes between 5 and 7 are compared to oxDNA simulation results in Figure \ref{fig::Energy}. oxDNA reproduces the experimentally determined dependence on bulge size and polarity of sticky end. 5$'$ T-motifs are generally more stable, especially for larger bulge sizes, because they span the wider major groove. Agreement between experiment and simulation is close, but there are uncertainties in both sets of data. Firstly, the temperature in oxDNA simulations is set to $25\degree$C but gel electrophoresis experiments are performed at $4\degree$C. The extrapolation of simulation results to $4\degree$C using single-histogram reweighting,\cite{kumar1992} is described in Supplementary Material SI D. Secondly, oxDNA simulations were carried out using different salt conditions ([Na$^+$] = 0.5M) to those used in experiments ([Mg$^{2+}$] = 12.5mM). As discussed in section \ref{sec::oxDNA}, the simple treatment of electrostatics in oxDNA does not allow simulation of arbitrary solution conditions but the conversion factor between monovalent and divalent salt concentrations used here is roughly consistent with the scale factors provided by Bosco \textit{et al}.\cite{bosco2014} Thirdly, although the hydrogen-bonding potentials for our simulations are sequence-specific, the coaxial stacking interactions in oxDNA are not. Our estimates of the binding strengths of T-motifs therefore do not capture their dependence on the coaxial stacking strengths of the specific nucleotides flanking stacking sites $s_A$, $s_B$ and $s_C$. Finally, as outlined in Supplementary Material SII B, there is experimental uncertainty associated with estimation of concentrations in the electrophoresis gel. This introduces an additional systematic error in experimentally-determined dissociation constants.

%%%%%%%%%%%%%%%%%%%%%%%%%	     CONCLUSIONS	   %%%%%%%%%%%%%%%%%%%%%%%%%%%

\section{Conclusions}\label{sec::conclusions}

The T-motif has recently become a popular building block in DNA nanostructures. T-motifs are formed by the interaction of a bulge loop in one duplex and a sticky-end of another, creating a characteristic 3-arm ``T'' shape. A T-motif incorporating a 3$^\prime$ sticky-end leads to a junction in which the side arm is connected across the minor groove of a quasi-continuous duplex; a 5$^\prime$ sticky-end leads to a junction that spans the major groove. We have used coarse-grained oxDNA simulations to investigate the geometries and stabilities of T-motifs of both polarities with a range of bulge-loop and sticky-end sizes. 

If the size of the duplex formed by hybridization of the bulge loop and sticky end is approximately equal to the groove width, coaxial stacking interactions at either side of this duplex provide additional stability and favour the formation of a junction with well-defined `T'  geometry. Given the larger size of the major groove, 5$^\prime$ T-motifs can incorporate a larger number of base pairs in the junction and can adopt stable geometries for a wider range of bulge sizes. If the bulge loop is too short, there is a high probability that coaxial stacking of base pairs across the bulge on the loop duplex destabilises the junction. If the bulge is too long, the desired stacking interactions are disrupted when the sticky end and loop are fully hybridized, again leading to undesired non-`T' geometries. Experimentally determined T-motif binding affinities show a remarkable correlation with oxDNA results. Our simulations show that optimal bulge loop lengths are in the ranges  $4 \leq L \leq 8$ nucleotides for 5$^\prime$ T-motifs and $4 \leq L \leq 6$ nucleotides for 3$^\prime$ T-motifs.

Previous work on T-motifs\cite{hamada2009,li2017,mao2012,mao2015,tandon2019} has made use of the 5$^\prime$ variant due to its higher stability. We have shown that $3^\prime$ T-motifs can also be stable and adopt well-defined geometries. If incorporated within larger structures, the reduced stability of the $3^\prime$ junction could be compensated for by cooperative assembly, in which the formation of other structural links increases the local concentration of $3^\prime$ sticky ends at their target bulge loops.

%%%%%%%%%%%%%%%%%%%%%%%%%	     REFERENCES	   %%%%%%%%%%%%%%%%%%%%%%%%%%%

\section{Acknowledgments}\label{sec::ref}

KGY was supported by Engineering and Physical Sciences Research Council Life Sciences Interface Doctoral Training Centre EP/F500394/1 and AJT by a Royal Society Wolfson Research Merit Award. BN thanks Dr. Thomas Ouldridge for helpful discussions and Domen Pre\v sern for providing Figure S1.

\bibliography{bibliography}{}

%merlin.mbs aipnum4-1.bst 2010-07-25 4.21a (PWD, AO, DPC) hacked
%Control: key (0)
%Control: author (8) initials jnrlst
%Control: editor formatted (1) identically to author
%Control: production of article title (-1) disabled
%Control: page (0) single
%Control: year (1) truncated
%Control: production of eprint (0) enabled
\begin{thebibliography}{63}%
\makeatletter
\providecommand \@ifxundefined [1]{%
 \@ifx{#1\undefined}
}%
\providecommand \@ifnum [1]{%
 \ifnum #1\expandafter \@firstoftwo
 \else \expandafter \@secondoftwo
 \fi
}%
\providecommand \@ifx [1]{%
 \ifx #1\expandafter \@firstoftwo
 \else \expandafter \@secondoftwo
 \fi
}%
\providecommand \natexlab [1]{#1}%
\providecommand \enquote  [1]{``#1''}%
\providecommand \bibnamefont  [1]{#1}%
\providecommand \bibfnamefont [1]{#1}%
\providecommand \citenamefont [1]{#1}%
\providecommand \href@noop [0]{\@secondoftwo}%
\providecommand \href [0]{\begingroup \@sanitize@url \@href}%
\providecommand \@href[1]{\@@startlink{#1}\@@href}%
\providecommand \@@href[1]{\endgroup#1\@@endlink}%
\providecommand \@sanitize@url [0]{\catcode `\\12\catcode `\$12\catcode
  `\&12\catcode `\#12\catcode `\^12\catcode `\_12\catcode `\%12\relax}%
\providecommand \@@startlink[1]{}%
\providecommand \@@endlink[0]{}%
\providecommand \url  [0]{\begingroup\@sanitize@url \@url }%
\providecommand \@url [1]{\endgroup\@href {#1}{\urlprefix }}%
\providecommand \urlprefix  [0]{URL }%
\providecommand \Eprint [0]{\href }%
\providecommand \doibase [0]{http://dx.doi.org/}%
\providecommand \selectlanguage [0]{\@gobble}%
\providecommand \bibinfo  [0]{\@secondoftwo}%
\providecommand \bibfield  [0]{\@secondoftwo}%
\providecommand \translation [1]{[#1]}%
\providecommand \BibitemOpen [0]{}%
\providecommand \bibitemStop [0]{}%
\providecommand \bibitemNoStop [0]{.\EOS\space}%
\providecommand \EOS [0]{\spacefactor3000\relax}%
\providecommand \BibitemShut  [1]{\csname bibitem#1\endcsname}%
\let\auto@bib@innerbib\@empty
%</preamble>
\bibitem [{\citenamefont {Seeman}\ and\ \citenamefont
  {Sleiman}(2017)}]{seeman2018}%
  \BibitemOpen
  \bibfield  {author} {\bibinfo {author} {\bibfnamefont {N.~C.}\ \bibnamefont
  {Seeman}}\ and\ \bibinfo {author} {\bibfnamefont {H.~F.}\ \bibnamefont
  {Sleiman}},\ }\href {\doibase 10.1038/natrevmats.2017.68} {\bibfield
  {journal} {\bibinfo  {journal} {Nat. Rev. Mater}\ }\textbf {\bibinfo {volume}
  {3}},\ \bibinfo {pages} {17068} (\bibinfo {year} {2017})}\BibitemShut
  {NoStop}%
\bibitem [{\citenamefont {Holliday}(1964)}]{holliday1964}%
  \BibitemOpen
  \bibfield  {author} {\bibinfo {author} {\bibfnamefont {R.}~\bibnamefont
  {Holliday}},\ }\href {\doibase 10.1017/S0016672300001233} {\bibfield
  {journal} {\bibinfo  {journal} {Genetics Research}\ }\textbf {\bibinfo
  {volume} {5}},\ \bibinfo {pages} {282} (\bibinfo {year} {1964})}\BibitemShut
  {NoStop}%
\bibitem [{\citenamefont {Seeman}(1982)}]{seeman1982}%
  \BibitemOpen
  \bibfield  {author} {\bibinfo {author} {\bibfnamefont {N.~C.}\ \bibnamefont
  {Seeman}},\ }\href {\doibase 10.1016/0022-5193(82)90002-9} {\bibfield
  {journal} {\bibinfo  {journal} {J. Theor. Biol}\ }\textbf {\bibinfo {volume}
  {99}},\ \bibinfo {pages} {237} (\bibinfo {year} {1982})}\BibitemShut
  {NoStop}%
\bibitem [{\citenamefont {Seeman}\ and\ \citenamefont
  {Kallenbach}(1983)}]{seeman1983}%
  \BibitemOpen
  \bibfield  {author} {\bibinfo {author} {\bibfnamefont {N.~C.}\ \bibnamefont
  {Seeman}}\ and\ \bibinfo {author} {\bibfnamefont {N.~R.}\ \bibnamefont
  {Kallenbach}},\ }\href {\doibase 10.1016/s0006-3495(83)84292-1} {\bibfield
  {journal} {\bibinfo  {journal} {Biophys. J}\ }\textbf {\bibinfo {volume}
  {44}},\ \bibinfo {pages} {201} (\bibinfo {year} {1983})}\BibitemShut
  {NoStop}%
\bibitem [{\citenamefont {Fu}\ and\ \citenamefont {Seeman}(1993)}]{seeman1993}%
  \BibitemOpen
  \bibfield  {author} {\bibinfo {author} {\bibfnamefont {T.~J.}\ \bibnamefont
  {Fu}}\ and\ \bibinfo {author} {\bibfnamefont {N.~C.}\ \bibnamefont
  {Seeman}},\ }\href {\doibase 10.1021/bi00064a003} {\bibfield  {journal}
  {\bibinfo  {journal} {Biochemistry}\ }\textbf {\bibinfo {volume} {32}},\
  \bibinfo {pages} {3211} (\bibinfo {year} {1993})}\BibitemShut {NoStop}%
\bibitem [{\citenamefont {Winfree}\ \emph {et~al.}(1998)\citenamefont
  {Winfree}, \citenamefont {Liu}, \citenamefont {Wenzler},\ and\ \citenamefont
  {Seeman}}]{winfree1998}%
  \BibitemOpen
  \bibfield  {author} {\bibinfo {author} {\bibfnamefont {E.}~\bibnamefont
  {Winfree}}, \bibinfo {author} {\bibfnamefont {F.}~\bibnamefont {Liu}},
  \bibinfo {author} {\bibfnamefont {L.~A.}\ \bibnamefont {Wenzler}}, \ and\
  \bibinfo {author} {\bibfnamefont {N.~C.}\ \bibnamefont {Seeman}},\ }\href
  {\doibase 10.1038/28998} {\bibfield  {journal} {\bibinfo  {journal} {Nature}\
  }\textbf {\bibinfo {volume} {394}},\ \bibinfo {pages} {539} (\bibinfo {year}
  {1998})}\BibitemShut {NoStop}%
\bibitem [{\citenamefont {He}\ \emph {et~al.}(2006)\citenamefont {He},
  \citenamefont {Tian}, \citenamefont {Ribbe},\ and\ \citenamefont
  {Mao}}]{liu2005}%
  \BibitemOpen
  \bibfield  {author} {\bibinfo {author} {\bibfnamefont {Y.}~\bibnamefont
  {He}}, \bibinfo {author} {\bibfnamefont {Y.}~\bibnamefont {Tian}}, \bibinfo
  {author} {\bibfnamefont {A.~E.}\ \bibnamefont {Ribbe}}, \ and\ \bibinfo
  {author} {\bibfnamefont {C.}~\bibnamefont {Mao}},\ }\href {\doibase
  10.1021/ja0665141} {\bibfield  {journal} {\bibinfo  {journal} {J. Am. Chem.
  Soc}\ }\textbf {\bibinfo {volume} {128}},\ \bibinfo {pages} {15978} (\bibinfo
  {year} {2006})}\BibitemShut {NoStop}%
\bibitem [{\citenamefont {Zheng}\ \emph {et~al.}(2009)\citenamefont {Zheng},
  \citenamefont {Birktoft}, \citenamefont {Chen}, \citenamefont {Wang},
  \citenamefont {Sha}, \citenamefont {Constantinou}, \citenamefont {Ginell},
  \citenamefont {Mao},\ and\ \citenamefont {Seeman}}]{zheng2009}%
  \BibitemOpen
  \bibfield  {author} {\bibinfo {author} {\bibfnamefont {J.}~\bibnamefont
  {Zheng}}, \bibinfo {author} {\bibfnamefont {J.~J.}\ \bibnamefont {Birktoft}},
  \bibinfo {author} {\bibfnamefont {Y.}~\bibnamefont {Chen}}, \bibinfo {author}
  {\bibfnamefont {T.}~\bibnamefont {Wang}}, \bibinfo {author} {\bibfnamefont
  {R.}~\bibnamefont {Sha}}, \bibinfo {author} {\bibfnamefont {P.~E.}\
  \bibnamefont {Constantinou}}, \bibinfo {author} {\bibfnamefont {S.~L.}\
  \bibnamefont {Ginell}}, \bibinfo {author} {\bibfnamefont {C.}~\bibnamefont
  {Mao}}, \ and\ \bibinfo {author} {\bibfnamefont {N.}~\bibnamefont {Seeman}},\
  }\href@noop {} {\bibfield  {journal} {\bibinfo  {journal} {Nature}\ }\textbf
  {\bibinfo {volume} {461}},\ \bibinfo {pages} {74} (\bibinfo {year}
  {2009})}\BibitemShut {NoStop}%
\bibitem [{\citenamefont {Shih}, \citenamefont {Quispe},\ and\ \citenamefont
  {Joyce}(2004)}]{shih2004}%
  \BibitemOpen
  \bibfield  {author} {\bibinfo {author} {\bibfnamefont {W.~M.}\ \bibnamefont
  {Shih}}, \bibinfo {author} {\bibfnamefont {J.~D.}\ \bibnamefont {Quispe}}, \
  and\ \bibinfo {author} {\bibfnamefont {G.~F.}\ \bibnamefont {Joyce}},\ }\href
  {\doibase 10.1038/nature02307} {\bibfield  {journal} {\bibinfo  {journal}
  {Nature}\ }\textbf {\bibinfo {volume} {427}},\ \bibinfo {pages} {618}
  (\bibinfo {year} {2004})}\BibitemShut {NoStop}%
\bibitem [{\citenamefont {Goodman}\ \emph {et~al.}(2005)\citenamefont
  {Goodman}, \citenamefont {Schaap}, \citenamefont {Tardin}, \citenamefont
  {Erben}, \citenamefont {Berry}, \citenamefont {Schmidt},\ and\ \citenamefont
  {Turberfield}}]{goodman2005}%
  \BibitemOpen
  \bibfield  {author} {\bibinfo {author} {\bibfnamefont {R.~P.}\ \bibnamefont
  {Goodman}}, \bibinfo {author} {\bibfnamefont {I.~A.~T.}\ \bibnamefont
  {Schaap}}, \bibinfo {author} {\bibfnamefont {C.~F.}\ \bibnamefont {Tardin}},
  \bibinfo {author} {\bibfnamefont {C.~M.}\ \bibnamefont {Erben}}, \bibinfo
  {author} {\bibfnamefont {R.~M.}\ \bibnamefont {Berry}}, \bibinfo {author}
  {\bibfnamefont {C.~F.}\ \bibnamefont {Schmidt}}, \ and\ \bibinfo {author}
  {\bibfnamefont {A.~J.}\ \bibnamefont {Turberfield}},\ }\href {\doibase
  10.1126/science.1120367} {\bibfield  {journal} {\bibinfo  {journal}
  {Science}\ }\textbf {\bibinfo {volume} {310}},\ \bibinfo {pages} {1661}
  (\bibinfo {year} {2005})}\BibitemShut {NoStop}%
\bibitem [{\citenamefont {Zhang}\ \emph {et~al.}(2008)\citenamefont {Zhang},
  \citenamefont {Su}, \citenamefont {He}, \citenamefont {Zhao}, \citenamefont
  {Fang}, \citenamefont {Ribbe}, \citenamefont {Jiang},\ and\ \citenamefont
  {Mao}}]{zhang2008}%
  \BibitemOpen
  \bibfield  {author} {\bibinfo {author} {\bibfnamefont {C.}~\bibnamefont
  {Zhang}}, \bibinfo {author} {\bibfnamefont {M.}~\bibnamefont {Su}}, \bibinfo
  {author} {\bibfnamefont {Y.}~\bibnamefont {He}}, \bibinfo {author}
  {\bibfnamefont {X.}~\bibnamefont {Zhao}}, \bibinfo {author} {\bibfnamefont
  {P.}~\bibnamefont {Fang}}, \bibinfo {author} {\bibfnamefont {A.~E.}\
  \bibnamefont {Ribbe}}, \bibinfo {author} {\bibfnamefont {W.}~\bibnamefont
  {Jiang}}, \ and\ \bibinfo {author} {\bibfnamefont {C.}~\bibnamefont {Mao}},\
  }\href {\doibase 10.1073/pnas.0803841105} {\bibfield  {journal} {\bibinfo
  {journal} {Proc. Natl. Acad. Sci. USA}\ }\textbf {\bibinfo {volume} {105}},\
  \bibinfo {pages} {10665} (\bibinfo {year} {2008})}\BibitemShut {NoStop}%
\bibitem [{\citenamefont {He}\ \emph {et~al.}(2008)\citenamefont {He},
  \citenamefont {Ye}, \citenamefont {Su}, \citenamefont {Zhang}, \citenamefont
  {Ribbe}, \citenamefont {Jiang},\ and\ \citenamefont {Mao}}]{he2008}%
  \BibitemOpen
  \bibfield  {author} {\bibinfo {author} {\bibfnamefont {Y.}~\bibnamefont
  {He}}, \bibinfo {author} {\bibfnamefont {T.}~\bibnamefont {Ye}}, \bibinfo
  {author} {\bibfnamefont {M.}~\bibnamefont {Su}}, \bibinfo {author}
  {\bibfnamefont {C.}~\bibnamefont {Zhang}}, \bibinfo {author} {\bibfnamefont
  {A.~E.}\ \bibnamefont {Ribbe}}, \bibinfo {author} {\bibfnamefont
  {W.}~\bibnamefont {Jiang}}, \ and\ \bibinfo {author} {\bibfnamefont
  {C.}~\bibnamefont {Mao}},\ }\href {\doibase 10.1038/nature06597} {\bibfield
  {journal} {\bibinfo  {journal} {Nature}\ }\textbf {\bibinfo {volume} {452}},\
  \bibinfo {pages} {198} (\bibinfo {year} {2008})}\BibitemShut {NoStop}%
\bibitem [{\citenamefont {Zhang}\ \emph {et~al.}(2009)\citenamefont {Zhang},
  \citenamefont {Ko}, \citenamefont {Su}, \citenamefont {Leng}, \citenamefont
  {Ribbe}, \citenamefont {Jiang},\ and\ \citenamefont {Mao}}]{zhang2009}%
  \BibitemOpen
  \bibfield  {author} {\bibinfo {author} {\bibfnamefont {C.}~\bibnamefont
  {Zhang}}, \bibinfo {author} {\bibfnamefont {S.}~\bibnamefont {Ko}}, \bibinfo
  {author} {\bibfnamefont {M.}~\bibnamefont {Su}}, \bibinfo {author}
  {\bibfnamefont {Y.}~\bibnamefont {Leng}}, \bibinfo {author} {\bibfnamefont
  {A.~E.}\ \bibnamefont {Ribbe}}, \bibinfo {author} {\bibfnamefont
  {W.}~\bibnamefont {Jiang}}, \ and\ \bibinfo {author} {\bibfnamefont
  {C.}~\bibnamefont {Mao}},\ }\href {\doibase 10.1021/ja809666h} {\bibfield
  {journal} {\bibinfo  {journal} {J. Am. Chem. Soc}\ }\textbf {\bibinfo
  {volume} {131}},\ \bibinfo {pages} {1413} (\bibinfo {year}
  {2009})}\BibitemShut {NoStop}%
\bibitem [{\citenamefont {He}\ \emph {et~al.}(2010)\citenamefont {He},
  \citenamefont {Su}, \citenamefont {Fang}, \citenamefont {Zhang},
  \citenamefont {Ribbe}, \citenamefont {Jiang},\ and\ \citenamefont
  {Mao}}]{he2010}%
  \BibitemOpen
  \bibfield  {author} {\bibinfo {author} {\bibfnamefont {Y.}~\bibnamefont
  {He}}, \bibinfo {author} {\bibfnamefont {M.}~\bibnamefont {Su}}, \bibinfo
  {author} {\bibfnamefont {P.}~\bibnamefont {Fang}}, \bibinfo {author}
  {\bibfnamefont {C.}~\bibnamefont {Zhang}}, \bibinfo {author} {\bibfnamefont
  {A.~E.}\ \bibnamefont {Ribbe}}, \bibinfo {author} {\bibfnamefont
  {W.}~\bibnamefont {Jiang}}, \ and\ \bibinfo {author} {\bibfnamefont
  {C.}~\bibnamefont {Mao}},\ }\href {\doibase 10.1002/anie.200904513}
  {\bibfield  {journal} {\bibinfo  {journal} {Angew. Chem. Int. Ed}\ }\textbf
  {\bibinfo {volume} {49}},\ \bibinfo {pages} {748} (\bibinfo {year}
  {2010})}\BibitemShut {NoStop}%
\bibitem [{\citenamefont {Iinuma}\ \emph {et~al.}(2014)\citenamefont {Iinuma},
  \citenamefont {Ke}, \citenamefont {Jungmann}, \citenamefont {Schlichthaerle},
  \citenamefont {Woehrstein},\ and\ \citenamefont {Yin}}]{Iinuma2014}%
  \BibitemOpen
  \bibfield  {author} {\bibinfo {author} {\bibfnamefont {R.}~\bibnamefont
  {Iinuma}}, \bibinfo {author} {\bibfnamefont {Y.}~\bibnamefont {Ke}}, \bibinfo
  {author} {\bibfnamefont {R.}~\bibnamefont {Jungmann}}, \bibinfo {author}
  {\bibfnamefont {T.}~\bibnamefont {Schlichthaerle}}, \bibinfo {author}
  {\bibfnamefont {J.~B.}\ \bibnamefont {Woehrstein}}, \ and\ \bibinfo {author}
  {\bibfnamefont {P.}~\bibnamefont {Yin}},\ }\href {\doibase
  10.1126/science.1250944} {\bibfield  {journal} {\bibinfo  {journal}
  {Science}\ }\textbf {\bibinfo {volume} {344}},\ \bibinfo {pages} {65}
  (\bibinfo {year} {2014})}\BibitemShut {NoStop}%
\bibitem [{\citenamefont {Rothemund}(2006)}]{rothemund2006}%
  \BibitemOpen
  \bibfield  {author} {\bibinfo {author} {\bibfnamefont {P.~W.~K.}\
  \bibnamefont {Rothemund}},\ }\href {\doibase 10.1038/nature04586} {\bibfield
  {journal} {\bibinfo  {journal} {Nature}\ }\textbf {\bibinfo {volume} {440}},\
  \bibinfo {pages} {297} (\bibinfo {year} {2006})}\BibitemShut {NoStop}%
\bibitem [{\citenamefont {Wei}, \citenamefont {Dai},\ and\ \citenamefont
  {Yin}(2012)}]{wei2012}%
  \BibitemOpen
  \bibfield  {author} {\bibinfo {author} {\bibfnamefont {B.}~\bibnamefont
  {Wei}}, \bibinfo {author} {\bibfnamefont {M.}~\bibnamefont {Dai}}, \ and\
  \bibinfo {author} {\bibfnamefont {P.}~\bibnamefont {Yin}},\ }\href {\doibase
  10.1038/nature11075} {\bibfield  {journal} {\bibinfo  {journal} {Nature}\
  }\textbf {\bibinfo {volume} {485}},\ \bibinfo {pages} {623} (\bibinfo {year}
  {2012})}\BibitemShut {NoStop}%
\bibitem [{\citenamefont {Ke}\ \emph {et~al.}(2012)\citenamefont {Ke},
  \citenamefont {Ong}, \citenamefont {Shih},\ and\ \citenamefont
  {Yin}}]{ke2012}%
  \BibitemOpen
  \bibfield  {author} {\bibinfo {author} {\bibfnamefont {Y.}~\bibnamefont
  {Ke}}, \bibinfo {author} {\bibfnamefont {L.~L.}\ \bibnamefont {Ong}},
  \bibinfo {author} {\bibfnamefont {W.~M.}\ \bibnamefont {Shih}}, \ and\
  \bibinfo {author} {\bibfnamefont {P.}~\bibnamefont {Yin}},\ }\href {\doibase
  10.1126/science.1227268} {\bibfield  {journal} {\bibinfo  {journal}
  {Science}\ }\textbf {\bibinfo {volume} {338}},\ \bibinfo {pages} {1177}
  (\bibinfo {year} {2012})}\BibitemShut {NoStop}%
\bibitem [{\citenamefont {Ke}\ \emph {et~al.}(2009)\citenamefont {Ke},
  \citenamefont {Sharma}, \citenamefont {Liu}, \citenamefont {Jahn},
  \citenamefont {Liu},\ and\ \citenamefont {Yan}}]{ke2009}%
  \BibitemOpen
  \bibfield  {author} {\bibinfo {author} {\bibfnamefont {Y.}~\bibnamefont
  {Ke}}, \bibinfo {author} {\bibfnamefont {J.}~\bibnamefont {Sharma}}, \bibinfo
  {author} {\bibfnamefont {M.}~\bibnamefont {Liu}}, \bibinfo {author}
  {\bibfnamefont {K.}~\bibnamefont {Jahn}}, \bibinfo {author} {\bibfnamefont
  {Y.}~\bibnamefont {Liu}}, \ and\ \bibinfo {author} {\bibfnamefont
  {H.}~\bibnamefont {Yan}},\ }\href {\doibase 10.1021/nl901165f} {\bibfield
  {journal} {\bibinfo  {journal} {Nano. Lett}\ }\textbf {\bibinfo {volume}
  {9}},\ \bibinfo {pages} {2445} (\bibinfo {year} {2009})}\BibitemShut
  {NoStop}%
\bibitem [{\citenamefont {Douglas}\ \emph {et~al.}(2009)\citenamefont
  {Douglas}, \citenamefont {Dietz}, \citenamefont {Liedl}, \citenamefont
  {H\"{o}gberg}, \citenamefont {Graf},\ and\ \citenamefont
  {Shih}}]{douglas2009}%
  \BibitemOpen
  \bibfield  {author} {\bibinfo {author} {\bibfnamefont {S.~M.}\ \bibnamefont
  {Douglas}}, \bibinfo {author} {\bibfnamefont {H.}~\bibnamefont {Dietz}},
  \bibinfo {author} {\bibfnamefont {T.}~\bibnamefont {Liedl}}, \bibinfo
  {author} {\bibfnamefont {B.}~\bibnamefont {H\"{o}gberg}}, \bibinfo {author}
  {\bibfnamefont {F.}~\bibnamefont {Graf}}, \ and\ \bibinfo {author}
  {\bibfnamefont {W.~M.}\ \bibnamefont {Shih}},\ }\href {\doibase
  10.1038/nature08016} {\bibfield  {journal} {\bibinfo  {journal} {Nature}\
  }\textbf {\bibinfo {volume} {459}},\ \bibinfo {pages} {414} (\bibinfo {year}
  {2009})}\BibitemShut {NoStop}%
\bibitem [{\citenamefont {Dietz}, \citenamefont {Douglas},\ and\ \citenamefont
  {Shih}(2009)}]{dietz2009}%
  \BibitemOpen
  \bibfield  {author} {\bibinfo {author} {\bibfnamefont {H.}~\bibnamefont
  {Dietz}}, \bibinfo {author} {\bibfnamefont {S.~M.}\ \bibnamefont {Douglas}},
  \ and\ \bibinfo {author} {\bibfnamefont {W.~M.}\ \bibnamefont {Shih}},\
  }\href {\doibase 10.1126/science.1174251} {\bibfield  {journal} {\bibinfo
  {journal} {Science}\ }\textbf {\bibinfo {volume} {325}},\ \bibinfo {pages}
  {725} (\bibinfo {year} {2009})}\BibitemShut {NoStop}%
\bibitem [{\citenamefont {Han}\ \emph {et~al.}(2011)\citenamefont {Han},
  \citenamefont {Pal}, \citenamefont {Nangreave}, \citenamefont {Deng},
  \citenamefont {Liu},\ and\ \citenamefont {Yan}}]{han2011}%
  \BibitemOpen
  \bibfield  {author} {\bibinfo {author} {\bibfnamefont {D.}~\bibnamefont
  {Han}}, \bibinfo {author} {\bibfnamefont {S.}~\bibnamefont {Pal}}, \bibinfo
  {author} {\bibfnamefont {J.}~\bibnamefont {Nangreave}}, \bibinfo {author}
  {\bibfnamefont {Z.}~\bibnamefont {Deng}}, \bibinfo {author} {\bibfnamefont
  {Y.}~\bibnamefont {Liu}}, \ and\ \bibinfo {author} {\bibfnamefont
  {H.}~\bibnamefont {Yan}},\ }\href@noop {} {\bibfield  {journal} {\bibinfo
  {journal} {Science}\ }\textbf {\bibinfo {volume} {332}},\ \bibinfo {pages}
  {342} (\bibinfo {year} {2011})}\BibitemShut {NoStop}%
\bibitem [{\citenamefont {Liedl}\ \emph {et~al.}(2010)\citenamefont {Liedl},
  \citenamefont {H{\"o}gberg}, \citenamefont {Tytell}, \citenamefont {Ingber},\
  and\ \citenamefont {Shih}}]{liedl2010}%
  \BibitemOpen
  \bibfield  {author} {\bibinfo {author} {\bibfnamefont {T.}~\bibnamefont
  {Liedl}}, \bibinfo {author} {\bibfnamefont {B.}~\bibnamefont {H{\"o}gberg}},
  \bibinfo {author} {\bibfnamefont {J.}~\bibnamefont {Tytell}}, \bibinfo
  {author} {\bibfnamefont {D.~E.}\ \bibnamefont {Ingber}}, \ and\ \bibinfo
  {author} {\bibfnamefont {W.~M.}\ \bibnamefont {Shih}},\ }\href@noop {}
  {\bibfield  {journal} {\bibinfo  {journal} {Nat. Nanotechnol}\ }\textbf
  {\bibinfo {volume} {5}},\ \bibinfo {pages} {520} (\bibinfo {year}
  {2010})}\BibitemShut {NoStop}%
\bibitem [{\citenamefont {Barish}\ \emph {et~al.}(2009)\citenamefont {Barish},
  \citenamefont {Schulman}, \citenamefont {Rothemund},\ and\ \citenamefont
  {E}}]{schulman2009}%
  \BibitemOpen
  \bibfield  {author} {\bibinfo {author} {\bibfnamefont {R.~D.}\ \bibnamefont
  {Barish}}, \bibinfo {author} {\bibfnamefont {R.}~\bibnamefont {Schulman}},
  \bibinfo {author} {\bibfnamefont {P.~W.~K.}\ \bibnamefont {Rothemund}}, \
  and\ \bibinfo {author} {\bibfnamefont {W.}~\bibnamefont {E}},\ }\href
  {\doibase 10.1073/pnas.0808736106} {\bibfield  {journal} {\bibinfo  {journal}
  {Proc. Natl. Acad. Sci. USA}\ }\textbf {\bibinfo {volume} {106}},\ \bibinfo
  {pages} {6054} (\bibinfo {year} {2009})}\BibitemShut {NoStop}%
\bibitem [{\citenamefont {LaBean}\ \emph {et~al.}(2000)\citenamefont {LaBean},
  \citenamefont {Yan}, \citenamefont {Kopatsch}, \citenamefont {Liu},
  \citenamefont {Winfree}, \citenamefont {Reif},\ and\ \citenamefont
  {Seeman}}]{labean2000}%
  \BibitemOpen
  \bibfield  {author} {\bibinfo {author} {\bibfnamefont {T.~H.}\ \bibnamefont
  {LaBean}}, \bibinfo {author} {\bibfnamefont {H.}~\bibnamefont {Yan}},
  \bibinfo {author} {\bibfnamefont {J.}~\bibnamefont {Kopatsch}}, \bibinfo
  {author} {\bibfnamefont {F.}~\bibnamefont {Liu}}, \bibinfo {author}
  {\bibfnamefont {E.}~\bibnamefont {Winfree}}, \bibinfo {author} {\bibfnamefont
  {J.~H.}\ \bibnamefont {Reif}}, \ and\ \bibinfo {author} {\bibfnamefont
  {N.~C.}\ \bibnamefont {Seeman}},\ }\href {\doibase 10.1021/ja993393e}
  {\bibfield  {journal} {\bibinfo  {journal} {J. Am. Chem. Soc}\ }\textbf
  {\bibinfo {volume} {122}},\ \bibinfo {pages} {1848} (\bibinfo {year}
  {2000})}\BibitemShut {NoStop}%
\bibitem [{\citenamefont {Wang}\ \emph {et~al.}(2019)\citenamefont {Wang},
  \citenamefont {Chen}, \citenamefont {Byoungkwon}, \citenamefont {Huang},
  \citenamefont {Bai}, \citenamefont {Xu}, \citenamefont {Bellot},
  \citenamefont {Ke}, \citenamefont {Xiang},\ and\ \citenamefont
  {Wei}}]{wang2019}%
  \BibitemOpen
  \bibfield  {author} {\bibinfo {author} {\bibfnamefont {W.}~\bibnamefont
  {Wang}}, \bibinfo {author} {\bibfnamefont {S.}~\bibnamefont {Chen}}, \bibinfo
  {author} {\bibfnamefont {A.}~\bibnamefont {Byoungkwon}}, \bibinfo {author}
  {\bibfnamefont {K.}~\bibnamefont {Huang}}, \bibinfo {author} {\bibfnamefont
  {T.}~\bibnamefont {Bai}}, \bibinfo {author} {\bibfnamefont {M.}~\bibnamefont
  {Xu}}, \bibinfo {author} {\bibfnamefont {G.}~\bibnamefont {Bellot}}, \bibinfo
  {author} {\bibfnamefont {Y.}~\bibnamefont {Ke}}, \bibinfo {author}
  {\bibfnamefont {Y.}~\bibnamefont {Xiang}}, \ and\ \bibinfo {author}
  {\bibfnamefont {B.}~\bibnamefont {Wei}},\ }\href {\doibase
  10.1038/s41467-019-08647-7} {\bibfield  {journal} {\bibinfo  {journal} {Nat.
  Commun}\ }\textbf {\bibinfo {volume} {10}},\ \bibinfo {pages} {1067}
  (\bibinfo {year} {2019})}\BibitemShut {NoStop}%
\bibitem [{\citenamefont {Zhang}\ \emph {et~al.}(2015)\citenamefont {Zhang},
  \citenamefont {Jiang}, \citenamefont {Wu}, \citenamefont {Li}, \citenamefont
  {Mao}, \citenamefont {Liu},\ and\ \citenamefont {Yan}}]{zhang2015}%
  \BibitemOpen
  \bibfield  {author} {\bibinfo {author} {\bibfnamefont {F.}~\bibnamefont
  {Zhang}}, \bibinfo {author} {\bibfnamefont {S.}~\bibnamefont {Jiang}},
  \bibinfo {author} {\bibfnamefont {S.}~\bibnamefont {Wu}}, \bibinfo {author}
  {\bibfnamefont {Y.}~\bibnamefont {Li}}, \bibinfo {author} {\bibfnamefont
  {C.}~\bibnamefont {Mao}}, \bibinfo {author} {\bibfnamefont {Y.}~\bibnamefont
  {Liu}}, \ and\ \bibinfo {author} {\bibfnamefont {H.}~\bibnamefont {Yan}},\
  }\href {\doibase 10.1038/nnano.2015.162} {\bibfield  {journal} {\bibinfo
  {journal} {Nature Nanotechnology}\ }\textbf {\bibinfo {volume} {10}},\
  \bibinfo {pages} {779} (\bibinfo {year} {2015})}\BibitemShut {NoStop}%
\bibitem [{\citenamefont {Benson}\ \emph {et~al.}(2015)\citenamefont {Benson},
  \citenamefont {Mohammed}, \citenamefont {Gardell}, \citenamefont {Masich},
  \citenamefont {Czeizler}, \citenamefont {Orponen},\ and\ \citenamefont
  {H\"ogberg}}]{benson2015}%
  \BibitemOpen
  \bibfield  {author} {\bibinfo {author} {\bibfnamefont {E.}~\bibnamefont
  {Benson}}, \bibinfo {author} {\bibfnamefont {A.}~\bibnamefont {Mohammed}},
  \bibinfo {author} {\bibfnamefont {J.}~\bibnamefont {Gardell}}, \bibinfo
  {author} {\bibfnamefont {S.}~\bibnamefont {Masich}}, \bibinfo {author}
  {\bibfnamefont {E.}~\bibnamefont {Czeizler}}, \bibinfo {author}
  {\bibfnamefont {P.}~\bibnamefont {Orponen}}, \ and\ \bibinfo {author}
  {\bibfnamefont {B.}~\bibnamefont {H\"ogberg}},\ }\href {\doibase
  10.1038/nature14586} {\bibfield  {journal} {\bibinfo  {journal} {Nature}\
  }\textbf {\bibinfo {volume} {523}},\ \bibinfo {pages} {441} (\bibinfo {year}
  {2015})}\BibitemShut {NoStop}%
\bibitem [{\citenamefont {Hong}\ \emph {et~al.}(2016)\citenamefont {Hong},
  \citenamefont {Jiang}, \citenamefont {Wang}, \citenamefont {Liu},\ and\
  \citenamefont {Yan}}]{hong2016}%
  \BibitemOpen
  \bibfield  {author} {\bibinfo {author} {\bibfnamefont {F.}~\bibnamefont
  {Hong}}, \bibinfo {author} {\bibfnamefont {S.}~\bibnamefont {Jiang}},
  \bibinfo {author} {\bibfnamefont {T.}~\bibnamefont {Wang}}, \bibinfo {author}
  {\bibfnamefont {Y.}~\bibnamefont {Liu}}, \ and\ \bibinfo {author}
  {\bibfnamefont {H.}~\bibnamefont {Yan}},\ }\href {\doibase
  10.1002/anie.201607050} {\bibfield  {journal} {\bibinfo  {journal} {Angew.
  Chem. Int. Ed}\ }\textbf {\bibinfo {volume} {55}},\ \bibinfo {pages} {12832}
  (\bibinfo {year} {2016})}\BibitemShut {NoStop}%
\bibitem [{\citenamefont {Zhang}\ \emph {et~al.}(2002)\citenamefont {Zhang},
  \citenamefont {Yan}, \citenamefont {Shen},\ and\ \citenamefont
  {Seeman}}]{zhang2002}%
  \BibitemOpen
  \bibfield  {author} {\bibinfo {author} {\bibfnamefont {X.}~\bibnamefont
  {Zhang}}, \bibinfo {author} {\bibfnamefont {H.}~\bibnamefont {Yan}}, \bibinfo
  {author} {\bibfnamefont {Z.}~\bibnamefont {Shen}}, \ and\ \bibinfo {author}
  {\bibfnamefont {N.~C.}\ \bibnamefont {Seeman}},\ }\href {\doibase
  10.1021/ja026973b} {\bibfield  {journal} {\bibinfo  {journal} {J. Am. Chem.
  Soc}\ }\textbf {\bibinfo {volume} {124}},\ \bibinfo {pages} {12940} (\bibinfo
  {year} {2002})}\BibitemShut {NoStop}%
\bibitem [{\citenamefont {Qian}\ \emph {et~al.}(2012)\citenamefont {Qian},
  \citenamefont {Yu}, \citenamefont {Wang}, \citenamefont {Dong},\ and\
  \citenamefont {C}}]{qian2012}%
  \BibitemOpen
  \bibfield  {author} {\bibinfo {author} {\bibfnamefont {H.}~\bibnamefont
  {Qian}}, \bibinfo {author} {\bibfnamefont {J.}~\bibnamefont {Yu}}, \bibinfo
  {author} {\bibfnamefont {P.}~\bibnamefont {Wang}}, \bibinfo {author}
  {\bibfnamefont {Q.~F.}\ \bibnamefont {Dong}}, \ and\ \bibinfo {author}
  {\bibfnamefont {M.}~\bibnamefont {C}},\ }\href {\doibase 10.1039/C2CC37106E}
  {\bibfield  {journal} {\bibinfo  {journal} {ChemComm}\ }\textbf {\bibinfo
  {volume} {48}},\ \bibinfo {pages} {12216} (\bibinfo {year}
  {2012})}\BibitemShut {NoStop}%
\bibitem [{\citenamefont {Yan}\ and\ \citenamefont {Seeman}(2001)}]{yan2001}%
  \BibitemOpen
  \bibfield  {author} {\bibinfo {author} {\bibfnamefont {H.}~\bibnamefont
  {Yan}}\ and\ \bibinfo {author} {\bibfnamefont {N.~C.}\ \bibnamefont
  {Seeman}},\ }\href {\doibase https://doi.org/10.1016/S1472-7862(02)00031-X}
  {\bibfield  {journal} {\bibinfo  {journal} {J. Supramol. Chem.}\ }\textbf
  {\bibinfo {volume} {1}},\ \bibinfo {pages} {229 } (\bibinfo {year}
  {2001})}\BibitemShut {NoStop}%
\bibitem [{\citenamefont {Gerling}\ \emph {et~al.}(2015)\citenamefont
  {Gerling}, \citenamefont {Wagenbauer}, \citenamefont {Neuner},\ and\
  \citenamefont {Dietz}}]{gerling2015}%
  \BibitemOpen
  \bibfield  {author} {\bibinfo {author} {\bibfnamefont {T.}~\bibnamefont
  {Gerling}}, \bibinfo {author} {\bibfnamefont {K.~F.}\ \bibnamefont
  {Wagenbauer}}, \bibinfo {author} {\bibfnamefont {A.}~\bibnamefont {Neuner}},
  \ and\ \bibinfo {author} {\bibfnamefont {H.}~\bibnamefont {Dietz}},\ }\href
  {\doibase 10.1080/07391102.2015.1032688} {\bibfield  {journal} {\bibinfo
  {journal} {J. Biomol. Struct. Dyn}\ }\textbf {\bibinfo {volume} {33}},\
  \bibinfo {pages} {46} (\bibinfo {year} {2015})}\BibitemShut {NoStop}%
\bibitem [{\citenamefont {Hamada}\ and\ \citenamefont
  {Murata}(2009)}]{hamada2009}%
  \BibitemOpen
  \bibfield  {author} {\bibinfo {author} {\bibfnamefont {S.}~\bibnamefont
  {Hamada}}\ and\ \bibinfo {author} {\bibfnamefont {S.}~\bibnamefont
  {Murata}},\ }\href {\doibase 10.1002/ange.200902662} {\bibfield  {journal}
  {\bibinfo  {journal} {Angew. Chem. Int. Ed}\ }\textbf {\bibinfo {volume}
  {121}},\ \bibinfo {pages} {6952} (\bibinfo {year} {2009})}\BibitemShut
  {NoStop}%
\bibitem [{\citenamefont {Li}\ \emph {et~al.}(2017)\citenamefont {Li},
  \citenamefont {Zuo}, \citenamefont {Yu}, \citenamefont {Zhao},\ and\
  \citenamefont {C}}]{li2017}%
  \BibitemOpen
  \bibfield  {author} {\bibinfo {author} {\bibfnamefont {M.}~\bibnamefont
  {Li}}, \bibinfo {author} {\bibfnamefont {H.}~\bibnamefont {Zuo}}, \bibinfo
  {author} {\bibfnamefont {J.}~\bibnamefont {Yu}}, \bibinfo {author}
  {\bibfnamefont {X.}~\bibnamefont {Zhao}}, \ and\ \bibinfo {author}
  {\bibfnamefont {M.}~\bibnamefont {C}},\ }\href {\doibase 10.1039/C7NR03640J}
  {\bibfield  {journal} {\bibinfo  {journal} {Nanoscale}\ }\textbf {\bibinfo
  {volume} {9}},\ \bibinfo {pages} {10601} (\bibinfo {year}
  {2017})}\BibitemShut {NoStop}%
\bibitem [{\citenamefont {Li}\ \emph {et~al.}(2012)\citenamefont {Li},
  \citenamefont {Zhang}, \citenamefont {Hao}, \citenamefont {Tian},
  \citenamefont {Wang},\ and\ \citenamefont {Mao}}]{mao2012}%
  \BibitemOpen
  \bibfield  {author} {\bibinfo {author} {\bibfnamefont {X.}~\bibnamefont
  {Li}}, \bibinfo {author} {\bibfnamefont {C.}~\bibnamefont {Zhang}}, \bibinfo
  {author} {\bibfnamefont {C.}~\bibnamefont {Hao}}, \bibinfo {author}
  {\bibfnamefont {C.}~\bibnamefont {Tian}}, \bibinfo {author} {\bibfnamefont
  {G.}~\bibnamefont {Wang}}, \ and\ \bibinfo {author} {\bibfnamefont
  {C.}~\bibnamefont {Mao}},\ }\href {\doibase 10.1021/nn300813w} {\bibfield
  {journal} {\bibinfo  {journal} {ACS Nano}\ }\textbf {\bibinfo {volume} {6}},\
  \bibinfo {pages} {5138} (\bibinfo {year} {2012})}\BibitemShut {NoStop}%
\bibitem [{\citenamefont {Zuo}\ \emph {et~al.}(2015)\citenamefont {Zuo},
  \citenamefont {Wu}, \citenamefont {Li}, \citenamefont {Li}, \citenamefont
  {Jiang},\ and\ \citenamefont {Mao}}]{mao2015}%
  \BibitemOpen
  \bibfield  {author} {\bibinfo {author} {\bibfnamefont {H.}~\bibnamefont
  {Zuo}}, \bibinfo {author} {\bibfnamefont {S.}~\bibnamefont {Wu}}, \bibinfo
  {author} {\bibfnamefont {M.}~\bibnamefont {Li}}, \bibinfo {author}
  {\bibfnamefont {Y.}~\bibnamefont {Li}}, \bibinfo {author} {\bibfnamefont
  {W.}~\bibnamefont {Jiang}}, \ and\ \bibinfo {author} {\bibfnamefont
  {C.}~\bibnamefont {Mao}},\ }\href {\doibase 10.1002/anie.201507375}
  {\bibfield  {journal} {\bibinfo  {journal} {Angew. Chem. Int. Ed}\ }\textbf
  {\bibinfo {volume} {54}},\ \bibinfo {pages} {15118} (\bibinfo {year}
  {2015})}\BibitemShut {NoStop}%
\bibitem [{\citenamefont {Tandon}\ \emph {et~al.}(2019)\citenamefont {Tandon},
  \citenamefont {Kim}, \citenamefont {Song}, \citenamefont {Cho}, \citenamefont
  {Bashar}, \citenamefont {Shin}, \citenamefont {Ha},\ and\ \citenamefont
  {Park}}]{tandon2019}%
  \BibitemOpen
  \bibfield  {author} {\bibinfo {author} {\bibfnamefont {A.}~\bibnamefont
  {Tandon}}, \bibinfo {author} {\bibfnamefont {S.}~\bibnamefont {Kim}},
  \bibinfo {author} {\bibfnamefont {Y.}~\bibnamefont {Song}}, \bibinfo {author}
  {\bibfnamefont {H.}~\bibnamefont {Cho}}, \bibinfo {author} {\bibfnamefont
  {S.}~\bibnamefont {Bashar}}, \bibinfo {author} {\bibfnamefont
  {J.}~\bibnamefont {Shin}}, \bibinfo {author} {\bibfnamefont {T.~H.}\
  \bibnamefont {Ha}}, \ and\ \bibinfo {author} {\bibfnamefont {S.~H.}\
  \bibnamefont {Park}},\ }\href {\doibase 10.1038/s41598-019-38699-0}
  {\bibfield  {journal} {\bibinfo  {journal} {Nat. Sci. Rep}\ }\textbf
  {\bibinfo {volume} {9}},\ \bibinfo {pages} {2252} (\bibinfo {year}
  {2019})}\BibitemShut {NoStop}%
\bibitem [{\citenamefont {Ouldridge}, \citenamefont {Louis},\ and\
  \citenamefont {Doye}(2011)}]{ouldridge2011}%
  \BibitemOpen
  \bibfield  {author} {\bibinfo {author} {\bibfnamefont {T.~E.}\ \bibnamefont
  {Ouldridge}}, \bibinfo {author} {\bibfnamefont {A.~A.}\ \bibnamefont
  {Louis}}, \ and\ \bibinfo {author} {\bibfnamefont {J.~P.~K.}\ \bibnamefont
  {Doye}},\ }\href {\doibase 10.1063/1.3552946} {\bibfield  {journal} {\bibinfo
   {journal} {J. Chem. Phys}\ }\textbf {\bibinfo {volume} {134}},\ \bibinfo
  {pages} {085101} (\bibinfo {year} {2011})}\BibitemShut {NoStop}%
\bibitem [{\citenamefont {\v{S}ulc}\ \emph {et~al.}(2012)\citenamefont
  {\v{S}ulc}, \citenamefont {Romano}, \citenamefont {Ouldridge}, \citenamefont
  {Rovigatti}, \citenamefont {Doye},\ and\ \citenamefont {Louis}}]{sulc2012}%
  \BibitemOpen
  \bibfield  {author} {\bibinfo {author} {\bibfnamefont {P.}~\bibnamefont
  {\v{S}ulc}}, \bibinfo {author} {\bibfnamefont {F.}~\bibnamefont {Romano}},
  \bibinfo {author} {\bibfnamefont {T.~E.}\ \bibnamefont {Ouldridge}}, \bibinfo
  {author} {\bibfnamefont {L.}~\bibnamefont {Rovigatti}}, \bibinfo {author}
  {\bibfnamefont {J.~P.~K.}\ \bibnamefont {Doye}}, \ and\ \bibinfo {author}
  {\bibfnamefont {A.~A.}\ \bibnamefont {Louis}},\ }\href {\doibase
  10.1063/1.4754132} {\bibfield  {journal} {\bibinfo  {journal} {J. Chem.
  Phys}\ }\textbf {\bibinfo {volume} {137}},\ \bibinfo {pages} {135101}
  (\bibinfo {year} {2012})}\BibitemShut {NoStop}%
\bibitem [{\citenamefont {Doye}\ \emph {et~al.}(2013)\citenamefont {Doye},
  \citenamefont {Ouldridge}, \citenamefont {Louis}, \citenamefont {Romano},
  \citenamefont {\v{S}ulc}, \citenamefont {Matek}, \citenamefont {Snodin},
  \citenamefont {Rovigatti}, \citenamefont {Schreck}, \citenamefont
  {Harrison},\ and\ \citenamefont {Smith}}]{doye2013}%
  \BibitemOpen
  \bibfield  {author} {\bibinfo {author} {\bibfnamefont {J.~P.~K.}\
  \bibnamefont {Doye}}, \bibinfo {author} {\bibfnamefont {T.~E.}\ \bibnamefont
  {Ouldridge}}, \bibinfo {author} {\bibfnamefont {A.~A.}\ \bibnamefont
  {Louis}}, \bibinfo {author} {\bibfnamefont {F.}~\bibnamefont {Romano}},
  \bibinfo {author} {\bibfnamefont {P.}~\bibnamefont {\v{S}ulc}}, \bibinfo
  {author} {\bibfnamefont {C.}~\bibnamefont {Matek}}, \bibinfo {author}
  {\bibfnamefont {B.~E.~K.}\ \bibnamefont {Snodin}}, \bibinfo {author}
  {\bibfnamefont {L.}~\bibnamefont {Rovigatti}}, \bibinfo {author}
  {\bibfnamefont {J.~S.}\ \bibnamefont {Schreck}}, \bibinfo {author}
  {\bibfnamefont {R.~M.}\ \bibnamefont {Harrison}}, \ and\ \bibinfo {author}
  {\bibfnamefont {W.~P.~J.}\ \bibnamefont {Smith}},\ }\href {\doibase
  10.1039/c3cp53545b} {\bibfield  {journal} {\bibinfo  {journal} {Phys. Chem.
  Chem. Phys}\ }\textbf {\bibinfo {volume} {15}},\ \bibinfo {pages} {20381}
  (\bibinfo {year} {2013})}\BibitemShut {NoStop}%
\bibitem [{\citenamefont {Snodin}\ \emph {et~al.}(2015)\citenamefont {Snodin},
  \citenamefont {Randisi}, \citenamefont {Mosayebi}, \citenamefont {\v{S}ulc},
  \citenamefont {Schreck}, \citenamefont {Romano}, \citenamefont {Ouldridge},
  \citenamefont {Tsukanov}, \citenamefont {Nir}, \citenamefont {Louis},\ and\
  \citenamefont {Doye}}]{snodin2015}%
  \BibitemOpen
  \bibfield  {author} {\bibinfo {author} {\bibfnamefont {B.~E.~K.}\
  \bibnamefont {Snodin}}, \bibinfo {author} {\bibfnamefont {F.}~\bibnamefont
  {Randisi}}, \bibinfo {author} {\bibfnamefont {M.}~\bibnamefont {Mosayebi}},
  \bibinfo {author} {\bibfnamefont {P.}~\bibnamefont {\v{S}ulc}}, \bibinfo
  {author} {\bibfnamefont {J.~S.}\ \bibnamefont {Schreck}}, \bibinfo {author}
  {\bibfnamefont {F.}~\bibnamefont {Romano}}, \bibinfo {author} {\bibfnamefont
  {T.~E.}\ \bibnamefont {Ouldridge}}, \bibinfo {author} {\bibfnamefont
  {R.}~\bibnamefont {Tsukanov}}, \bibinfo {author} {\bibfnamefont
  {E.}~\bibnamefont {Nir}}, \bibinfo {author} {\bibfnamefont {A.~A.}\
  \bibnamefont {Louis}}, \ and\ \bibinfo {author} {\bibfnamefont {J.~P.~K.}\
  \bibnamefont {Doye}},\ }\href {\doibase 10.1063/1.4921957} {\bibfield
  {journal} {\bibinfo  {journal} {J. Chem. Phys.}\ }\textbf {\bibinfo {volume}
  {142}},\ \bibinfo {pages} {234901} (\bibinfo {year} {2015})}\BibitemShut
  {NoStop}%
\bibitem [{\citenamefont {Yoo}\ and\ \citenamefont
  {Aksimentiev}(2013)}]{yoo2013}%
  \BibitemOpen
  \bibfield  {author} {\bibinfo {author} {\bibfnamefont {J.}~\bibnamefont
  {Yoo}}\ and\ \bibinfo {author} {\bibfnamefont {A.}~\bibnamefont
  {Aksimentiev}},\ }\href {\doibase 10.1073/pnas.1316521110} {\bibfield
  {journal} {\bibinfo  {journal} {Proc. Natl. Acad. Sci. USA}\ }\textbf
  {\bibinfo {volume} {110}},\ \bibinfo {pages} {20099} (\bibinfo {year}
  {2013})}\BibitemShut {NoStop}%
\bibitem [{\citenamefont {G\"opfrich}\ \emph {et~al.}(2016)\citenamefont
  {G\"opfrich}, \citenamefont {Li}, \citenamefont {Ricci}, \citenamefont
  {Bhamidimarri}, \citenamefont {Yoo}, \citenamefont {Gyenes}, \citenamefont
  {Ohmann}, \citenamefont {Winterhalter}, \citenamefont {Aksimentiev},\ and\
  \citenamefont {Keyser}}]{gopfrich2016}%
  \BibitemOpen
  \bibfield  {author} {\bibinfo {author} {\bibfnamefont {K.}~\bibnamefont
  {G\"opfrich}}, \bibinfo {author} {\bibfnamefont {C.~Y.}\ \bibnamefont {Li}},
  \bibinfo {author} {\bibfnamefont {M.}~\bibnamefont {Ricci}}, \bibinfo
  {author} {\bibfnamefont {S.~P.}\ \bibnamefont {Bhamidimarri}}, \bibinfo
  {author} {\bibfnamefont {J.}~\bibnamefont {Yoo}}, \bibinfo {author}
  {\bibfnamefont {B.}~\bibnamefont {Gyenes}}, \bibinfo {author} {\bibfnamefont
  {A.}~\bibnamefont {Ohmann}}, \bibinfo {author} {\bibfnamefont
  {M.}~\bibnamefont {Winterhalter}}, \bibinfo {author} {\bibfnamefont
  {A.}~\bibnamefont {Aksimentiev}}, \ and\ \bibinfo {author} {\bibfnamefont
  {U.~F.}\ \bibnamefont {Keyser}},\ }\href {\doibase 10.1021/acsnano.6b03759}
  {\bibfield  {journal} {\bibinfo  {journal} {ACS Nano}\ }\textbf {\bibinfo
  {volume} {10}},\ \bibinfo {pages} {8207} (\bibinfo {year}
  {2016})}\BibitemShut {NoStop}%
\bibitem [{\citenamefont {Wu}\ \emph {et~al.}(2013)\citenamefont {Wu},
  \citenamefont {Czajkowsky}, \citenamefont {Zhang}, \citenamefont {Qu},
  \citenamefont {Ye}, \citenamefont {Zeng}, \citenamefont {Zhou}, \citenamefont
  {Hu}, \citenamefont {Shao}, \citenamefont {Li},\ and\ \citenamefont
  {Fan}}]{wu2013}%
  \BibitemOpen
  \bibfield  {author} {\bibinfo {author} {\bibfnamefont {N.}~\bibnamefont
  {Wu}}, \bibinfo {author} {\bibfnamefont {D.~M.}\ \bibnamefont {Czajkowsky}},
  \bibinfo {author} {\bibfnamefont {J.}~\bibnamefont {Zhang}}, \bibinfo
  {author} {\bibfnamefont {J.}~\bibnamefont {Qu}}, \bibinfo {author}
  {\bibfnamefont {M.}~\bibnamefont {Ye}}, \bibinfo {author} {\bibfnamefont
  {D.}~\bibnamefont {Zeng}}, \bibinfo {author} {\bibfnamefont {X.}~\bibnamefont
  {Zhou}}, \bibinfo {author} {\bibfnamefont {J.}~\bibnamefont {Hu}}, \bibinfo
  {author} {\bibfnamefont {Z.}~\bibnamefont {Shao}}, \bibinfo {author}
  {\bibfnamefont {B.}~\bibnamefont {Li}}, \ and\ \bibinfo {author}
  {\bibfnamefont {C.}~\bibnamefont {Fan}},\ }\href {\doibase 10.1021/ja403863a}
  {\bibfield  {journal} {\bibinfo  {journal} {J. Am. Chem. Soc.}\ }\textbf
  {\bibinfo {volume} {135}},\ \bibinfo {pages} {12172} (\bibinfo {year}
  {2013})}\BibitemShut {NoStop}%
\bibitem [{\citenamefont {Lee}, \citenamefont {Lee},\ and\ \citenamefont
  {Kim}(2017)}]{lee2017}%
  \BibitemOpen
  \bibfield  {author} {\bibinfo {author} {\bibfnamefont {C.}~\bibnamefont
  {Lee}}, \bibinfo {author} {\bibfnamefont {J.~Y.}\ \bibnamefont {Lee}}, \ and\
  \bibinfo {author} {\bibfnamefont {D.~N.}\ \bibnamefont {Kim}},\ }\href
  {\doibase 10.1038/s41467-017-02127-6} {\bibfield  {journal} {\bibinfo
  {journal} {Nat. Commun}\ }\textbf {\bibinfo {volume} {8}},\ \bibinfo {pages}
  {2067} (\bibinfo {year} {2017})}\BibitemShut {NoStop}%
\bibitem [{\citenamefont {Penna}\ and\ \citenamefont
  {Perico}(2010)}]{penna2010}%
  \BibitemOpen
  \bibfield  {author} {\bibinfo {author} {\bibfnamefont {G.~L.}\ \bibnamefont
  {Penna}}\ and\ \bibinfo {author} {\bibfnamefont {A.}~\bibnamefont {Perico}},\
  }\href {\doibase 10.1016/j.bpj.2010.03.024} {\bibfield  {journal} {\bibinfo
  {journal} {Biophys. J}\ }\textbf {\bibinfo {volume} {98}},\ \bibinfo {pages}
  {2964} (\bibinfo {year} {2010})}\BibitemShut {NoStop}%
\bibitem [{\citenamefont {Castro}\ \emph {et~al.}(2011)\citenamefont {Castro},
  \citenamefont {Kilchherr}, \citenamefont {Kim}, \citenamefont {Shiao},
  \citenamefont {Wauer}, \citenamefont {Wortmann}, \citenamefont {Bathe},\ and\
  \citenamefont {Dietz}}]{castro2011}%
  \BibitemOpen
  \bibfield  {author} {\bibinfo {author} {\bibfnamefont {C.~E.}\ \bibnamefont
  {Castro}}, \bibinfo {author} {\bibfnamefont {F.}~\bibnamefont {Kilchherr}},
  \bibinfo {author} {\bibfnamefont {D.~N.}\ \bibnamefont {Kim}}, \bibinfo
  {author} {\bibfnamefont {E.~L.}\ \bibnamefont {Shiao}}, \bibinfo {author}
  {\bibfnamefont {T.}~\bibnamefont {Wauer}}, \bibinfo {author} {\bibfnamefont
  {P.}~\bibnamefont {Wortmann}}, \bibinfo {author} {\bibfnamefont
  {M.}~\bibnamefont {Bathe}}, \ and\ \bibinfo {author} {\bibfnamefont
  {H.}~\bibnamefont {Dietz}},\ }\href {\doibase 10.1038/nmeth.1570} {\bibfield
  {journal} {\bibinfo  {journal} {Nat. Methods}\ }\textbf {\bibinfo {volume}
  {8}},\ \bibinfo {pages} {221} (\bibinfo {year} {2011})}\BibitemShut {NoStop}%
\bibitem [{\citenamefont {Ouldridge}\ \emph
  {et~al.}(2013{\natexlab{a}})\citenamefont {Ouldridge}, \citenamefont
  {\v{S}ulc}, \citenamefont {Romano}, \citenamefont {Doye},\ and\ \citenamefont
  {Louis}}]{ouldridge2013}%
  \BibitemOpen
  \bibfield  {author} {\bibinfo {author} {\bibfnamefont {T.~E.}\ \bibnamefont
  {Ouldridge}}, \bibinfo {author} {\bibfnamefont {P.}~\bibnamefont {\v{S}ulc}},
  \bibinfo {author} {\bibfnamefont {F.}~\bibnamefont {Romano}}, \bibinfo
  {author} {\bibfnamefont {J.~P.~K.}\ \bibnamefont {Doye}}, \ and\ \bibinfo
  {author} {\bibfnamefont {A.~A.}\ \bibnamefont {Louis}},\ }\href {\doibase
  10.1093/nar/gkt687} {\bibfield  {journal} {\bibinfo  {journal} {Nucleic Acids
  Res}\ }\textbf {\bibinfo {volume} {41}},\ \bibinfo {pages} {8886} (\bibinfo
  {year} {2013}{\natexlab{a}})}\BibitemShut {NoStop}%
\bibitem [{\citenamefont {Srinivas}\ \emph {et~al.}(2013)\citenamefont
  {Srinivas}, \citenamefont {Ouldridge}, \citenamefont {\v{S}ulc},
  \citenamefont {Schaeffer}, \citenamefont {Yurke}, \citenamefont {Louis},
  \citenamefont {Doye},\ and\ \citenamefont {Winfree}}]{srinivas2013}%
  \BibitemOpen
  \bibfield  {author} {\bibinfo {author} {\bibfnamefont {N.}~\bibnamefont
  {Srinivas}}, \bibinfo {author} {\bibfnamefont {T.~E.}\ \bibnamefont
  {Ouldridge}}, \bibinfo {author} {\bibfnamefont {P.}~\bibnamefont {\v{S}ulc}},
  \bibinfo {author} {\bibfnamefont {J.~M.}\ \bibnamefont {Schaeffer}}, \bibinfo
  {author} {\bibfnamefont {B.}~\bibnamefont {Yurke}}, \bibinfo {author}
  {\bibfnamefont {A.~A.}\ \bibnamefont {Louis}}, \bibinfo {author}
  {\bibfnamefont {J.~P.~K.}\ \bibnamefont {Doye}}, \ and\ \bibinfo {author}
  {\bibfnamefont {E.}~\bibnamefont {Winfree}},\ }\href@noop {} {\bibfield
  {journal} {\bibinfo  {journal} {Nucleic Acids Res}\ }\textbf {\bibinfo
  {volume} {41}},\ \bibinfo {pages} {10641} (\bibinfo {year}
  {2013})}\BibitemShut {NoStop}%
\bibitem [{\citenamefont {Mosayebi}\ \emph {et~al.}(2014)\citenamefont
  {Mosayebi}, \citenamefont {Romano}, \citenamefont {Ouldridge}, \citenamefont
  {Doye},\ and\ \citenamefont {Louis}}]{mosayebi2014}%
  \BibitemOpen
  \bibfield  {author} {\bibinfo {author} {\bibfnamefont {M.}~\bibnamefont
  {Mosayebi}}, \bibinfo {author} {\bibfnamefont {F.}~\bibnamefont {Romano}},
  \bibinfo {author} {\bibfnamefont {T.~E.}\ \bibnamefont {Ouldridge}}, \bibinfo
  {author} {\bibfnamefont {J.~P.~K.}\ \bibnamefont {Doye}}, \ and\ \bibinfo
  {author} {\bibfnamefont {A.~A.}\ \bibnamefont {Louis}},\ }\href {\doibase
  10.1021/jp510061f} {\bibfield  {journal} {\bibinfo  {journal} {J. Chem. Phys.
  B.}\ }\textbf {\bibinfo {volume} {118}},\ \bibinfo {pages} {14326} (\bibinfo
  {year} {2014})}\BibitemShut {NoStop}%
\bibitem [{\citenamefont {Ouldridge}, \citenamefont {Louis},\ and\
  \citenamefont {K}(2010)}]{ouldridge2010}%
  \BibitemOpen
  \bibfield  {author} {\bibinfo {author} {\bibfnamefont {T.~E.}\ \bibnamefont
  {Ouldridge}}, \bibinfo {author} {\bibfnamefont {A.~A.}\ \bibnamefont
  {Louis}}, \ and\ \bibinfo {author} {\bibfnamefont {D.~J.~P.}\ \bibnamefont
  {K}},\ }\href@noop {} {\bibfield  {journal} {\bibinfo  {journal} {Phys. Rev.
  Lett.}\ }\textbf {\bibinfo {volume} {104}},\ \bibinfo {pages} {178101}
  (\bibinfo {year} {2010})}\BibitemShut {NoStop}%
\bibitem [{\citenamefont {Ouldridge}\ \emph
  {et~al.}(2013{\natexlab{b}})\citenamefont {Ouldridge}, \citenamefont {Hoare},
  \citenamefont {Louis}, \citenamefont {Doye}, \citenamefont {Bath},\ and\
  \citenamefont {Turberfield}}]{ouldridge2013_walker}%
  \BibitemOpen
  \bibfield  {author} {\bibinfo {author} {\bibfnamefont {T.~E.}\ \bibnamefont
  {Ouldridge}}, \bibinfo {author} {\bibfnamefont {R.~L.}\ \bibnamefont
  {Hoare}}, \bibinfo {author} {\bibfnamefont {A.~A.}\ \bibnamefont {Louis}},
  \bibinfo {author} {\bibfnamefont {J.~P.~K.}\ \bibnamefont {Doye}}, \bibinfo
  {author} {\bibfnamefont {J.}~\bibnamefont {Bath}}, \ and\ \bibinfo {author}
  {\bibfnamefont {A.~J.}\ \bibnamefont {Turberfield}},\ }\href {\doibase
  10.1021/nn3058483} {\bibfield  {journal} {\bibinfo  {journal} {ACS Nano}\
  }\textbf {\bibinfo {volume} {7}},\ \bibinfo {pages} {2479} (\bibinfo {year}
  {2013}{\natexlab{b}})}\BibitemShut {NoStop}%
\bibitem [{\citenamefont {\v{S}ulc}\ \emph {et~al.}(2014)\citenamefont
  {\v{S}ulc}, \citenamefont {Ouldridge}, \citenamefont {Romano}, \citenamefont
  {Doye},\ and\ \citenamefont {Louis}}]{sulc2014}%
  \BibitemOpen
  \bibfield  {author} {\bibinfo {author} {\bibfnamefont {P.}~\bibnamefont
  {\v{S}ulc}}, \bibinfo {author} {\bibfnamefont {T.~E.}\ \bibnamefont
  {Ouldridge}}, \bibinfo {author} {\bibfnamefont {F.}~\bibnamefont {Romano}},
  \bibinfo {author} {\bibfnamefont {J.~P.~K.}\ \bibnamefont {Doye}}, \ and\
  \bibinfo {author} {\bibfnamefont {A.~A.}\ \bibnamefont {Louis}},\ }\href@noop
  {} {\bibfield  {journal} {\bibinfo  {journal} {Natural Computing}\ }\textbf
  {\bibinfo {volume} {13}},\ \bibinfo {pages} {535} (\bibinfo {year}
  {2014})}\BibitemShut {NoStop}%
\bibitem [{\citenamefont {Snodin}\ \emph {et~al.}(2016)\citenamefont {Snodin},
  \citenamefont {Romano}, \citenamefont {Rovigatti}, \citenamefont {Ouldridge},
  \citenamefont {Louis},\ and\ \citenamefont {Doye}}]{snodin2016}%
  \BibitemOpen
  \bibfield  {author} {\bibinfo {author} {\bibfnamefont {B.~E.~K.}\
  \bibnamefont {Snodin}}, \bibinfo {author} {\bibfnamefont {F.}~\bibnamefont
  {Romano}}, \bibinfo {author} {\bibfnamefont {L.}~\bibnamefont {Rovigatti}},
  \bibinfo {author} {\bibfnamefont {T.~E.}\ \bibnamefont {Ouldridge}}, \bibinfo
  {author} {\bibfnamefont {A.~A.}\ \bibnamefont {Louis}}, \ and\ \bibinfo
  {author} {\bibfnamefont {J.~P.~K.}\ \bibnamefont {Doye}},\ }\href {\doibase
  10.1021/acsnano.5b05865} {\bibfield  {journal} {\bibinfo  {journal} {ACS
  Nano}\ }\textbf {\bibinfo {volume} {10}},\ \bibinfo {pages} {1724} (\bibinfo
  {year} {2016})}\BibitemShut {NoStop}%
\bibitem [{\citenamefont {Fonseca}\ \emph {et~al.}()\citenamefont {Fonseca},
  \citenamefont {Romano}, \citenamefont {Schreck}, \citenamefont {Ouldridge},
  \citenamefont {Doye},\ and\ \citenamefont {Louis}}]{fonseca2018}%
  \BibitemOpen
  \bibfield  {author} {\bibinfo {author} {\bibfnamefont {P.}~\bibnamefont
  {Fonseca}}, \bibinfo {author} {\bibfnamefont {F.}~\bibnamefont {Romano}},
  \bibinfo {author} {\bibfnamefont {J.~S.}\ \bibnamefont {Schreck}}, \bibinfo
  {author} {\bibfnamefont {T.~E.}\ \bibnamefont {Ouldridge}}, \bibinfo {author}
  {\bibfnamefont {J.~P.~K.}\ \bibnamefont {Doye}}, \ and\ \bibinfo {author}
  {\bibfnamefont {A.~A.}\ \bibnamefont {Louis}},\ }\href@noop {} {\bibfield
  {journal} {\bibinfo  {journal} {J. Chem. Phys.}\ }\textbf {\bibinfo {volume}
  {148}},\ \bibinfo {pages} {134910}}\BibitemShut {NoStop}%
\bibitem [{\citenamefont {Schreck}\ \emph {et~al.}(2015)\citenamefont
  {Schreck}, \citenamefont {Ouldridge}, \citenamefont {Romano}, \citenamefont
  {Louis},\ and\ \citenamefont {Doye}}]{schreck2015}%
  \BibitemOpen
  \bibfield  {author} {\bibinfo {author} {\bibfnamefont {J.~S.}\ \bibnamefont
  {Schreck}}, \bibinfo {author} {\bibfnamefont {T.~E.}\ \bibnamefont
  {Ouldridge}}, \bibinfo {author} {\bibfnamefont {F.}~\bibnamefont {Romano}},
  \bibinfo {author} {\bibfnamefont {A.~A.}\ \bibnamefont {Louis}}, \ and\
  \bibinfo {author} {\bibfnamefont {J.~P.~K.}\ \bibnamefont {Doye}},\ }\href
  {\doibase 10.1063/1.4917199} {\bibfield  {journal} {\bibinfo  {journal} {J.
  Chem. Phys}\ }\textbf {\bibinfo {volume} {142}},\ \bibinfo {pages} {165101}
  (\bibinfo {year} {2015})}\BibitemShut {NoStop}%
\bibitem [{\citenamefont {Macchion}, \citenamefont {Str\"omberg},\ and\
  \citenamefont {Nilsson}(2008)}]{macchion2008}%
  \BibitemOpen
  \bibfield  {author} {\bibinfo {author} {\bibfnamefont {B.~N.}\ \bibnamefont
  {Macchion}}, \bibinfo {author} {\bibfnamefont {R.}~\bibnamefont
  {Str\"omberg}}, \ and\ \bibinfo {author} {\bibfnamefont {L.}~\bibnamefont
  {Nilsson}},\ }\href {\doibase 10.1080/07391102.2008.10507232} {\bibfield
  {journal} {\bibinfo  {journal} {J. Biomol. Struct. Dyn}\ }\textbf {\bibinfo
  {volume} {26}},\ \bibinfo {pages} {163} (\bibinfo {year} {2008})}\BibitemShut
  {NoStop}%
\bibitem [{\citenamefont {Whitelam}\ \emph {et~al.}(2008)\citenamefont
  {Whitelam}, \citenamefont {Feng}, \citenamefont {Hagan},\ and\ \citenamefont
  {Geissler}}]{whitelam2008}%
  \BibitemOpen
  \bibfield  {author} {\bibinfo {author} {\bibfnamefont {S.}~\bibnamefont
  {Whitelam}}, \bibinfo {author} {\bibfnamefont {E.~H.}\ \bibnamefont {Feng}},
  \bibinfo {author} {\bibfnamefont {M.~F.}\ \bibnamefont {Hagan}}, \ and\
  \bibinfo {author} {\bibfnamefont {P.~L.}\ \bibnamefont {Geissler}},\ }\href
  {\doibase 10.1039/b810031d} {\bibfield  {journal} {\bibinfo  {journal} {Soft
  Matter}\ }\textbf {\bibinfo {volume} {5}},\ \bibinfo {pages} {1251} (\bibinfo
  {year} {2008})}\BibitemShut {NoStop}%
\bibitem [{\citenamefont {Torrie}\ and\ \citenamefont
  {Valleau}(1977)}]{torrie1977}%
  \BibitemOpen
  \bibfield  {author} {\bibinfo {author} {\bibfnamefont {G.~M.}\ \bibnamefont
  {Torrie}}\ and\ \bibinfo {author} {\bibfnamefont {J.~P.}\ \bibnamefont
  {Valleau}},\ }\href {\doibase 10.1016/0021-9991(77)90121-8} {\bibfield
  {journal} {\bibinfo  {journal} {J. Comput. Phys}\ }\textbf {\bibinfo {volume}
  {23}},\ \bibinfo {pages} {187} (\bibinfo {year} {1977})}\BibitemShut
  {NoStop}%
\bibitem [{\citenamefont {Ouldridge}, \citenamefont {Louis},\ and\
  \citenamefont {Doye}(2010)}]{Ouldridge2010bulk}%
  \BibitemOpen
  \bibfield  {author} {\bibinfo {author} {\bibfnamefont {T.~E.}\ \bibnamefont
  {Ouldridge}}, \bibinfo {author} {\bibfnamefont {A.~A.}\ \bibnamefont
  {Louis}}, \ and\ \bibinfo {author} {\bibfnamefont {J.~P.~K.}\ \bibnamefont
  {Doye}},\ }\href {\doibase 10.1088/0953-8984/22/10/104102} {\bibfield
  {journal} {\bibinfo  {journal} {Journal of Physics: Condensed Matter}\
  }\textbf {\bibinfo {volume} {22}},\ \bibinfo {pages} {104102} (\bibinfo
  {year} {2010})}\BibitemShut {NoStop}%
\bibitem [{\citenamefont {Kumar}\ \emph {et~al.}(1992)\citenamefont {Kumar},
  \citenamefont {Rosenberg}, \citenamefont {Bouzida}, \citenamefont
  {Swendsen},\ and\ \citenamefont {Kollman}}]{kumar1992}%
  \BibitemOpen
  \bibfield  {author} {\bibinfo {author} {\bibfnamefont {S.}~\bibnamefont
  {Kumar}}, \bibinfo {author} {\bibfnamefont {J.~M.}\ \bibnamefont
  {Rosenberg}}, \bibinfo {author} {\bibfnamefont {D.}~\bibnamefont {Bouzida}},
  \bibinfo {author} {\bibfnamefont {R.~H.}\ \bibnamefont {Swendsen}}, \ and\
  \bibinfo {author} {\bibfnamefont {P.~A.}\ \bibnamefont {Kollman}},\ }\href
  {\doibase 10.1002/jcc.540130812} {\bibfield  {journal} {\bibinfo  {journal}
  {J. Comput. Chem}\ }\textbf {\bibinfo {volume} {13}},\ \bibinfo {pages}
  {1011} (\bibinfo {year} {1992})}\BibitemShut {NoStop}%
\bibitem [{\citenamefont {Bosco}, \citenamefont {Camunas-Soler},\ and\
  \citenamefont {Ritort}(2014)}]{bosco2014}%
  \BibitemOpen
  \bibfield  {author} {\bibinfo {author} {\bibfnamefont {A.}~\bibnamefont
  {Bosco}}, \bibinfo {author} {\bibfnamefont {J.}~\bibnamefont
  {Camunas-Soler}}, \ and\ \bibinfo {author} {\bibfnamefont {F.}~\bibnamefont
  {Ritort}},\ }\href {\doibase 10.1093/nar/gkt1089} {\bibfield  {journal}
  {\bibinfo  {journal} {Nucleic Acids Res}\ }\textbf {\bibinfo {volume} {42}},\
  \bibinfo {pages} {2064} (\bibinfo {year} {2014})}\BibitemShut {NoStop}%
\end{thebibliography}%


%merlin.mbs aipnum4-1.bst 2010-07-25 4.21a (PWD, AO, DPC) hacked
%Control: key (0)
%Control: author (8) initials jnrlst
%Control: editor formatted (1) identically to author
%Control: production of article title (-1) disabled
%Control: page (0) single
%Control: year (1) truncated
%Control: production of eprint (0) enabled
\begin{thebibliography}{6}%
\makeatletter
\providecommand \@ifxundefined [1]{%
 \@ifx{#1\undefined}
}%
\providecommand \@ifnum [1]{%
 \ifnum #1\expandafter \@firstoftwo
 \else \expandafter \@secondoftwo
 \fi
}%
\providecommand \@ifx [1]{%
 \ifx #1\expandafter \@firstoftwo
 \else \expandafter \@secondoftwo
 \fi
}%
\providecommand \natexlab [1]{#1}%
\providecommand \enquote  [1]{``#1''}%
\providecommand \bibnamefont  [1]{#1}%
\providecommand \bibfnamefont [1]{#1}%
\providecommand \citenamefont [1]{#1}%
\providecommand \href@noop [0]{\@secondoftwo}%
\providecommand \href [0]{\begingroup \@sanitize@url \@href}%
\providecommand \@href[1]{\@@startlink{#1}\@@href}%
\providecommand \@@href[1]{\endgroup#1\@@endlink}%
\providecommand \@sanitize@url [0]{\catcode `\\12\catcode `\$12\catcode
  `\&12\catcode `\#12\catcode `\^12\catcode `\_12\catcode `\%12\relax}%
\providecommand \@@startlink[1]{}%
\providecommand \@@endlink[0]{}%
\providecommand \url  [0]{\begingroup\@sanitize@url \@url }%
\providecommand \@url [1]{\endgroup\@href {#1}{\urlprefix }}%
\providecommand \urlprefix  [0]{URL }%
\providecommand \Eprint [0]{\href }%
\providecommand \doibase [0]{http://dx.doi.org/}%
\providecommand \selectlanguage [0]{\@gobble}%
\providecommand \bibinfo  [0]{\@secondoftwo}%
\providecommand \bibfield  [0]{\@secondoftwo}%
\providecommand \translation [1]{[#1]}%
\providecommand \BibitemOpen [0]{}%
\providecommand \bibitemStop [0]{}%
\providecommand \bibitemNoStop [0]{.\EOS\space}%
\providecommand \EOS [0]{\spacefactor3000\relax}%
\providecommand \BibitemShut  [1]{\csname bibitem#1\endcsname}%
\let\auto@bib@innerbib\@empty
%</preamble>
\bibitem [{\citenamefont {Ouldridge}, \citenamefont {Louis},\ and\
  \citenamefont {Doye}(2011)}]{ouldridge2011}%
  \BibitemOpen
  \bibfield  {author} {\bibinfo {author} {\bibfnamefont {T.~E.}\ \bibnamefont
  {Ouldridge}}, \bibinfo {author} {\bibfnamefont {A.~A.}\ \bibnamefont
  {Louis}}, \ and\ \bibinfo {author} {\bibfnamefont {J.~P.~K.}\ \bibnamefont
  {Doye}},\ }\href {\doibase 10.1063/1.3552946} {\bibfield  {journal} {\bibinfo
   {journal} {J. Chem. Phys}\ }\textbf {\bibinfo {volume} {134}},\ \bibinfo
  {pages} {085101} (\bibinfo {year} {2011})}\BibitemShut {NoStop}%
\bibitem [{\citenamefont {\v{S}ulc}\ \emph {et~al.}(2012)\citenamefont
  {\v{S}ulc}, \citenamefont {Romano}, \citenamefont {Ouldridge}, \citenamefont
  {Rovigatti}, \citenamefont {Doye},\ and\ \citenamefont {Louis}}]{sulc2012}%
  \BibitemOpen
  \bibfield  {author} {\bibinfo {author} {\bibfnamefont {P.}~\bibnamefont
  {\v{S}ulc}}, \bibinfo {author} {\bibfnamefont {F.}~\bibnamefont {Romano}},
  \bibinfo {author} {\bibfnamefont {T.~E.}\ \bibnamefont {Ouldridge}}, \bibinfo
  {author} {\bibfnamefont {L.}~\bibnamefont {Rovigatti}}, \bibinfo {author}
  {\bibfnamefont {J.~P.~K.}\ \bibnamefont {Doye}}, \ and\ \bibinfo {author}
  {\bibfnamefont {A.~A.}\ \bibnamefont {Louis}},\ }\href {\doibase
  10.1063/1.4754132} {\bibfield  {journal} {\bibinfo  {journal} {J. Chem.
  Phys}\ }\textbf {\bibinfo {volume} {137}},\ \bibinfo {pages} {135101}
  (\bibinfo {year} {2012})}\BibitemShut {NoStop}%
\bibitem [{\citenamefont {Snodin}\ \emph {et~al.}(2015)\citenamefont {Snodin},
  \citenamefont {Randisi}, \citenamefont {Mosayebi}, \citenamefont {\v{S}ulc},
  \citenamefont {Schreck}, \citenamefont {Romano}, \citenamefont {Ouldridge},
  \citenamefont {Tsukanov}, \citenamefont {Nir}, \citenamefont {Louis},\ and\
  \citenamefont {Doye}}]{snodin2015}%
  \BibitemOpen
  \bibfield  {author} {\bibinfo {author} {\bibfnamefont {B.~E.~K.}\
  \bibnamefont {Snodin}}, \bibinfo {author} {\bibfnamefont {F.}~\bibnamefont
  {Randisi}}, \bibinfo {author} {\bibfnamefont {M.}~\bibnamefont {Mosayebi}},
  \bibinfo {author} {\bibfnamefont {P.}~\bibnamefont {\v{S}ulc}}, \bibinfo
  {author} {\bibfnamefont {J.~S.}\ \bibnamefont {Schreck}}, \bibinfo {author}
  {\bibfnamefont {F.}~\bibnamefont {Romano}}, \bibinfo {author} {\bibfnamefont
  {T.~E.}\ \bibnamefont {Ouldridge}}, \bibinfo {author} {\bibfnamefont
  {R.}~\bibnamefont {Tsukanov}}, \bibinfo {author} {\bibfnamefont
  {E.}~\bibnamefont {Nir}}, \bibinfo {author} {\bibfnamefont {A.~A.}\
  \bibnamefont {Louis}}, \ and\ \bibinfo {author} {\bibfnamefont {J.~P.~K.}\
  \bibnamefont {Doye}},\ }\href {\doibase 10.1063/1.4921957} {\bibfield
  {journal} {\bibinfo  {journal} {J. Chem. Phys.}\ }\textbf {\bibinfo {volume}
  {142}},\ \bibinfo {pages} {234901} (\bibinfo {year} {2015})}\BibitemShut
  {NoStop}%
\bibitem [{\citenamefont {Torrie}\ and\ \citenamefont
  {Valleau}(1977)}]{torrie1977}%
  \BibitemOpen
  \bibfield  {author} {\bibinfo {author} {\bibfnamefont {G.~M.}\ \bibnamefont
  {Torrie}}\ and\ \bibinfo {author} {\bibfnamefont {J.~P.}\ \bibnamefont
  {Valleau}},\ }\href {\doibase 10.1016/0021-9991(77)90121-8} {\bibfield
  {journal} {\bibinfo  {journal} {J. Comput. Phys}\ }\textbf {\bibinfo {volume}
  {23}},\ \bibinfo {pages} {187} (\bibinfo {year} {1977})}\BibitemShut
  {NoStop}%
\bibitem [{\citenamefont {Whitelam}\ \emph {et~al.}(2008)\citenamefont
  {Whitelam}, \citenamefont {Feng}, \citenamefont {Hagan},\ and\ \citenamefont
  {Geissler}}]{whitelam2008}%
  \BibitemOpen
  \bibfield  {author} {\bibinfo {author} {\bibfnamefont {S.}~\bibnamefont
  {Whitelam}}, \bibinfo {author} {\bibfnamefont {E.~H.}\ \bibnamefont {Feng}},
  \bibinfo {author} {\bibfnamefont {M.~F.}\ \bibnamefont {Hagan}}, \ and\
  \bibinfo {author} {\bibfnamefont {P.~L.}\ \bibnamefont {Geissler}},\ }\href
  {\doibase 10.1039/b810031d} {\bibfield  {journal} {\bibinfo  {journal} {Soft
  Matter}\ }\textbf {\bibinfo {volume} {5}},\ \bibinfo {pages} {1251} (\bibinfo
  {year} {2008})}\BibitemShut {NoStop}%
\bibitem [{\citenamefont {Kumar}\ \emph {et~al.}(1992)\citenamefont {Kumar},
  \citenamefont {Rosenberg}, \citenamefont {Bouzida}, \citenamefont
  {Swendsen},\ and\ \citenamefont {Kollman}}]{kumar1992}%
  \BibitemOpen
  \bibfield  {author} {\bibinfo {author} {\bibfnamefont {S.}~\bibnamefont
  {Kumar}}, \bibinfo {author} {\bibfnamefont {J.~M.}\ \bibnamefont
  {Rosenberg}}, \bibinfo {author} {\bibfnamefont {D.}~\bibnamefont {Bouzida}},
  \bibinfo {author} {\bibfnamefont {R.~H.}\ \bibnamefont {Swendsen}}, \ and\
  \bibinfo {author} {\bibfnamefont {P.~A.}\ \bibnamefont {Kollman}},\ }\href
  {\doibase 10.1002/jcc.540130812} {\bibfield  {journal} {\bibinfo  {journal}
  {J. Comput. Chem}\ }\textbf {\bibinfo {volume} {13}},\ \bibinfo {pages}
  {1011} (\bibinfo {year} {1992})}\BibitemShut {NoStop}%
\end{thebibliography}%

\end{document}

%\preprint{AIP/123-QED}

\title[Characterising DNA T-motifs by Simulation and Experiment]{Supplementary Material for Characterising DNA T-motifs by Simulation and Experiment\vspace{0.5em}}% Force line breaks with \\

%\thanks{Footnote to title of article.}

\author{Behnam Najafi}
%\email{behnam.najafi@physics.ox.ac.uk}
\affiliation{ 
Clarendon Laboratory, Department of Physics, University of Oxford, Parks Rd, Oxford OX1 3PU, United Kingdom
}
\author{Katherine G. Young}
%\email{katherine.young@physics.ox.ac.uk}
\affiliation{ 
Clarendon Laboratory, Department of Physics, University of Oxford, Parks Rd, Oxford OX1 3PU, United Kingdom
}
\author{Jonathan Bath}
\affiliation{ 
Clarendon Laboratory, Department of Physics, University of Oxford, Parks Rd, Oxford OX1 3PU, United Kingdom
}
\author{Ard A. Louis}
\affiliation{ 
Rudolph Peierls Centre for Theoretical Physics, University of Oxford, Keble Road, Oxford OX1 3NP, United Kingdom
}
\author{Jonathan P. K. Doye}
\affiliation{ 
Physical and Theoretical Chemistry Laboratory, Department of Chemistry, University of Oxford, South Parks Road, Oxford OX1 3QZ, United Kingdom
}
\author{Andrew J. Turberfield}
\email{andrew.turberfield@physics.ox.ac.uk}
\affiliation{ 
Clarendon Laboratory, Department of Physics, University of Oxford, Parks Rd, Oxford OX1 3PU, United Kingdom
}
 \homepage{}

\date{\today}% It is always \today, today,
             %  but any date may be explicitly specified

%\pacs{Valid PACS appear here}% PACS, the Physics and Astronomy
                             % Classification Scheme.
%\keywords{DNA Nanotechnology, DNA Origami, Self-assembly, DNA T-junctions}%Use showkeys class option if keyword
                              %display desired
\maketitle

\renewcommand{\thesection}{S\Roman{section}}
\renewcommand\thefigure{S\arabic{figure}}
\renewcommand\thetable{S\arabic{table}}

\tableofcontents

%%%%%%%%%%%%%%%%%%%%%%%%%%%%%%%%%%%%%%%%%%%%%%%%%%%%%%%%%%%%
%%%%%%%%%%%%%%%%%%%%%%%%%%%%%%%%%%%%%%%%%%%%%%%%%%%%%%%%%%%%
\section{Theoretical Methods}

%%%%%%%%%%%%%%%%%%%%%%%%%%%%%%%%%%%%%%%%%%%%%%%%%%%%%%%%%%%%
\subsection{The oxDNA Model}\label{app::oxDNA}

\begin{figure}[h]
	\centering
		\includegraphics[width=0.5\textwidth]{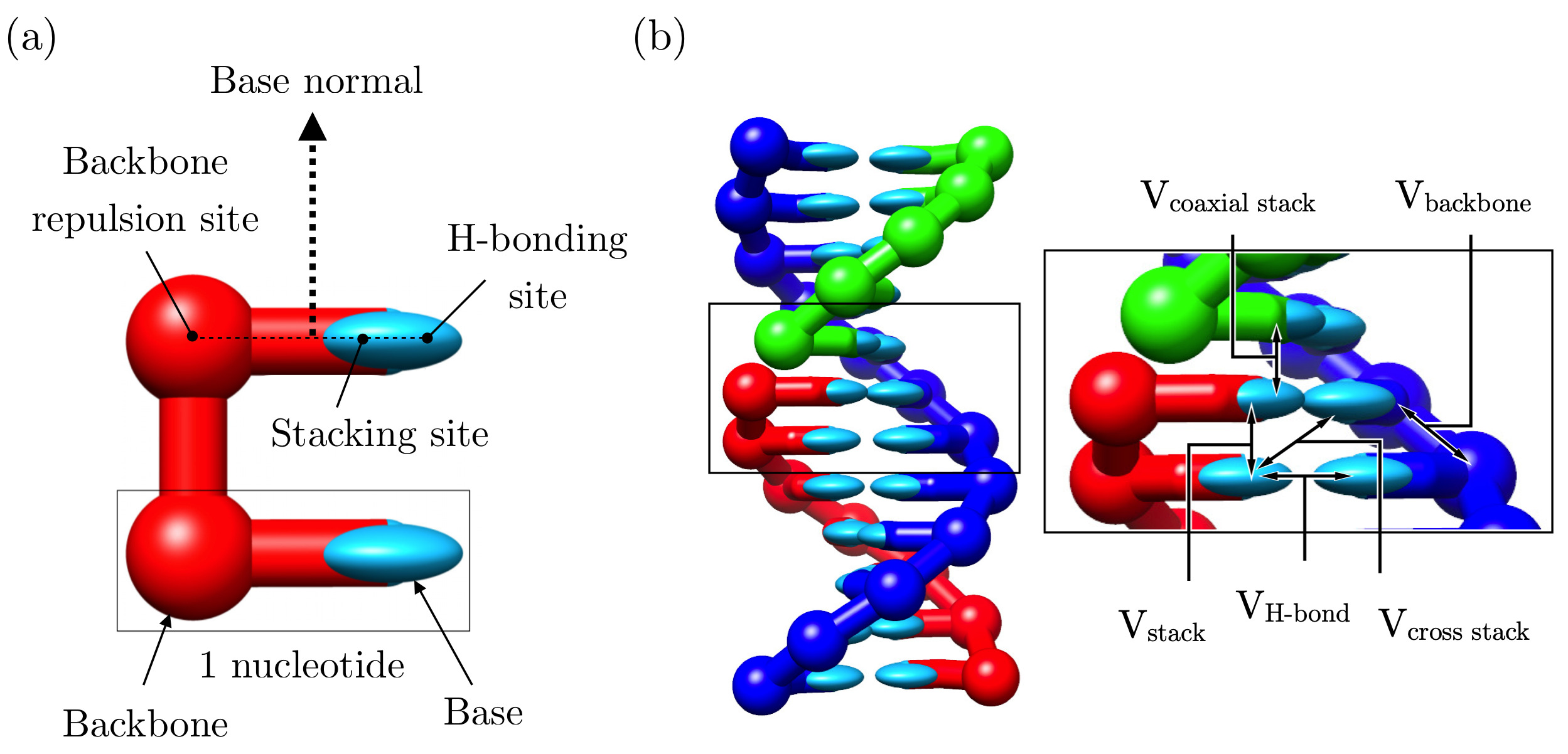}
	\caption{Simplified representation of (a) the rigid nucleotides that are the basic units in oxDNA and (b) an 11 base-pair double helix that illustrates the various interactions in the oxDNA model. Produced by Domen Pre\v sern.}\label{fig::oxDNA}
\end{figure}

The fundamental unit of the coarse-grained oxDNA model\cite{ouldridge2011,sulc2012,snodin2015} is a rigid nucleotide composed of three interaction sites (Figure \ref{fig::oxDNA}). Interaction potentials representing backbone linkages, hydrogen bonds and stacking interactions are designed to provide a physically realistic description of the effects of the underlying chemical interactions; interaction energies have been fitted to experimental data in order to reproduce much of the structural, mechanical and thermodynamic properties of single- and double-stranded DNA,\cite{ouldridge2011} and to incorporate the sequence-dependence of Watson-Crick base paring interaction strengths.\cite{sulc2012} 

The orientation of each base is specified by a normal vector, giving a notional plane for the base. The relative angles of these planes are used to modulate interactions. For example, $V_\text{H-Bond}$ is modulated by angular terms that favour co-linear alignment of of all four backbone and hydrogen-bonding sites and anti-alignment of the normal vectors, while $V_\text{stack}$ is modulated by angular terms that favour the alignment of base normals with the vector between stacking sites, encouraging co-planar stacks.

In this work, we use the most recent version of oxDNA, described in Snodin et al\cite{snodin2015} (sometimes called ``oxDNA2''), in which different widths for the major and minor double helical grooves were introduced resulting in a more realistic description of duplex DNA geometry. This was achieved by adjusting the position of the backbone site without changing the duplex radius, causing the three interaction sites to lie in a plane rather than a straight line. Additionally, new experimental data was used to re-parametrise the oxDNA potential to account for the different stacking strengths of consecutive adenine/thymine bases along single strands. However, coaxial stacking interaction strengths in the model remain independent of base identity.

%%%%%%%%%%%%%%%%%%%%%%%%%%%%%%%%%%%%%%%%%%%%%%%%%%%%%%%%%%%%
\subsection{Umbrella Sampling} \label{app::US}

To calculate free-energy profiles during formation of the T-motifs, we use the umbrella sampling (US) technique.\cite{torrie1977} This allows us to efficiently sample states involved in hybridization events by biasing simulations towards meta stable states, reducing waiting times by lowering energy barriers. Typically, an order parameter $Q(\mathbf{x}^N)$, where $\mathbf{x}^N$ represents the coordinates of the system, is used to define macrostates of the system. A bias energy $W(Q)$ is applied to the sampling distribution, leading to biased partial partition functions of the form:
\begin{equation}\label{eq::zbiased}
Z_Q^{\text{biased}}= \int d\mathbf{x}^N \delta_{Q,Q'}e^{-\beta[V(\mathbf{x}^N) - W(Q'(\mathbf{x}^N))]} \text{,}
\end{equation}
where $\beta=1/k_BT$ and $V(\mathbf{x}^N))$ is the potential energy. The biasing weights $e^{\beta W(Q(\mathbf{x}^N))}$ are chosen such that $Z_Q^{\text{biased}}$ remains approximately constant in $Q$. The unbiased partial partition function can be recovered as $Z_Q = Z_Q^{\text{biased}} / e^{\beta W(Q(\mathbf{x}^N))}$.

%%%%%%%%%%%%%%%%%%%%%%%%%%%%%%%%%%%%%%%%%%%%%%%%%%%%%%%%%%%%
\subsection{Simulation Details} \label{app::sim}

All simulations are performed at 25$\degree$C in cubic boxes with periodic boundary conditions corresponding to strand concentration of 42 $\mu$M. The sequence of bulge loops and sticky ends with lengths $L\in\{5,6,7\}$ are the same as those used in experimental measurements of $K_d$; nucleotides were added/removed for other bulge lengths as shown in Table \ref{table::oxseq}.

For each T-motif, the system is prepared in states in which all native base pairs in the bulge-loop duplex and the sticky-end duplex have formed. Virtual Move Monte Carlo (VMMC)\cite{whitelam2008} trajectories were simulated in two umbrella sampling windows (Figure \ref{fig::US}). In the first window, until the first base pair is formed, the minimum distance between the nucleotides on the sticky end and those of the bulge loop is used as an order parameter and biasing weights are applied to the distribution to increase sampling of states with small separation and to efficiently sample diffusive states. In the second window, the number of base-pairs between the bulge-loop and the sticky end is used as an order parameter. Each simulation window was repeated five times, with each run consisting of $\sim10^9$ VMMC steps. The weighted histogram analysis method (WHAM)\cite{kumar1992} was used to link the US windows through shared states and obtain free energy profiles. To obtain distributions for geometric quantities and stacking states, a subset of configurations was selected for analysis, allowing $\sim10^4$ steps between each configuration to minimise correlations.

%%%%%%%%%%%%%%%%%%%%%%%%%%%%%%%%%%%%%%%%%%%%%%%%%%%%%%%%%%%%
\subsection{Extrapolation to Low Temperatures} \label{app::extrapolate}

\begin{figure}[]
	\centering
		\includegraphics[trim=1cm 1cm 0.3cm 0cm, width=1\linewidth]{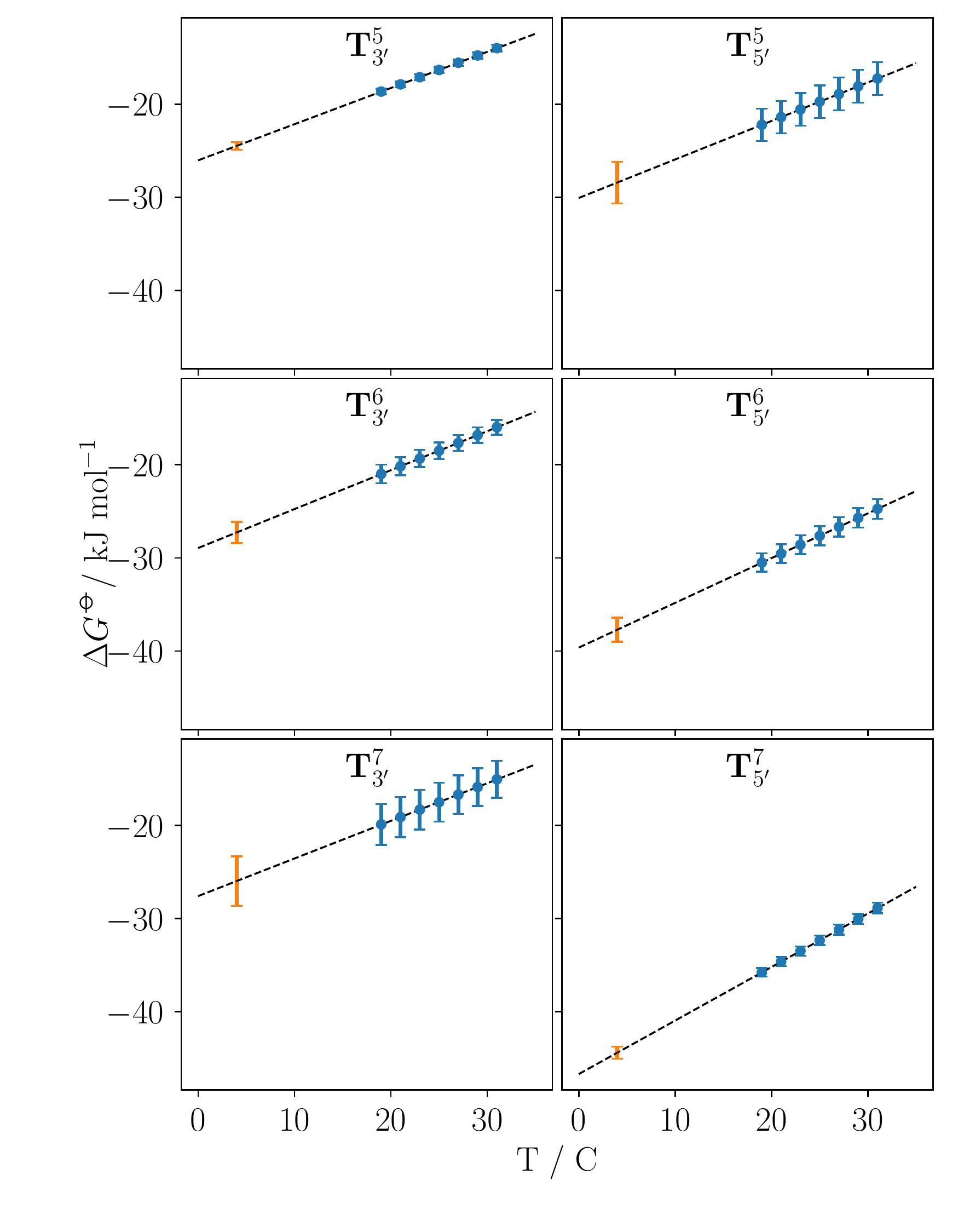}	
	\caption{
		Extrapolation of simulation results to lower temperature. During oxDNA simulations at 25$\degree$C and concentration of 42$\mu$M, changes in Gibbs free energy on T-motif formation at nearby temperatures were estimated. These results were scaled to molar concentration and extrapolated to 4$\degree$C, the temperature at which experiments were performed.
		}
	\label{fig::extrapolate}
\end{figure}

\begin{figure}[h]
	\centering
		\includegraphics[width=0.5\textwidth]{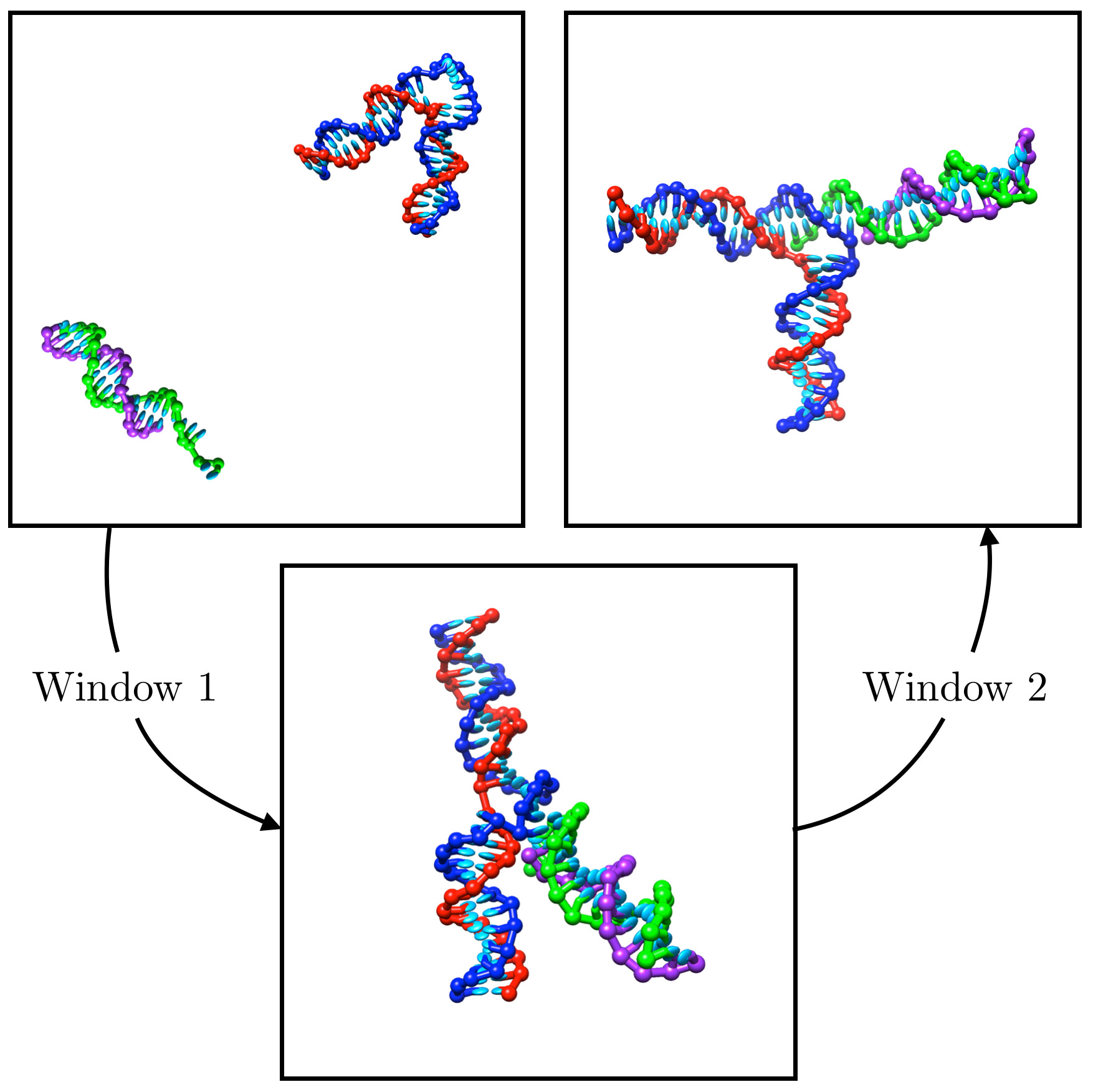}
	\caption{Sample snapshots from oxDNA simulations illustrating umbrella sampling windows.}\label{fig::US}
\end{figure}

The single-histogram re-weighting method\cite{kumar1992} was used to estimate the change in Gibbs free energy of T-motif formation at 4$\degree$C, the temperature of electrophoresis experiments, from oxDNA simulations at 25$\degree$C. During the simulation, states were grouped according to their energies $E$ and values of the order parameter of interest $Q$. We assume that the density of states is independent of temperature: the probability distribution $p(E,Q;T_0)$, where $T_0$ is the simulation temperature, can then be used to obtain the probability distribution $p(E,Q;T)\propto p(E,Q;T_0) e^{(\beta - \beta_0)E}$ at another temperature $T$. The extrapolation process is more accurate if the target temperature T is close to $T_0$ due to the increased number of shared states with energy E. We extrapolated simulation results to $T\in \{19,21,23,27,29,31\}$C to estimate the dependence of $\Delta G(n_{bp}) = G(n_{bp}) - G(n_{bp}=0)$ on temperature. Results for $\Delta G(L)$ for T-motifs of loop length $L$ are shown in Figure \ref{fig::extrapolate}. Assuming the change in entropy $\Delta S$ is constant over the range of temperatures, a linear fit was applied to the data. Extrapolated values at 4$\degree$C are used in Figure 8 of the main text.

\renewcommand{\arraystretch}{1.5}
\begin{table}[h]
\resizebox{\linewidth}{!} {\begin{tabular}{| l | l | l |}
\hline
$L$ & Strand & Sequence \\
\hline
3 & Blue & \bluestrand{ATCCTTGGGCAGCTA\underline{CTA}ATCTATAGTTCAAT} \\
   & Green 5$^\prime$ & \greenstrand{\underline{TAG}AATAAGAGACTATAC} \\ 
   & Green 3$^\prime$ & \greenstrand{AATAAGAGACTATAC\underline{TAG}} \\
\hline
4 & Blue & \bluestrand{ATCCTTGGGCAGCTA\underline{CTAC}ATCTATAGTTCAAT} \\
   & Green 5$^\prime$ & \greenstrand{\underline{GTAG}AATAAGAGACTATAC} \\ 
   & Green 3$^\prime$ & \greenstrand{AATAAGAGACTATAC\underline{GTAG}} \\ 
\hline
5 & Blue & \bluestrand{ATCCTTGGGCAGCTA\underline{CTACA}ATCTATAGTTCAAT} \\
   & Green 5$^\prime$ & \greenstrand{\underline{TGTAG}AATAAGAGACTATAC} \\ 
   & Green 3$^\prime$ & \greenstrand{AATAAGAGACTATAC\underline{TGTAG}} \\ 
\hline
6 & Blue & \bluestrand{ATCCTTGGGCAGCTA\underline{CTACAG}ATCTATAGTTCAAT} \\
   & Green 5$^\prime$ & \greenstrand{\underline{CTGTAG}AATAAGAGACTATAC} \\ 
   & Green 3$^\prime$ & \greenstrand{AATAAGAGACTATAC\underline{CTGTAG}} \\ 
\hline
7 & Blue & \bluestrand{ATCCTTGGGCAGCTA\underline{CTACAGG}ATCTATAGTTCAAT} \\
   & Green 5$^\prime$ & \greenstrand{\underline{CCTGTAG}AATAAGAGACTATAC} \\ 
   & Green 3$^\prime$ & \greenstrand{AATAAGAGACTATAC\underline{CCTGTAG}} \\ 
\hline
8 & Blue & \bluestrand{ATCCTTGGGCAGCTA\underline{CTACAGGA}ATCTATAGTTCAAT} \\
   & Green 5$^\prime$ & \greenstrand{\underline{TCCTGTAG}AATAAGAGACTATAC} \\ 
   & Green 3$^\prime$ & \greenstrand{AATAAGAGACTATAC\underline{TCCTGTAG}} \\ 
\hline
9 & Blue & \bluestrand{ATCCTTGGGCAGCTA\underline{CTACAGGAA}ATCTATAGTTCAAT} \\
   & Green 5$^\prime$ & \greenstrand{\underline{TTCCTGTAG}AATAAGAGACTATAC} \\ 
   & Green 3$^\prime$ & \greenstrand{AATAAGAGACTATAC\underline{TTCCTGTAG}} \\ 
\hline
\end{tabular}}
\caption[Table caption text]{
Strand sequences used in oxDNA simulations of T-motifs, specified from 5$^\prime$ to 3$^\prime$ direction. The blue and green strands contain the $L$ bulge-loop and sticky-end nucleotides, respectively. The two green sequences correspond to the two possible junction polarities. The strand colours are based on T-motifs shown in Figure 1 (the red and purple strands have complementary sequences to the blue and green strands). The nucleotides involved in bulge-loop / sticky-end interaction are underlined for clarity.
}
\label{table::oxseq}
\end{table}

%%%%%%%%%%%%%%%%%%%%%%%%%%%%%%%%%%%%%%%%%%%%%%%%%%%%%%%%%%%%

\section{Experimental Methods}\label{app::Exp}

%%%%%%%%%%%%%%%%%%%%%%%%%%%%%%%%%%%%%%%%%%%%%%%%%%%%%%%%%%%%
\subsection{Material Used}\label{app::Material}

\begin{figure}[h]
	\centering
		\includegraphics[width=1\linewidth]{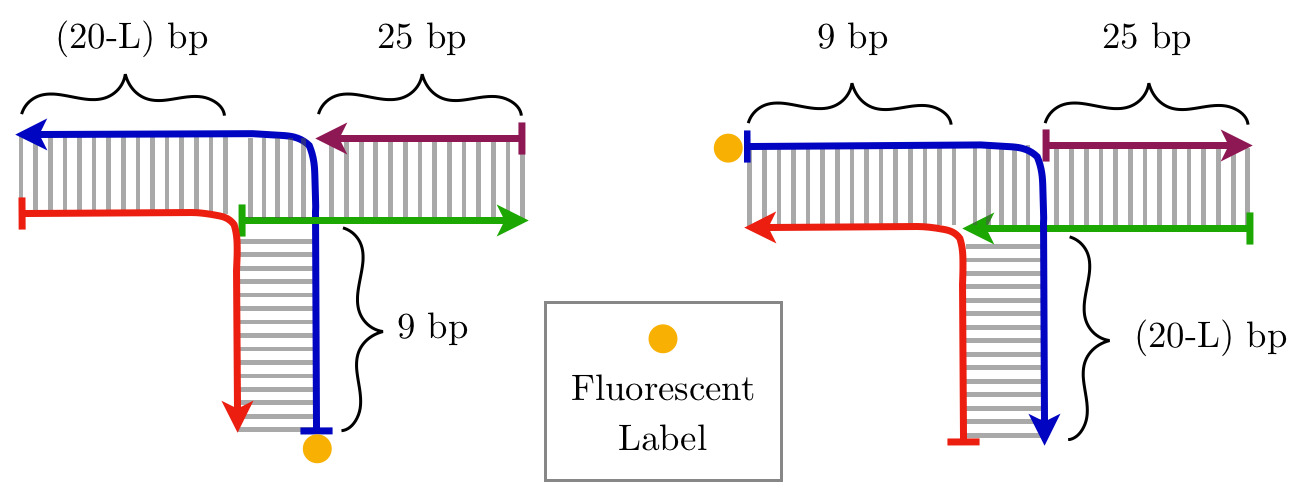}
	\caption{
	DNA strands used in experiments to measure binding strengths of $5'$ (left) and $3'$ (right) T-motifs. $L$ is the length of the bulge loop. See Table \ref{table::seq} for sequences.
}\label{fig::ExpSetup}
\end{figure}

All strands were purchased from Integrated DNA Technologies, Inc. (IDT)\cite{} and stored as 100$\mu$M stocks in TE buffer (10mM Tris base, 1mM EDTA-Na$_2$, adjusted to pH 8.0 with HCl). Each T-motif consists of four strands. The dimensions of the T-motifs used for measurements of $K_d$ are shown in Figure \ref{fig::ExpSetup} and the strand sequences are shown in Table \ref{table::seq}. The fluorescent label (FAM) is 6-carboxyfluorescein amidite.

\renewcommand{\arraystretch}{1.5}
\begin{table}[h]
\resizebox{\linewidth}{!} {\begin{tabular}{| l | l | l |}
\hline
T-motif & Strand & Sequence \\
\hline
All & Blue & \bluestrand{/6-FAM/TGGGCAGCT\underline{CTACAGG}AAATCTATAGTTC} \\
All & Purple & \purplestrand{CGTCGTTCTAGTATAGTCTCTTATT} \\
$\mathbf{T}_{5'}^{5}$,$\mathbf{T}_{3'}^{5}$ & Red & \redstrand{GAACTATAGATTTCCAGCTGCCCA} \\ 
$\mathbf{T}_{5'}^{6}$,$\mathbf{T}_{3'}^{6}$ & Red & \redstrand{GAACTATAGATTTCAGCTGCCCA} \\ 
$\mathbf{T}_{5'}^{7}$,$\mathbf{T}_{3'}^{7}$ & Red & \redstrand{GAACTATAGATTTAGCTGCCCA} \\ 
$\mathbf{T}_{3'}^{5}$ & Green & \greenstrand{AATAAGAGACTATACTAGAACGACG\underline{TGTAG}} \\ 
$\mathbf{T}_{3'}^{6}$ & Green & \greenstrand{AATAAGAGACTATACTAGAACGACG\underline{CTGTAG}} \\ 
$\mathbf{T}_{3'}^{7}$ & Green & \greenstrand{AATAAGAGACTATACTAGAACGACG\underline{CCTGTAG}} \\ 
$\mathbf{T}_{5'}^{5}$ & Green & \greenstrand{\underline{TGTAG}AATAAGAGACTATACTAGAACGACG} \\ 
$\mathbf{T}_{5'}^{6}$ & Green & \greenstrand{\underline{CTGTAG}AATAAGAGACTATACTAGAACGACG} \\ 
$\mathbf{T}_{5'}^{7}$ & Green & \greenstrand{\underline{CCTGTAG}AATAAGAGACTATACTAGAACGACG} \\ 
\hline
\end{tabular}}
\caption[Table caption text]{
	Strand sequences (5$^\prime$ to 3$^\prime$ direction) used for measurements of $K_d$, leading to free energy values in Figure 8. The nucleotides that are (potentially) involved in the hybridization of the bulge loop to the sticky end are underlined for clarity.
}
\label{table::seq}
\end{table}

%%%%%%%%%%%%%%%%%%%%%%%%%%%%%%%%%%%%%%%%%%%%%%%%%%%%%%%%%%%%
%\FloatBarrier
\subsection{Determining Electrophoretic Mobilities}\label{app::marker}

\begin{figure}[h]
	\centering
		\includegraphics[width=1\linewidth]{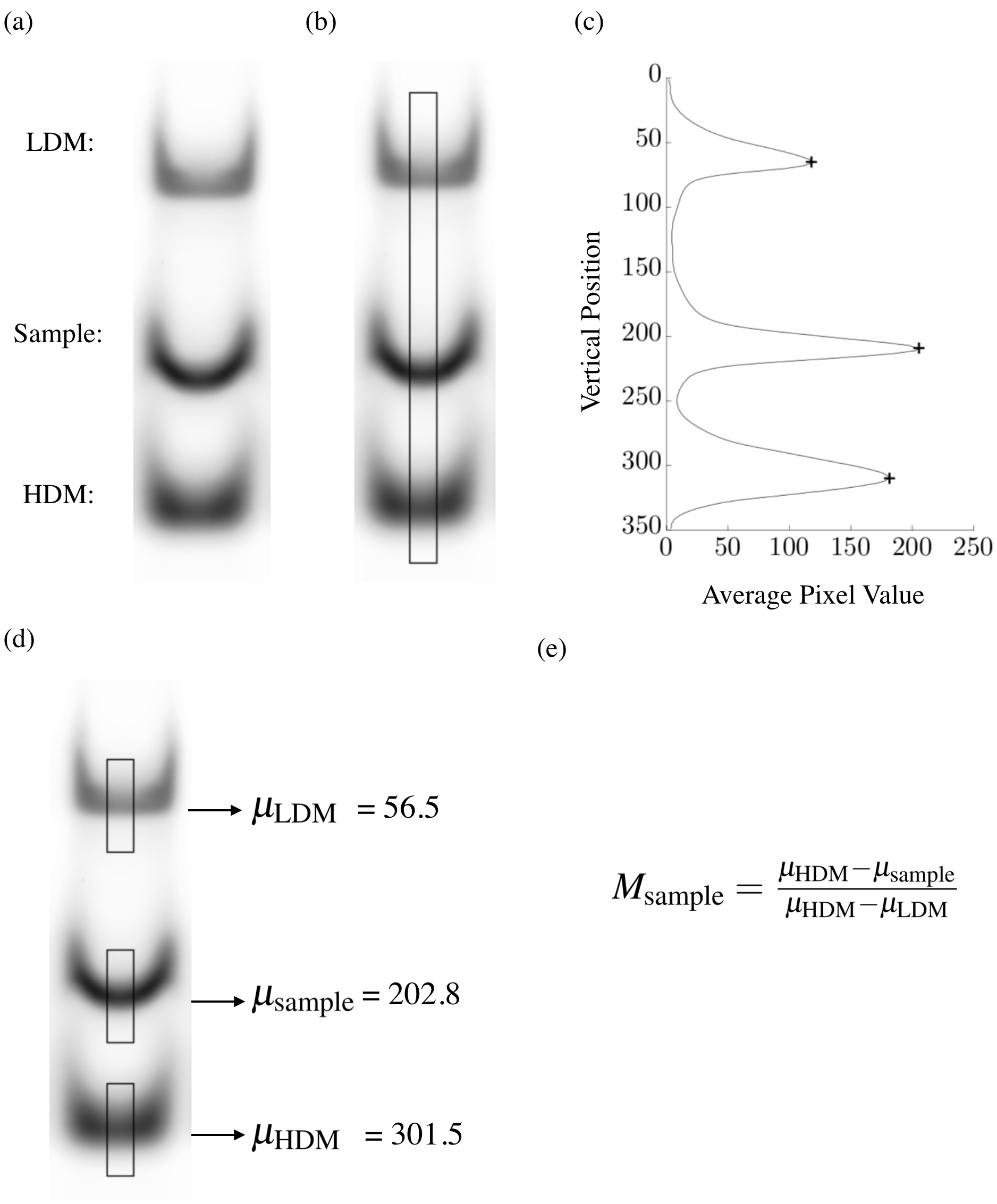}
	\caption{
	Process for measuring the relative electrophoretic mobility shifts of samples with the assistance of high mobility (HDM) and low mobility (LDM) reference markers. The process is semi-automated using a Fiji \cite{} macro with visual verification of boxes. (a) The user specifies the position of the lane containing the LDM, HDM and the sample. (b) The program draws a box through all bands in the lane with verification from the user. (c) Density maxima for each band are identified. (d) A box is drawn with a height of 70 pixels around each peak position and verified by the user. The centre of mass of each band is determined. (e) The relative electrophoretic mobility of the sample, $M_{\text{sample}}$, is calculated.
	}\label{fig::ExpMarker}
\end{figure}

The protocol used to measure relative electrophoretic mobilities of the samples is shown is Figure \ref{fig::ExpMarker}. Determination of binding strengths for T-motifs using PAGE requires accurate measurements of relative changes in electrophoretic mobility that are not dependent on the frequently-observed variations of the gel across lanes. To correct for differences between the positions of equivalent bands in different lanes, two reference markers were used: a high-mobility marker (HDM) and a low-mobility marker (LDM). Each marker is a duplex with a fluorescent label at the $5'$ end of one of its strands: sequences are given in Table \ref{table::marker}.

For each sample, all strands (including reference marker strands) were combined in a TAE/Mg$^{2+}$ buffer (40mM Tris base, 1mM EDTA-Na$_2$, 20mM acetic acid, 12.5mM MgCl$_2$, pH 8.5$\pm$0.2) and annealed in a thermocycler from 95 to 25$\degree$C at a rate of 1$\degree$C/min. Throughout our experiments, the concentration of loop duplex strands was kept constant at 0.2$\mu$M while the concentration of the sticky-end duplex was varied. The FAM-labelled marker strands were at 0.2$\mu$M and the unlabelled marker strands at 0.3$\mu$M. All gel images shown were recorded in the FAM fluorescence channel. Bands therefore correspond to the labelled loop duplex and (where present) the HDM and LDM marker duplexes.

Prior to loading onto gels, all samples were diluted with 0.25 volumes of 5$\times$ loading buffer (containing Bromophenol Blue and glycerol). All gels were 8\% 29:1 (mono:bis) acrylamide gels in TAE/Mg$^{2+}$, and run using the Mini-Protean Tetra Cell at 150V and 4$\degree$C for 90 minutes. The running buffer used was the same as that in the gels. 8$\mu$L of each sample with 2$\mu$L of 5$\times$ loading buffer was loaded per polyacrylamide gel lane.

To estimate the concentrations of strands in the gel environment, we assume that all strands are confined to the same constant volume during electrophoresis. The height of each observed band was estimated to be 1.5mm using the full width at half maximum value of the measured density profile. The dimensions of the gel lane (width 3.35mm, thickness 0.75mm) were used to estimate the occupied band volume $\sim$3.8$\mu$L. Since 8$\mu$L of sample was loaded into each lane, the strands are $\sim$2.1 times more concentrated within the gel than in the original sample. All concentrations given in this paper are corrected to reflect this change and are estimates of strand concentrations within the gel environment. The uncertainty in estimating the effective DNA concentration in a gel band creates an additional systematic error in our deduced binding affinities. A conservative limit of a factor of 2 on this uncertainty corresponds to a systematic error of 1.7 kJ mol$^{-1}$.

\begin{table}[h]
\resizebox{\linewidth}{!} {\begin{tabular}{| l | l | p{0.8\linewidth} |}
\hline
Marker & Strand & Sequence \\
\hline
HDM & Labelled & /6-FAM/AGTTTTAATCGCTGCCCA \\
HDM & Unlabelled & TGGGCAGCGATTAAAACT \\
LDM & Labelled & /6-FAM/CGGACGAGGCAGCTATGGTCGTGGGC-GATTTCAACATGACGTCAGTCAATACGGGGAG-AAGACCAGCGTCCG \\
LDM & Unlabelled & CGGACGCTGGTCTTCTCCCCGTATTGACTGAC-GTCATGTTGAAATCGCCCACGACCATAGCTGC-CTCGTCCG \\
\hline
\end{tabular}}
\caption[Table caption text]{Strand sequences of low mobility (LDM) and high mobility (HDM) reference markers.}
\label{table::marker}
\end{table}

%%%%%%%%%%%%%%%%%%%%%%%%%%%%%%%%%%%%%%%%%%%%%%%%%%%%%%%%%%%%
\subsection{Control Experiments}

Two sets of control experiments were performed on the $\mathbf{T}_{3'}^{5}$ T-motifs used in Figure 7.

\begin{figure}[h]
	\centering
		\includegraphics[trim=0cm 1cm 0cm 0cm, width=1.0\linewidth]{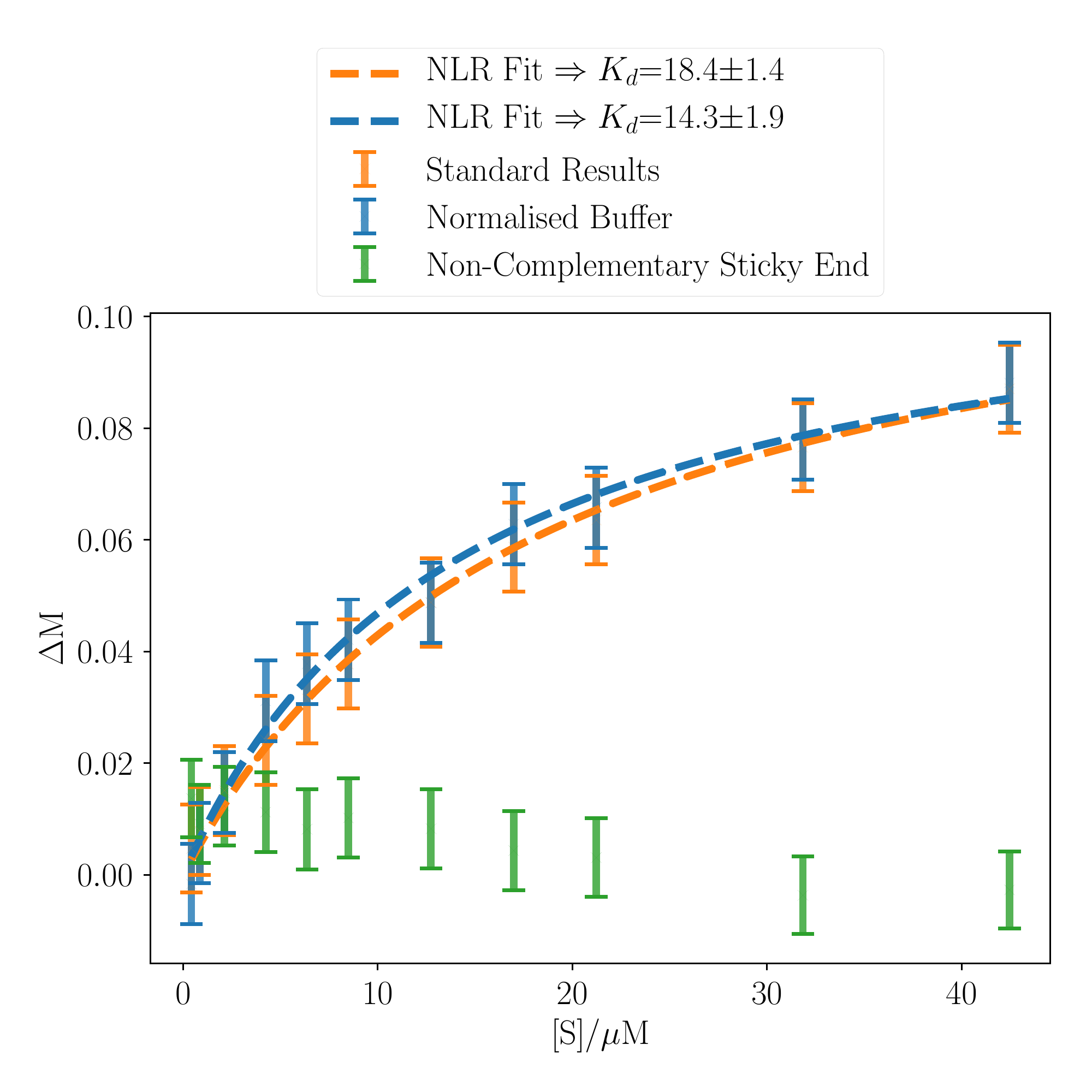}
	\caption{
		Control experiments on $\mathbf{T}_{3'}^{5}$. Changes in relative electrophoretic mobility for samples prepared with identical buffer compositions (blue) are compared with data presented in Figure 7 (orange). Data from control experiments in which the sticky end and bulge loop have non-complementary sequences are also shown (green). 
		}\label{fig::ExpControl}
\end{figure}

As a result of the difference between the TE storage buffer and the nominal composition of the  buffer in which the T-motifs and marker strands were annealed (TAE/Mg$^{2+}$), the precise composition of the annealing buffer depended on the concentration of sticky-end duplex strands. Controls were run with identical buffer compositions in all lanes to check that variations in buffer composition during annealing did not contribute significantly to the observed dependence of electrophoretic mobility on sticky-end duplex  concentration. As shown in Figure \ref{fig::ExpControl}, differences between the two sets of results were within experimental uncertainties.

A second set of control experiments were carried out with sticky-end sequences that are not complementary to the bulge-loops in order to exclude the possibility that the band shifts were due to non-specific interactions at high DNA concentrations. As shown in Figure \ref{fig::ExpControl}, in the presence of non-complementary linkers  there is no significant band shift ($\Delta M \sim 0$). We can therefore safely assume that the band shifts we observe are due to hybridization of the sticky end with bulge loops.

%%%%%%%%%%%%%%%%%%%%%%%%%%%%%%%%%%%%%%%%%%%%%%%%%%%%%%%%%%%%
\subsection{Effect of Rapid Dissociation} \label{app::gelshifts}

\begin{figure}[h]
	\centering
		\includegraphics[trim=0cm 0cm 0cm -1cm, width=1\linewidth]{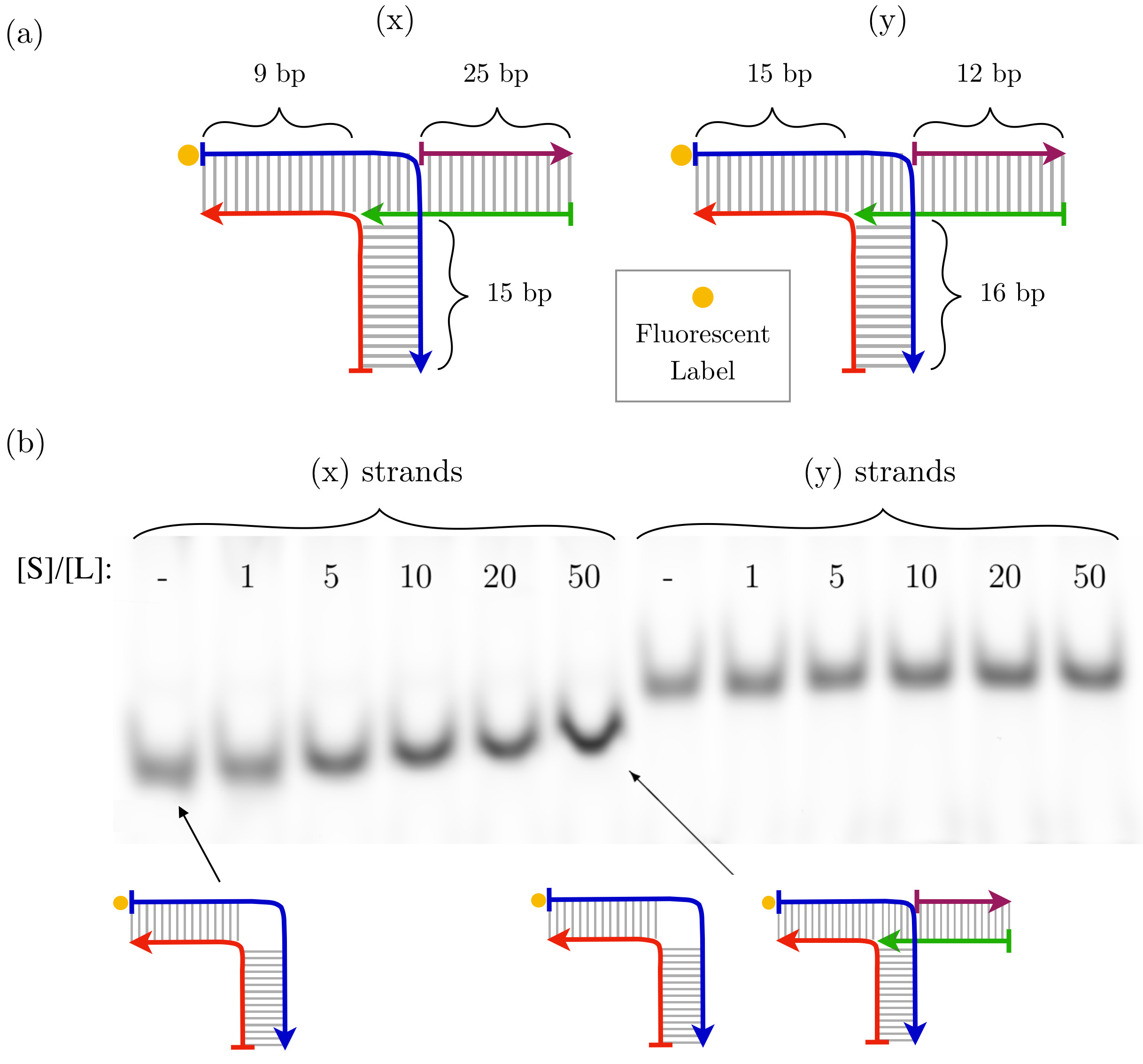}
	\caption{
	Gel electrophoresis experiments on 3$\prime$ T-motifs with 5-nucleotide bulge loops ($\mathbf{T}_{3'}^{5}$) with different arm lengths. (a) Two sets of strands were used: in (x), the loop duplex has similar dimensions and mobility to the sticky-end duplex; in y the sticky-end duplex migrates more quickly. (b) Gel lanes containing mixtures of the loop and sticky-end duplexes. The concentration of the loop duplex, [$L$]=0.42 $\mu$M, is kept the same across all lanes. The concentration of the sticky-end duplex, [$S$], is varied. A single band with mobility shift proportional to [S] is observed for (x) strands but no change in the mobility of the loop duplex is observed for (y) strands.
}\label{fig::ExpLength}
\end{figure}

Gel images presented in Figure \ref{fig::ExpLength} demonstrate that, when a loop duplex and sticky-end duplex with similar electrophoretic mobilities are combined and run in a gel, the mobility of the observed band (i.e. the band containing the labelled loop duplex) depends on the concentration of the sticky-end duplex. This band therefore contains both components and corresponds to the assembled T-motif. The dependence of its position on the concentration of the sticky-end duplex is consistent with co-migration and rapid equilibration between the individual loop and sticky-end duplexes and the T-motif. However, when the mobility difference between the loop duplex and sticky-end duplex is large, the difference between their mobilities causes them to separate in the gel and inhibits T-motif formation: in this case, the mobility of the band containing the loop duplex is independent of the concentration of sticky-end duplex. We conclude that the on- and off-rates for hybridisation of the bulge loop with the sticky-end are rapid in comparison to the time taken to run an electrophoresis gel.

\begin{figure}[h]
	\centering
		\includegraphics[trim=0cm 0cm 0cm -1cm, width=1\linewidth]{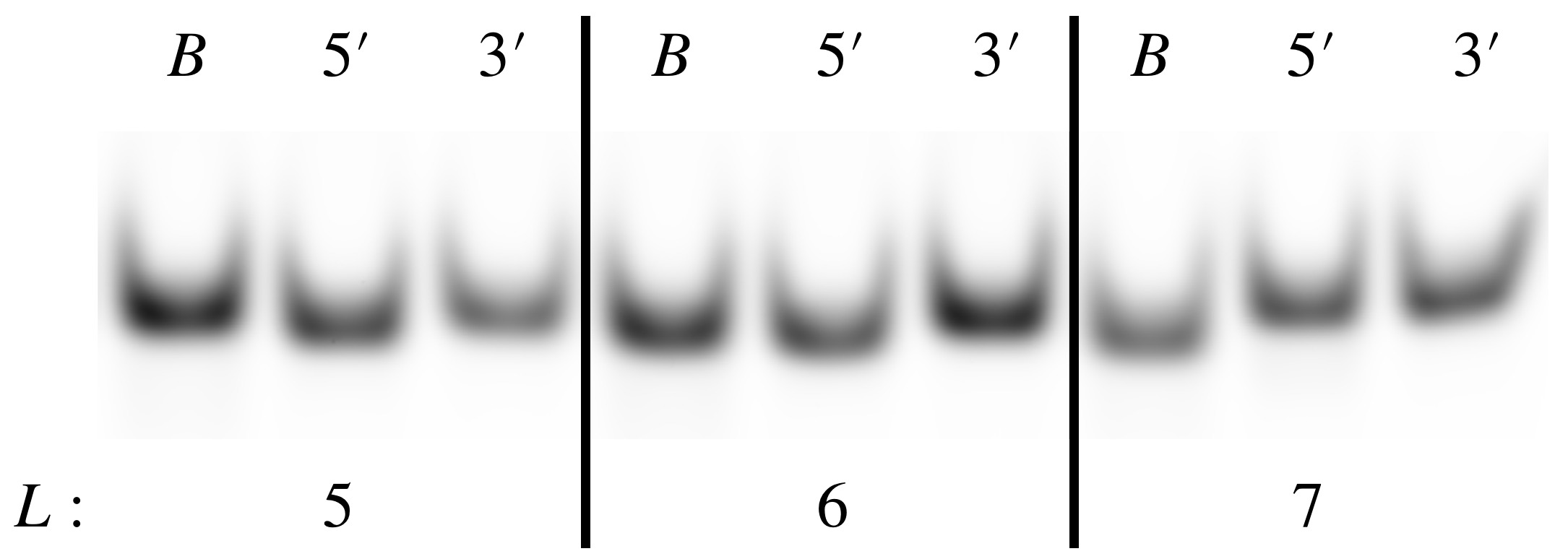}
	\caption{
	Mobilities of bulge-loop ($B$) and sticky-end (3$^\prime$, 5$^\prime$) duplexes that hybridize to form T-motifs shown in Figure \ref{fig::ExpSetup}.
	}\label{fig::ExpMob}
\end{figure}

To prevent separation of the bulge-loop and sticky-end duplexes within the gel, it is necessary that they migrate with similar mobilities. Figure \ref{fig::ExpMob} shows the differences in mobility between the bulge loop duplex and the (3$^\prime$ or 5$^\prime$) sticky-end duplex for loop lengths ($L$= 5,6,7 nt) used in this work. The mobility difference of the interacting duplexes for T-motifs with $L=7$ nt is greater than those with $L$=5 or 6 nt. Separation between the component duplexes during gel electrophoresis may result in an overestimation of $K_d$ for T-motifs with 7-nt loops.

%%%%%%%%%%%%%%%%%%%%%%%%%%%%%%%%%%%%%%%%%%%%%%%%%%%%%%%%%%%%
%%%%%%%%%%%%%%%%%%%%%%%%%%%%%%%%%%%%%%%%%%%%%%%%%%%%%%%%%%%%
%\FloatBarrier
%\newpage
%\newpage
\section{Additional Results}

%%%%%%%%%%%%%%%%%%%%%%%%%%%%%%%%%%%%%%%%%%%%%%%%%%%%%%%%%%%%
\subsection{Gels and Binding Curves}

In Figure 8 of the main text, experimentally determined and calculated binding affinities for T-motifs $\mathbf{T}_{\mathcal{P}}^{L}$, with bulge loop sizes $L\in\{5,6,7\}$nt and sticky-end polarities $\mathcal{P}\in\{5',3'\}$, are compared. Here, we present the gel electrophoresis results from which the experimental binding affinities were derived.

\begin{figure*}[h]
	\centering
		\includegraphics[trim=0cm 0cm 0cm 0cm, width=0.85\textwidth]{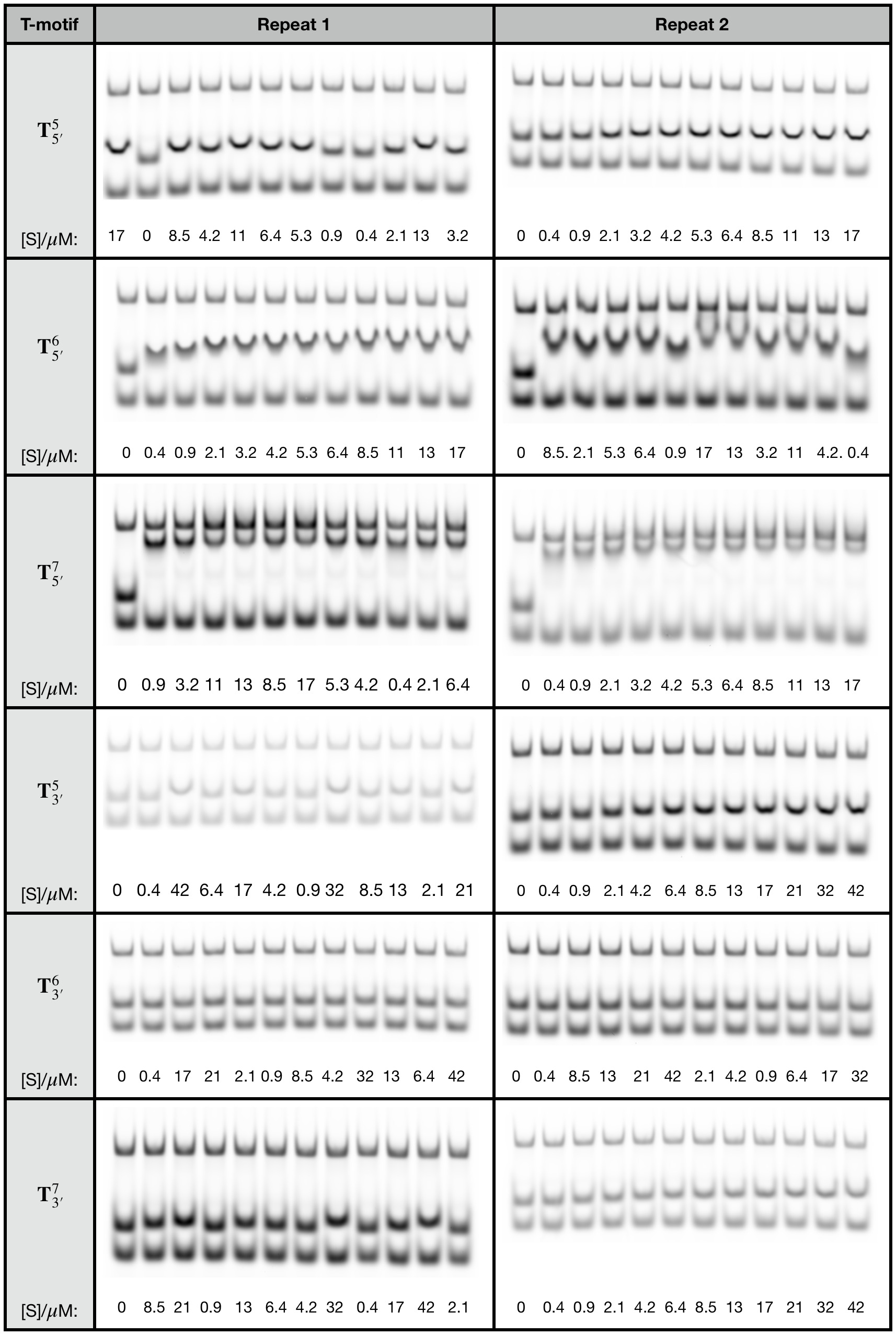}
	\caption{
	Gel electrophoresis results used to measure $K_d$ values. In each experiment, all lanes contain the bulge-loop duplex at a concentration of 0.42$\mu$M. The concentration of the sticky-end duplex, [S], varies across lanes as indicated below each figure. The experiments were carried out twice for each T-motif with different gel lane orders to minimise the errors arising from variability between gel lanes. %The lane order for repeat 2 of $\mathbf{T}_{5'}^{5}$, $\mathbf{T}_{5'}^{7}$, $\mathbf{T}_{3'}^{7}$ and $\mathbf{T}_{3'}^{7}$, and repeat 1 of $\mathbf{T}_{5'}^{6}$ follow increasing sticky-end duplex concentration. In all other experiments, the sticky-end duplex concentration varies across gel lanes in known random orders.
	}\label{fig::ExpGels}
\end{figure*}

As shown in Figure \ref{fig::ExpGels}, we performed two independent gel electrophoresis experiments for each T-motif. In each experiment, the concentration of the bulge-loop duplex was kept constant across all gel lanes at 0.42$\mu$M. The concentration of the sticky-end duplex, [S], was varied across gel lanes and the shift in relative electrophoretic mobility due to T-motif formation, $M$([S]), was recorded with the aid of reference markers. The distribution of different concentrations [S] across gel lanes was changed in individual repetitions of each experiment to minimise the errors arising from variability between gel lanes (for example due to the centre of the gel becoming hotter during electrophoresis).

\begin{figure*}[h]
	\centering
		\includegraphics[trim=2cm 0cm 2cm 0cm, width=1\textwidth]{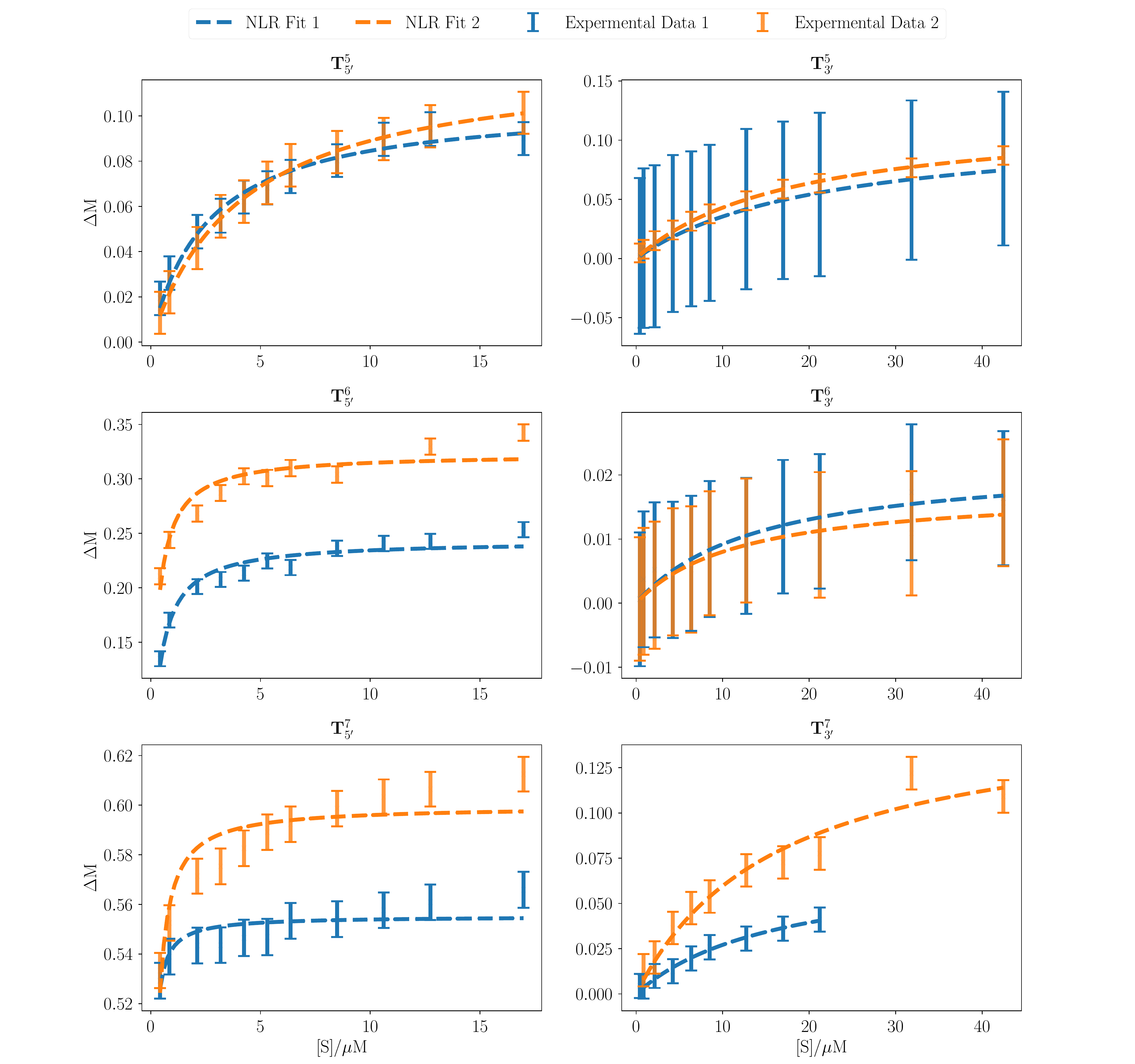}
	\caption{Binding curves constructed from gel electrophoresis data derived from experiments shown in Figure \ref{fig::ExpGels}. Non-linear regression fits are used to determine $K_d$ values shown in Table \ref{table::kd}.}\label{fig::ExpCurves}
\end{figure*}

Figure \ref{fig::ExpCurves} show the binding curves constructed from the gel data: $\Delta M(\text{[S]}) = M(\text{[S]}) - M(\text{[S]=0})$ and the error bars reflect the uncertainty in estimating the peak using the protocol outlined in Figure \ref{fig::ExpMarker}. Dissociation constants were extracted from each experiment by fitting our data points to the function $\Delta M(\text{[S]})=\Delta M_{max}\text{[S]}/(K_d+\text{[S]})$: fitted $K_d$ values are tabulated in Table \ref{table::kd}.

\begin{table*}[b]
\resizebox{0.35\linewidth}{!} {
\begin{tabular}{| l | l | l | l |}
\hline
$L$ & Repeat & $\mathcal{P}$=5$^\prime$ & $\mathcal{P}$=3$^\prime$ \\
\hline
5 nt & 1 & $2.59\pm0.32$ $\mu$M & $21.81\pm2.22$ $\mu$M\\ 
   & 2 & $4.14\pm0.15$ $\mu$M & $18.42\pm1.37$ $\mu$M\\ 
\hline
6 nt & 1 & $0.38\pm0.05$ $\mu$M & $14.44\pm4.20$ $\mu$M\\ 
   & 2 & $0.27\pm0.05$ $\mu$M & $12.19\pm3.21$ $\mu$M\\ 
\hline
7 nt & 1 & $0.02\pm0.01$ $\mu$M & $16.67\pm3.29$ $\mu$M\\  
   & 2 & $0.06\pm0.01$ $\mu$M & $16.55\pm4.42$ $\mu$M\\ 
\hline
\end{tabular}}
\caption[Table caption text]{Dissociation constant ($K_d$) values for T-motifs with bulge loop size, $L$, and sticky-end polarity $\mathcal{P}$.}
\label{table::kd}
\end{table*}

%%%%%%%%%%%%%%%%%%%%%%%%%%%%%%%%%%%%%%%%%%%%%%%%%%%%%%%%%%%%
\subsection{Thermodynamic Results}

\begin{figure*}[h]
	\centering
		\includegraphics[width=1\linewidth]{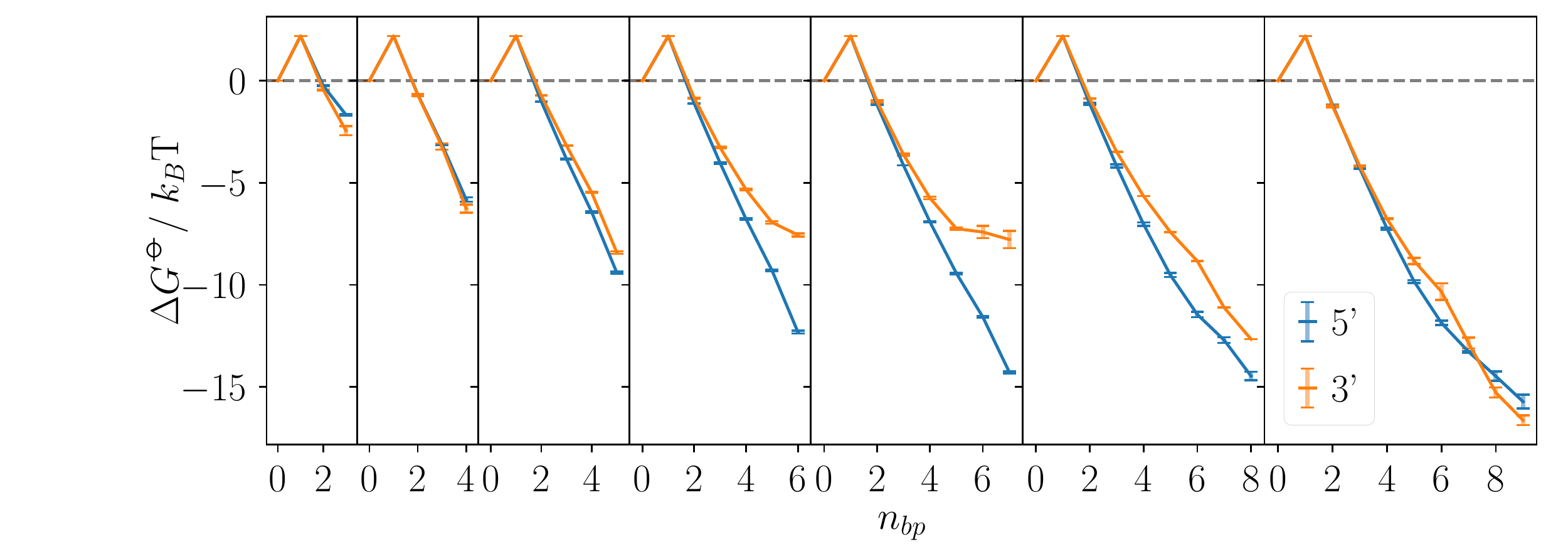}
	\caption{Free-energy profiles for all of the T-motif variations simulated using oxDNA.}\label{fig::all_dG}
\end{figure*}

Figure \ref{fig::all_dG} shows the free-energy profiles (Gibbs free energy as a function of the number of base pairs formed between bulge and sticky end) of all T-motifs simulated using oxDNA. Selected cases are presented in Figure 6 of the main text.

%%%%%%%%%%%%%%%%%%%%%%%%%%%%%%%%%%%%%%%%%%%%%%%%%%%%%%%%%%%%
\subsection{Additional Stacking Results}

\begin{figure*}[h]
	\centering
		\includegraphics[width=0.75\linewidth]{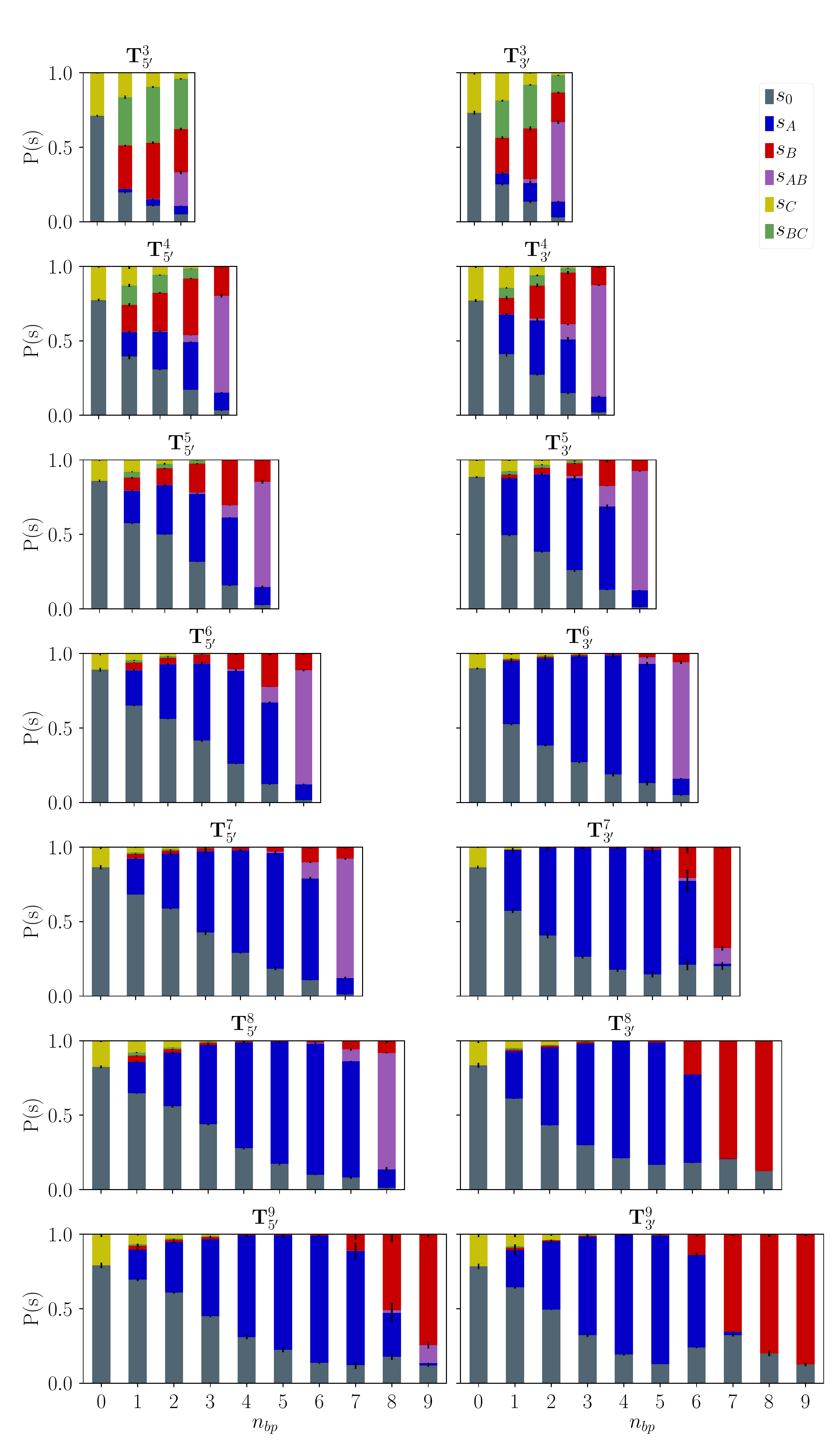}
	\caption{Probabilities of different stacking configurations for all T-motifs simulated. See main text for definitions.}\label{fig::all_stack}
\end{figure*}

Figure \ref{fig::all_stack} shows the probabilities of coaxial stacking states for equilibrium states as a function of the number of base-pairs, $n_{bp}$, for all T-motifs simulated using oxDNA. Special cases are discussed in the main text (see Figure 2).

%%%%%%%%%%%%%%%%%%%%%%%%%%%%%%%%%%%%%%%%%%%%%%%%%%%%%%%%%%%%
\subsection{Additional Geometric Results}\label{app::Geometry}

\begin{figure*}[h]
	\centering
		\includegraphics[width=1\textwidth]{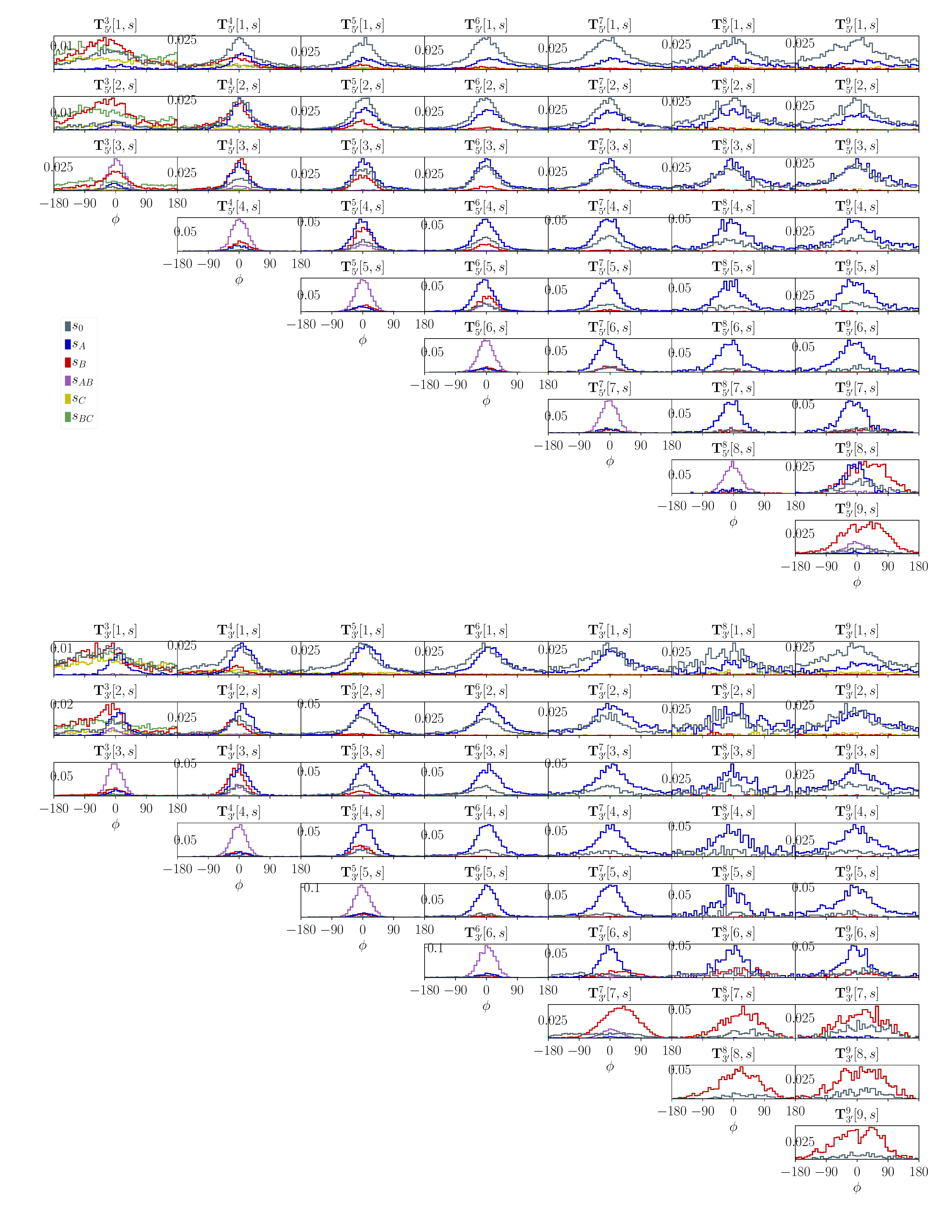}
	\caption{Probability distributions for the dihedral angle, $\phi$, defined in the scheme presented in Figure 1.}\label{fig::Phi}
\end{figure*}

\begin{figure*}[h]
	\centering
		\includegraphics[width=1\textwidth]{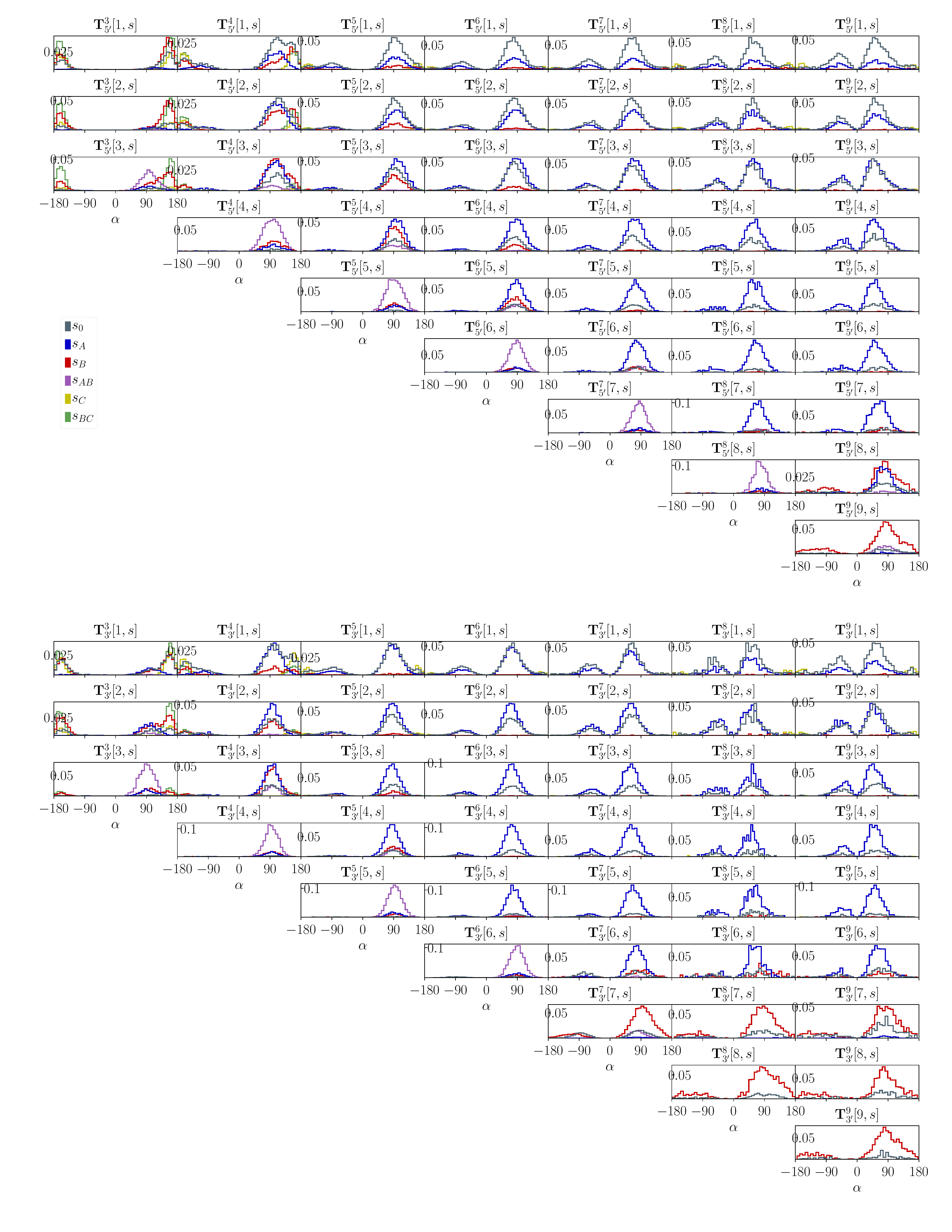}
	\caption{Probability distributions for angle $\alpha$ as defined in the scheme presented in Figure 1.}\label{fig::Alpha}
\end{figure*}

\begin{figure*}[h]
	\centering
		\includegraphics[width=1\textwidth]{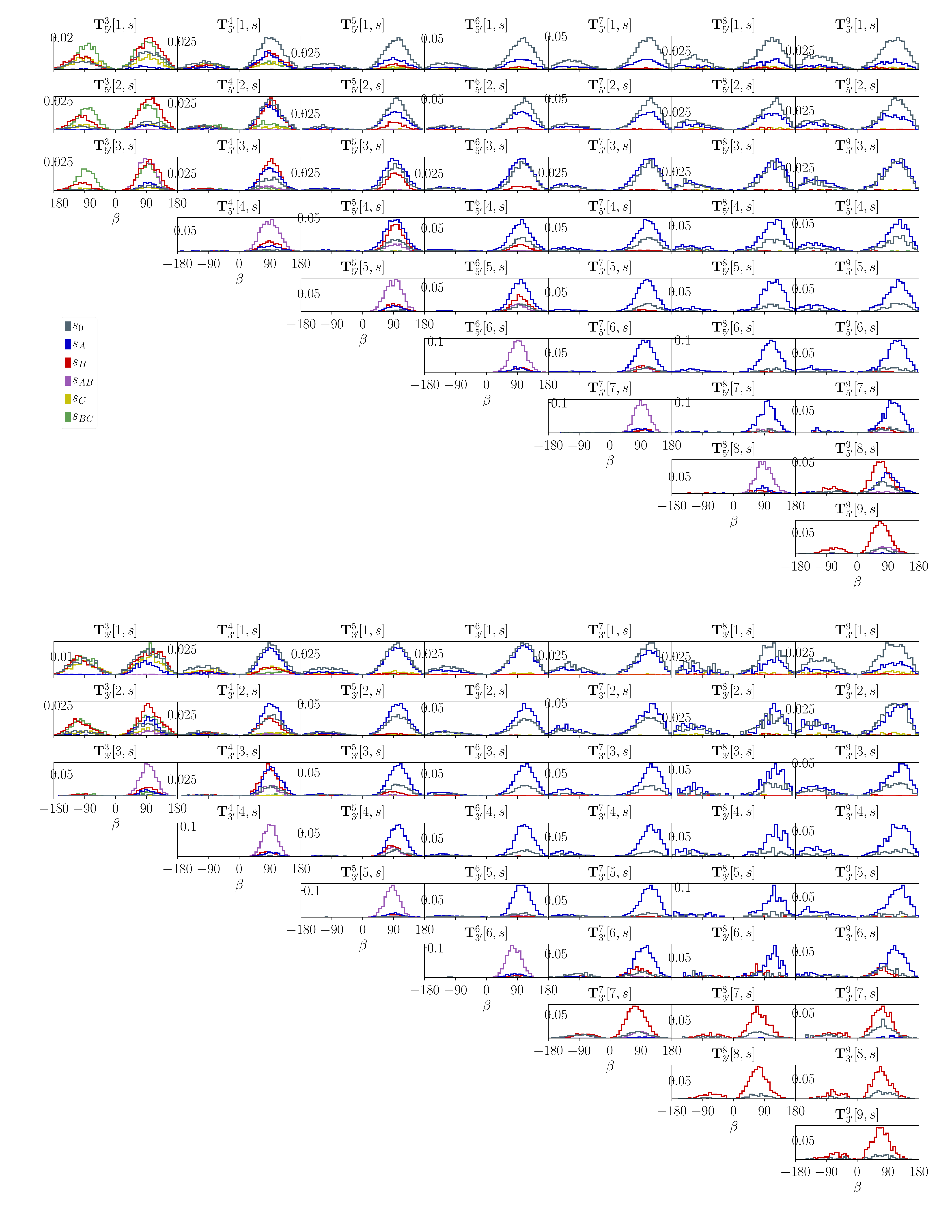}
	\caption{Probability distributions for angle $\beta$ as defined in the scheme presented in Figure 1.}\label{fig::Beta}
\end{figure*}

\begin{figure*}[h]
	\centering
		\includegraphics[width=1\textwidth]{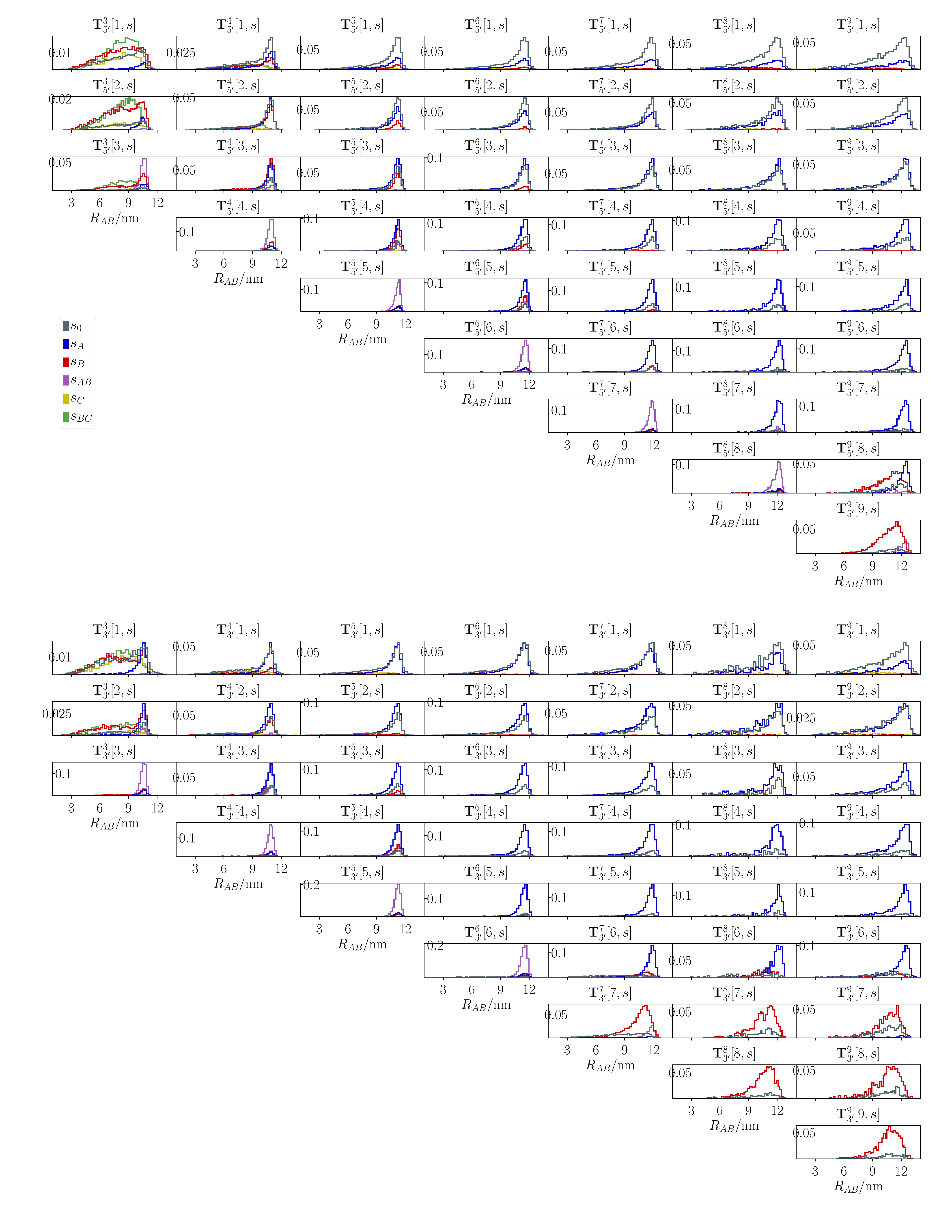}
	\caption{Probability distributions for the end-to-end distance, $R_{AB}$, between the two ends of the nicked duplex. See Figure 1 for schematic.}\label{fig::R_AB}
\end{figure*}

\begin{figure*}[h]
	\centering
		\includegraphics[width=1\textwidth]{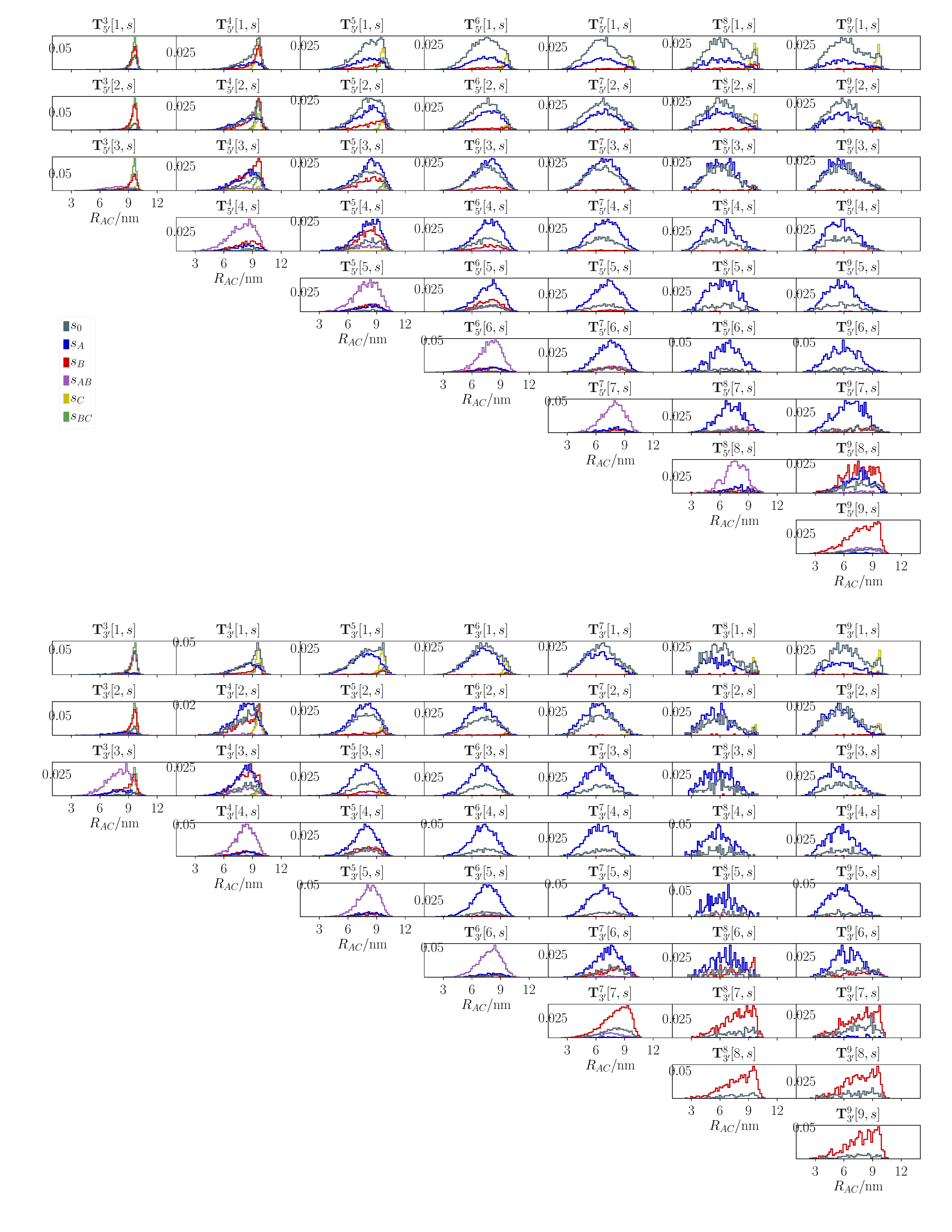}
	\caption{Probability distributions for the end-to-end distance, $R_{AC}$, between the two ends of the loop duplex. See Figure 1 for schematic.}\label{fig::R_AC}
\end{figure*}

\begin{figure*}[h]
	\centering
		\includegraphics[width=1\textwidth]{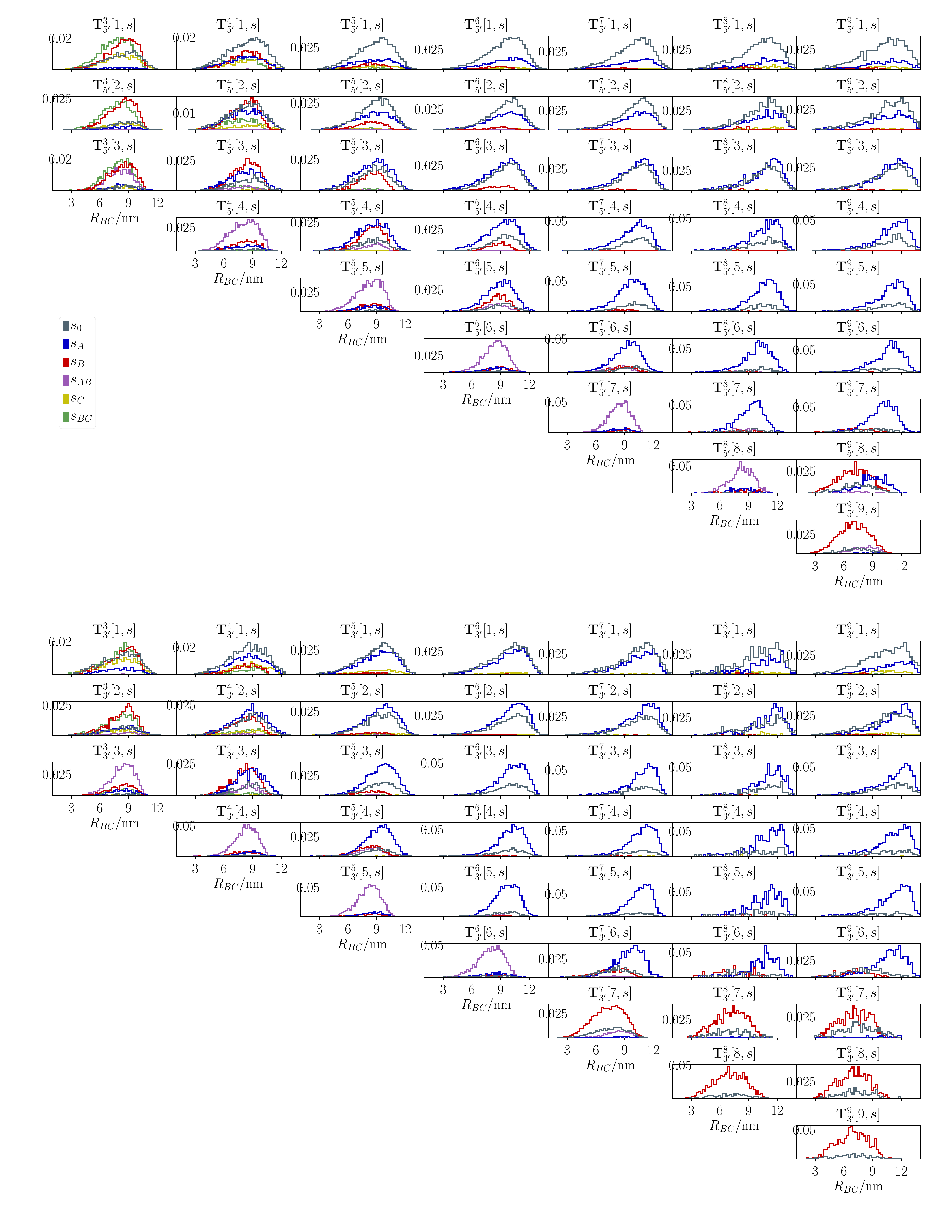}
	\caption{Probability distributions for the end-to-end distance, $R_{BC}$, between the end of sticky end duplex and the hanging arm of the T-motif. See Figure 1 for schematic.}\label{fig::R_BC}
\end{figure*}

\begin{figure*}[h]
	\centering
		\includegraphics[width=1\textwidth]{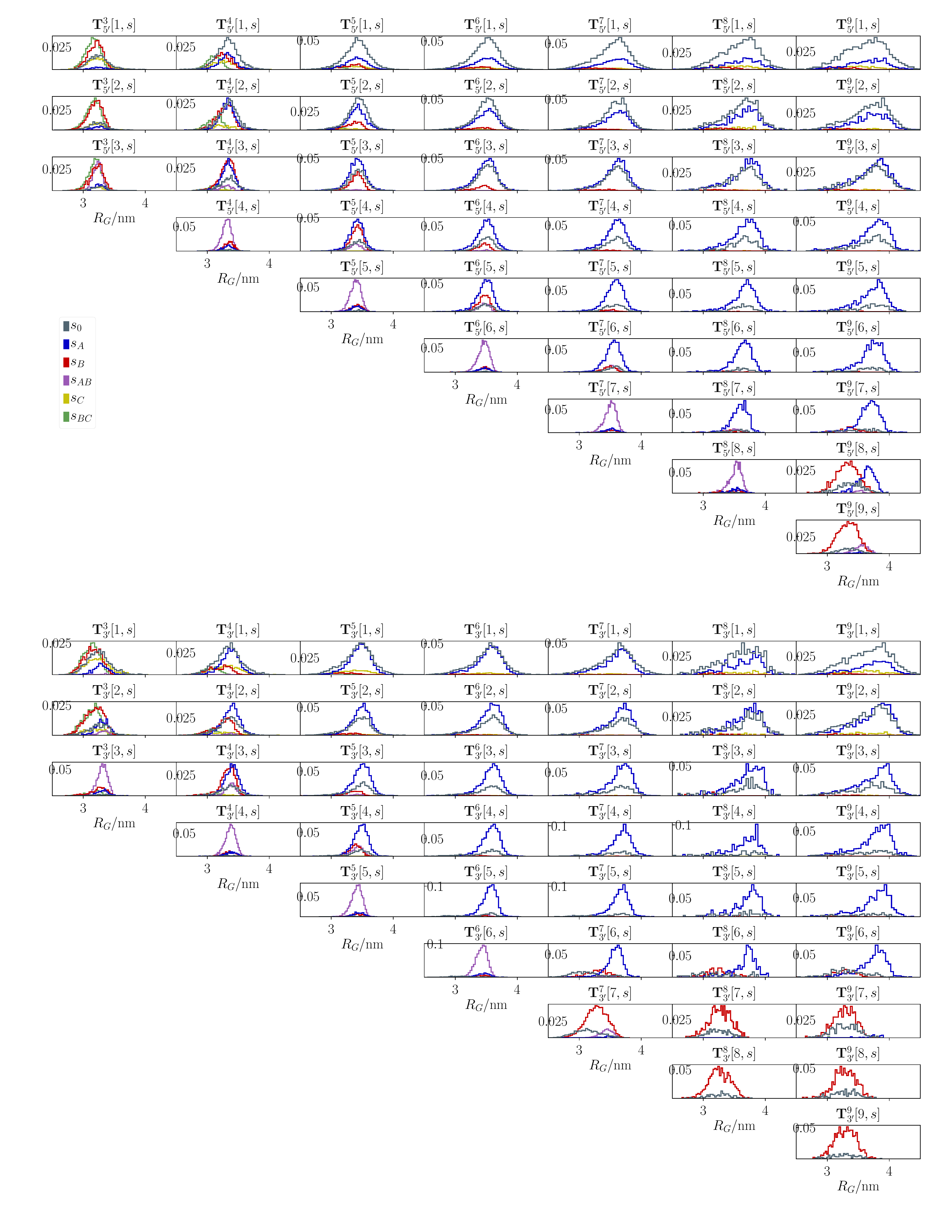}
	\caption{Probability distributions for the radius of gyration, $R_G$, shown for all T-motifs simulated using oxDNA.}\label{fig::R_G}
\end{figure*}

\begin{figure*}[h]
	\centering
		\includegraphics[width=1\textwidth]{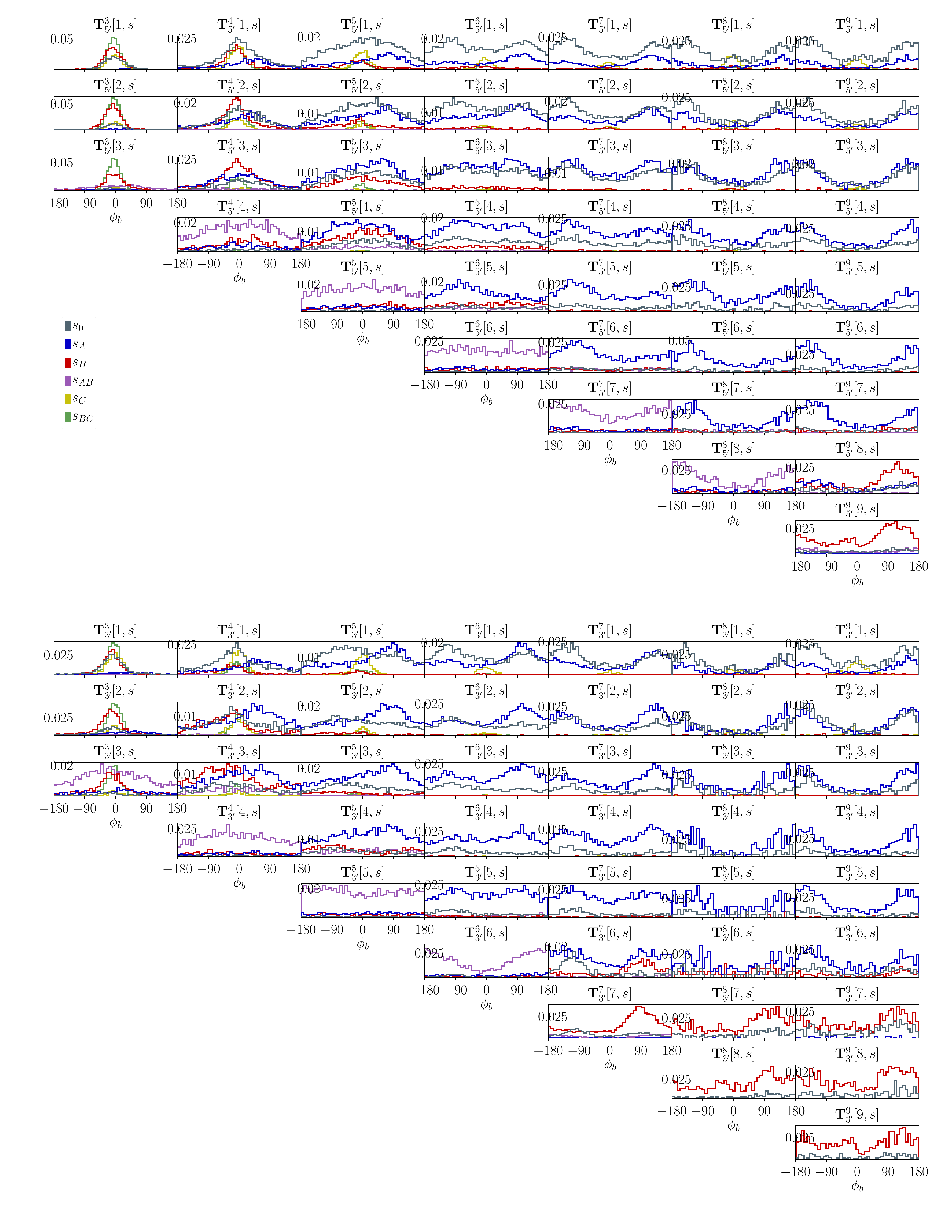}
	\caption{Probability distributions for the dihedral angle, $\phi_b$, between $\bold{a} \times \bold{b}$ and $\bold{b} \times \bold{c}$, taken to be positive for $[(\bold{a} \times \bold{b}) \times (\bold{b} \times \bold{c})] \cdot \bold{b} < 0$. This dihedral angle is better suited to configurations with stacking states $s_C$ and $s_{BC}$, in which the original loop duplex is extended.}\label{fig::Phi_b}
\end{figure*}

Figures \ref{fig::Phi}, \ref{fig::Alpha} and \ref{fig::Beta} show the probability distributions of angles $\phi$, $\alpha$ and $\beta$ for all simulated T-motifs. Figures \ref{fig::R_AB}, \ref{fig::R_AC} and \ref{fig::R_BC} show the probability distributions of end-to-end distances $R_{AB}$, $R_{AC}$ and $R_{BC}$, which measure the distances between the three ends of the T-motif arms, for all simulated T-motifs. For this purpose, the ends of the arms were defined to be the centres of mass of the two complementary nucleotides at the ends of each duplex arm. Figure \ref{fig::R_G} shows the probability distribution for the radius of gyration for each T-motif. Figure \ref{fig::Phi_b} shows the probability distribution of dihedral angle $\phi_b$ between the two planes with normal vectors $\bold{a} \times \bold{b}$ and $\bold{b} \times \bold{c}$. This dihedral angle is more suitable for configurations in stacking states $s_C$ and $s_{BC}$ in which the structure does not resemble a T-motif. In such cases, $\phi$ follows an approximately uniform distribution, the distribution of $\alpha$ is bimodal with peaks at $\pm 170 \degree$, and $\phi_b$ is unimodal, following roughly the same distribution of $\phi$ for stacking class $s_{AB}$.

%\begin{figure*}[h]
%	\centering
%		\includegraphics[width=1\textwidth]{Phi_a.pdf}
%	\caption{}\label{fig::Phi_a}
%\end{figure*}

%%%%%%%%%%%%%%%%%%%%%%%%%%%%%%%%%%%%%%%%%%%%%%%%%%%%%%%%%%%%
%%%%%%%%%%%%%%%%%%%%%%%%%%%%%%%%%%%%%%%%%%%%%%%%%%%%%%%%%%%%
\section{References}\label{app::ref}
\bibliography{bibliography}{}